\documentclass[letterpaper, 11pt]{article}

\usepackage{mathtools}
\usepackage{array}
\usepackage{enumitem}
\usepackage{amsmath,amsfonts,amssymb,bm}
\usepackage{titlesec}
\usepackage{pgfgantt,rotating}
\usepackage{float}
\usepackage{graphicx}
\usepackage{sidecap}
\usepackage{wrapfig}
\usepackage{bold-extra}
\usepackage{authblk}
\usepackage{appendix}
\usepackage{makecell}
\usepackage{dirtytalk}
\usepackage{placeins}
\usepackage{cite}

\usepackage[margin=1in]{geometry}

\newcommand{\noin}{\noindent}
\newcommand{\beq}{\begin{equation}}
\newcommand{\eeq}{\end{equation}}
\newcommand{\bgqar}{\begin{eqnarray}}
\newcommand{\enqar}{\end{eqnarray}}
\newcommand{\bgqarn}{\begin{eqnarray*}}
\newcommand{\enqarn}{\end{eqnarray*}}
\newcommand{\bgary}{\begin{array}}
\newcommand{\enary}{\end{array}}
\newcommand{\df}{:=}
\newcommand{\bld}[1]{\mbox{\boldmath $#1$}}

\graphicspath{{figures/}}

\title{Stochastic Identification-based Active Sensing Acousto-Ultrasound SHM Using Stationary Time Series Models}

\author{Shabbir Ahmed}
\author{Fotis Kopsaftopoulos\footnote{Corresponding author.}}

\affil{\small Intelligent Structural Systems Laboratory (ISSL) \\ Department of Mechanical, Aerospace and Nuclear Engineering \\ Rensselaer Polytechnic Institute, Troy, NY, USA \\ Email: \{ahmeds6,kopsaf\}@rpi.edu }

\date{\today}

\begin{document}

\maketitle


\begin{abstract}

In this work, a probabilistic damage detection and identification scheme using stochastic time series models in the context of acousto-ultrasound guided wave-based SHM is proposed, and its performance is assessed experimentally. In order to simplify the damage detection and identification process, model parameters are modified based on the singular value decomposition (SVD) as well as the principal component analysis (PCA)-based truncation approach. The modified model parameters are then used to estimate a statistical characteristic quantity that follows a chi-squared distribution. A probabilistic threshold is used instead of a user-defined margin to facilitate automatic damage detection. The method's effectiveness is assessed via multiple experiments using both metallic and composite coupons and under various damage scenarios using damage intersecting and damage non-intersecting paths. The results of the study confirm the high potential and effectiveness of the stochastic time series methods for guided wave-based damage detection and identification in a potentially automated way.


\end{abstract}


\clearpage

\section*{Important Conventions and Symbols}

\noin Definition is indicated by $\df$. Matrix transposition is indicated by the superscript $T$.

\noin Bold-face upper/lower case symbols designate matrix/column-vector quantities, respectively.
 
\noin A functional argument in parentheses designates function of a real variable; for instance $P(x)$ is a function of the real variable $x$.

\noin A functional argument in brackets designates function of an integer variable; for instance $x[t]$ is a function of normalized discrete time $(t=1,2,\ldots)$. The conversion from discrete normalized time to analog time is based on $(t-1)T_s$, with $T_s$ designating the sampling period.

\noin A hat designates estimator/estimate; for instance $\widehat{\bld{\theta}}$ is an estimator/estimate of $\bld{\theta}$.


\section*{Acronyms}

\noin\begin{tabular}{lcl} 
AIC  & : & Akaike information criteria \\
AR   & : & Autoregressive  \\
ARMA & : & Autoregressive moving average \\
ARMAX  & : & Autoregressive moving average with exogenous excitation  \\

BIC  & : & Bayesian information criterion  \\

FEM  & : & Finite element model \\
FRF  & : & Frequency response function  \\
iid  & : & identically independently distributed  \\
MA   & : & Moving average \\
OLS  & : & Ordinary least squares  \\
PE   & : & Prediction error  \\
PSD    & : & Power spectral density \\
PZT  & : & Lead zirconate titanate \\
RSS  & : & Residual sum of squares  \\
SHM  & : & Structural health monitoring  \\
SSS  & : & Signal sum of squares  \\
WLS  & : & Weighted least squares  \\

\end{tabular} 

\newpage\pagebreak 

\tableofcontents 


\section{Introduction} \label{sec:intro}

Structural health monitoring (SHM) refers to the process of damage diagnosis within a structure without affecting their integrity or performance \cite{Farrar-Worden07}. It ensures enhanced reliability and increased safety of a structure. An SHM process involves automatic extraction of damage-sensitive features or quantities from a series of periodic measurements coming from an array of permanently installed sensors on a structure/system and performing statistical analysis of these quantities to establish the current structural state of the system. With the recent emphasis on cyber-physical systems paradigm, incorporating an SHM system in the civil, mechanical, and aerospace structures have become essential to ensure safe operation of the safety-critical parts and to automatically inspect for flaws during in-service use, that is, to impart diagnostic and self-sensing capabilities \cite{Kopsaftopoulos-etal18-DDDAS}. This approach will obviate the need for periodic or routine maintenance and provide room for shifting towards a condition-based monitoring approach, ensuring safety with a significant amount of time and money saved.

An SHM approach may be broadly classified as a local or global approach \cite{kopsaftopoulos2010vibration}. For the local monitoring of a structure, a wide variety of methods are available based on ultrasound \cite{Ihn-Chang08}, eddy current \cite{sodano2007development}, acoustic emission \cite{dubuc2020acoustic}, and thermal field principles \cite{reilly2019temperature}. On the other hand, for the global monitoring of a structure, the vibration-based family of methods is usually employed, which utilizes random excitation/response signal, statistical model building, and statistical decision making to infer the current health state of the structure/system \cite{Fassois01,fassois2007time}. The fundamental premise is that small changes (cause or damage) in a structure force its vibration response to be changed (effect), which may be detected and associated with a specific cause (damage type) \cite{kopsaftopoulos2010vibration,avendano2017gaussian,Kopsaftopoulos-Fassois13}. However, a tiny change or damage may not be manifested in the vibration response signal, thus may remain undetected. Although vibration-based methods can detect global changes and are more robust in the face of environmental and operational conditions (EOC), they may be less sensitive to the local effects.

On the contrary, some of the local active sensing SHM approaches, such as ultrasonic guided wave-based methods, which use Lamb wave propagation within a thin structure, are extremely sensitive to local changes and can detect tiny changes or damages within a structure or on the surface \cite{roy2014novel}. These waves can easily be generated/detected by piezoelectric transducers in the form of an applied strain/voltage \cite{Ihn-Chang04a,Ihn-Chang04b}. The most widely used method for damage detection using guided waves is the concept of damage/health indices/indicators(D/HI). The idea is that the features of the signal from an unknown structural state are compared to that coming from the healthy structure \cite{amer2021statistical,Amer-etal21a,song2008smart,wang2017piezoelectric,tibaduiza2016structural,lize2018optimal,qiu2019baseline}. These features may be based on the specific mode wave packets of the guided wave signal, the amplitude/magnitude, or the energy content of the signal. These conventional DI-based approaches have been extensively used in the literature for their simplicity, damage/no-damage binary detection paradigm, and ease of decision making.

However, the selection of an appropriate DI equation may influence the performance of the damage detection algorithm of an SHM system. The DI should be chosen in such a way that it is highly sensitive to the growth, size, and orientation of the damage and less sensitive to other external factors such as material property variation, the effect of adhesive, and PZT placement \cite{janapati2016damage}. It was found that the sensor locations with respect to the direction of the crack propagation may significantly influence the DI evolution if the environmental and boundary conditions remain unaltered. To eliminate the limitations of conventional time-domain DIs, frequency-domain DIs \cite{jin2019monitoring,xu2013active} or a combination of time-frequency (mixed-domain) DIs \cite{hua2020time} have been proposed. By capitalizing on some non-linear features of the guided wave signals and thus formulating the non-linear DIs, barely visible fatigue cracks can be detected \cite{su2014acousto}. It has been reported that wavelet entropy-based detection and localization algorithm may perform better than the conventional time-of-arrival-based algorithm under the presence of damage \cite{ibanez2015detection,rojas2015damage}. The above-mentioned methods are deterministic and do not account for uncertainty nor allow for the extraction of the appropriate confidence intervals for damage detection \cite{amer2021statistical}. Recently, steps have been taken towards formulating probabilistic DIs using the Gaussian mixture model and other probabilistic and statistical tools \cite{qiu2017adaptive,haynes2013statistically,su2004lamb,peng2013novel,flynn2011maximum}. However, in the face of the signal's stochasticity and varying environmental and operational states, these DI-based methods may become ineffective \cite{Ahmed-Kopsaftopoulos19a,Ahmed-Kopsaftopoulos19b,Amer-etal20}. In order to circumvent these difficulties, the use of stochastic time series models may be an appropriate option \cite{ahmed2022time,wave21,DDDAS_shabbir}.

Stochastic time series models are probabilistic models and are usually used in the context of the vibration-based damage diagnosis process \cite{Kopsaftopoulos-Fassois08,Kopsaftopoulos-Fassois07,Kopsaftopoulos-Fassois06,Kopsaftopoulos-Fassois06ewshm,Kopsaftopoulos-etal10,Kopsaftopoulos-Fassois13,kopsaftopoulos2010vibration,kopsaftopoulos2018stochastic}. These are data-based rather than physics-based models, automatically accounting for uncertainty in the system, and have provision for statistical decision making. In addition, these models are easy to identify, have compact representation, and the same models can be used for both metallic and composite materials. In reality, ultrasonic guided waves are non-stationary signals as their variance change over time, and in order to properly represent them, time-varying parametric time series models are required \cite{wave21,Spiridonakos-Fassois2014,spiridonakos2013fs,poulimenos2009output}. However, in the context of damage diagnosis, more straightforward stationary representations (time-invariant) can also be used. Time-invariant parametric methods are based upon auto-regressive moving average (ARMA) or related types and their extensions. These methods have attracted considerable attention and have been used extensively in analyzing low-frequency vibration responses excited by a random white noise actuation. These methods remain unexplored in the context of ultrasonic guided wave-based SHM, which traditionally exploits a narrow-band high-frequency excitation.

The main objective of this work is the investigation and performance assessment of a novel damage detection and identification scheme using stationary stochastic time series models such as autoregressive (AR) models in the context of the guided wave-based SHM. Traditional deterministic tone-burst actuation was used in the present study to compare with the traditional DI-based damage detection performance. According to the authors' best of knowledge, this is the first study that explores the use of stochastic time series models in the context of active sensing acousto-ultrasound-based SHM. The main novel aspects of this study include:

\begin{itemize}
    \item Introduction of stochastic time series models for damage detection and identification in active sensing guided wave-based SHM,
    \item Use the complete signal, including the reflection part rather than using only the $S_0$ or $A_0$ mode or the non-reflecting part of the signal,
    \item Use of SVD and PCA on model parameters for reducing the dimensionality of the parameters,
    \item Extraction of confidence bounds of the model parameters and use those confidence bounds to formulate a probabilistic damage diagnosis scheme (damage detection and identification),
    \item Application of the proposed method in two different types of coupons: an aluminum plate and a composite plate. That is, the same method is applicable for metals and composites alike,
    \item Possibility of automating the whole damage diagnosis scheme without the need for any user intervention or user expertise.
    
\end{itemize}

The remainder of this article is organized as follows: Section 2 introduces the stochastic modeling of guided wave signal using the AR model and the process of AR model identification, Section 3 presents the theory of the probabilistic damage diagnosis scheme utilizing AR model parameters, Section 4 presents the implementation method of the proposed probabilistic damage detection and identification framework. Then the experimental setup, path selection, results, and discussion are presented for aluminum and composite plate in Section 5 and 6, respectively. Finally, Section 7 summarizes this study and comments on future research directions.

\section{Guided Wave Signal~Representation}\label{sec2}

Guided waves are inherently non-stationary due to their time-dependent (evolutionary) characteristics and are heavily influenced by environmental and operating conditions. However, in the case of weak non-stationarity, an auto-regressive (AR) model can be used to represent a guided wave signal for damage detection and identification.

An AR$(n)$ model is of the following form \cite{Ljung99}:
\begin{equation}
    y[t] + \sum_{i=1}^{n} a_i \cdot y[t-i] =  e[t] \qquad e[t] \sim \, \mbox{iid} \, \mathcal{N} \bigl( 0,\sigma^2_e \bigr) \label{eq:ar-model} 
\end{equation}
with $t$ designating the normalized discrete time ($t=1,2,3,\ldots$ with absolute time being $(t-1) T_s$, where $T_s$ stands for the sampling period), $y[t]$ the measured guided wave response signals as generated by the piezoelectric sensors of the structure, $n$ the AR polynomial order, and $e[t]$ the stochastic model residual (one-step-ahead prediction error) sequence, that is a white (serially uncorrelated), Gaussian, zero mean with variance $\sigma^2_e$ sequence. The symbol $\mathcal{N}(\cdot,\cdot)$ designates Gaussian distribution with the indicated mean and variance, and iid stands for identically independently distributed.

It can be shown that the minimum mean square error (MMSE) one-step-ahead prediction $ \widehat{y}[t/t-1]$ of the signal value $y[t]$ made at time $t-1$ (that is for given values of the signal up to time $t-1$) is\footnote{A hat designates estimator/estimate; for instance $\widehat{\theta}$ is an estimator/estimate of $\theta$.}:
\begin{equation}
     \widehat{y}[t/t-1] =- \sum_{i=1}^{na} a_i \cdot y[t-i] \label{eq:prediction} 
\end{equation}
Comparing this with the AR model of Equation (\ref{eq:ar-model}), it is evident that the one-step-ahead prediction error is equal to $e[t]$, that is:
\begin{equation}
     \widehat{e}[t/t-1] \triangleq y[t] - \widehat{y}[t/t-1] = e[t]  \label{eq:prediction error} 
\end{equation}
This is an important observation, as it indicates that the model's one-step-ahead prediction error (also referred to as the residual) coincides with the (uncorrelated) innovations generating the signal. This is valid as long as the true model parameters of Equation (\ref{eq:ar-model}) are used in the predictor Equation (\ref{eq:prediction}).

Using the backshift operator $\mathcal{B} (\mathcal{B}^i \cdot y[t]\triangleq y[t-i])$, the AR representation of Equation (\ref{eq:ar-model}) may be compactly re-written as: 
\begin{equation}
    y[t] + \sum_{i=1}^{na} a_i \cdot \mathcal{B}^i \cdot y[t] = e[t] \quad \iff  \quad A[\mathcal{B}] \cdot y[t] = e[t], \quad e[t] \sim \, \mbox{iid} \, \mathcal{N} \bigl( 0,\sigma^2_e \bigr)
     \label{eq:ar-model compact} 
\end{equation}
%
%
with
 \begin{equation}
       A[\mathcal{B}] = 1 + \sum_{i=1}^{na} a_i \cdot \mathcal{B}^i  \label{eq:ar-model compact 3}  
\end{equation}

The model identification problem is usually distinguished into two subproblems: (i) the \textit{{parameter estimation} 
} subproblem, and~(ii) the \textit{{model structure selection}} subproblem, presented in Sections~\ref{sec: model par est} and \ref{sec: model struc sel}, respectively.

\subsection{Model Parameter~Estimation \label{sec: model par est}}

The AR model of Equation (\ref{eq:ar-model}) can be parameterized in terms of the parameter vector $\bld{\Bar{\theta}}$, which has to be estimated from the measured signal, where
\begin{equation*}
    \bld{\Bar{\theta}} = [a_1 \; \ldots \; a_{na} \; \vdots \; \sigma^2_e]^T
\end{equation*}
and may be written in linear regression form as:
\begin{equation}
      y[t] = \bld{\phi}^T[t] \cdot \bld{\theta} + e[t]
\end{equation}
with 
\begin{equation*}
    \bld{\phi}[t] = \bigl[ -y[t-1] \; -y[t-2] \; \ldots \; -y[t-na] \bigr]^T
\end{equation*}

\begin{equation*}
    \bld{\theta} = [a_1 \; \ldots \; a_{na}]_{[na\times1]}^T
\end{equation*}
and $^T$ designating transposition. Then following substitution of the data for $t = 1,2,\cdots,N$, the following expression is obtained:

\begin{equation}
    \bld{y} = \bld{\phi}\cdot \bld{\theta} + \bld{e}
\end{equation}
where

\begin{equation}
    \bld{y} \coloneqq \begin{bmatrix} y[1] \\\vdots \\ y[N]\end{bmatrix} \qquad  \bld{\phi} \coloneqq \begin{bmatrix}\bld{\phi[1]} \\\vdots \\ \bld{\phi[N]}\end{bmatrix} \qquad  \bld{e} \coloneqq \begin{bmatrix}e[1] \\\vdots \\ e[N]\end{bmatrix} 
\end{equation}
Using the above linear regression framework, the simplest approach for estimating the AR parameter vector $\bld{\theta}$ is based on minimization of the ordinary least squares (OLS) criterion:

\begin{equation}
    J^{OLS} = \frac{1}{N}\sum_{t=1}^{N} e^{T}[t]e[t]
\end{equation}
A more appropriate criterion is (in view of the Gauss-Markov theorem) the weighted least squares (WLS) criterion:

\begin{equation}
    J^{WLS} = \frac{1}{N}\sum_{t=1}^{N} e^{T}[t]\Gamma_{e[t]}^{-1} e[t] = \frac{1}{N}\bld{e}^T\bld{\Gamma}_{e}^{-1}\bld{e}
\end{equation}
which leads to the weighted least squares (WLS) estimator:

\begin{equation}
    \bld{\widehat{\theta}}^{WLS} = [\bld{\phi}^{T}\bld{\Gamma}_{e}^{-1}\bld{\phi}]^{-1}[\bld{\phi}^{T}\bld{\Gamma}_{e}^{-1}\bld{y}]
\end{equation}
In these expressions, $\bld{\Gamma_e}= E\{ \bld{e}\bld{e^{T}}\}$ is the residual covariance matrix, which is practically unavailable. Nevertheless, it may be consistently estimated by applying ordinary least squares. Once $\bld{\widehat{\theta}}^{WLS}$ has been obtained, the final residual and residual variance can be obtained by:

\begin{equation}
    e[t] = y[t]-\phi^T[t]\cdot \bld{\widehat{\theta}}^{WLS}
\end{equation}

\begin{equation}
    \widehat{\sigma}_e^2(\bld{\widehat{\theta}}^{WLS}) = \frac{1}{N}\sum_{t=1}^{N}e^{2}[t,\bld{\widehat{\theta}}^{WLS}]
\end{equation}

The estimator $\bld{\widehat{\theta}}^{WLS}$ may, under mild conditions be shown to be asymptotically Gaussian distributed with mean coinciding with the true parameter vector $\bld{\theta^o}$ and covariance matrix $\bld{P_{\theta}}$:

\begin{equation}
    \sqrt{N}(\bld{\widehat{\theta}_N}-\bld{\theta^o}) \sim \mathcal{N}(\bld{\theta},\bld{P_{\theta}}) \qquad \text{as}\quad N \to \infty
\end{equation}



%
%
%

%
%

\subsection{Model Structure~Selection \label{sec: model struc sel}}

Model structure selection refers to the selection of the AR model order $na$. It is generally based on trial-and-error or successive fitting schemes~\cite{poulimenos2006parametric}, where models corresponding to various candidate structures are estimated, and the one providing the best fitness to the signal is selected. The fitness function may be the Gaussian log-likelihood function of each candidate model. The particular model that maximizes it is the most likely to be the actual underlying model responsible for the generation of the measured signal, in the sense that it maximizes the probability of having provided the measured signal values, and is thus selected. A problem with this approach is that the log–likelihood may be monotonically increasing with increasing model orders, and as a result, the over fitting of the measured signal occurs. For this reason, criteria such as the AIC (Akaike information criterion~\cite{akaike}) or the BIC (Bayesian information criterion~\cite{schwarz}) are generally used and can be represented as follows:
\begin{equation}
    \text{AIC} = -2 \cdot \ln{\mathcal{L}(\mathcal{M}(\bld{\theta},\sigma^2_e)|y^N)} + 2\cdot d
\end{equation}
\begin{equation}
\text{BIC} = -\ln{\mathcal{L}(\mathcal{M}(\bld{\theta},\sigma^2_e)|y^N)} +\frac{\ln{N}}{2}\cdot d
\end{equation}
with $\mathcal{L}$ designating the model likelihood, $N$ the number of signal samples, and $d$ the number of independently estimated model parameters. As it may be observed, both criteria consist of a superposition of the negative log-likelihood function and a term that penalizes the model order, or structural complexity, and thus discourages the model over fitting. Accordingly, the model that minimizes the AIC or the BIC is selected. The ratio of the residual sum of squares versus the signal sum of squares (RSS/SSS) may also be used as another fitness criterion for selecting the best model.

\section{Damage Diagnosis Using AR Model}

The damage detection and identification of a structure can be based on a characteristic quantity $Q = f(\bld{\theta})$, which is a function of the parameter vector $\bld{\theta}$ of an AR model. Three approaches that were taken in this work are described below:

\subsection{Standard AR Approach}
Let $\bld{\widehat{\theta}}$ designate a proper estimator of the parameter vector $\bld{\theta}$. For a sufficiently long signal, the estimator is (under mild assumptions) Gaussian distributed with mean equal to its true value $\bld{\theta}$ and a certain covariance $\bld{P}_{\theta}$, hence $\bld{\widehat{\theta}} \sim \mathcal{N}(\bld{\theta},\bld{P}_{\theta})$.
Damage detection is based on testing for statistically significant changes in the parameter vector $\bld{P}_{\theta}$ between the nominal and current state of the structure through the hypothesis testing problem.
\begin{align*}
   & H_0 : \delta \bld{\theta} = \bld{\theta}_{o}-\bld{\theta}_{u} = 0 \qquad \text{null hypothesis--healthy structure}\\
   & H_1 : \delta \bld{\theta} = \bld{\theta}_{o}-\bld{\theta}_{u} \neq 0 \qquad \text{alternative hypothesis -- damaged structure}
\end{align*}
The difference between the two parameter vector estimators also follows Gaussian distribution, that is, $\delta\bld{\widehat{\theta}} = \bld{\widehat{\theta}}_{o}-\bld{\widehat{\theta}}_{u} \sim \mathcal{N}(\delta\bld{\theta},\delta\bld{P})$, with $\delta\bld{\theta} = \bld{\theta}_{o}-\bld{\theta}_{u}$ and $\delta\bld{P} = \bld{P}_{o}+\bld{P}_{u}$, where $\bld{P}_{o}, \bld{P}_{u}$ designate the corresponding covariance matrices. Under the null ($H_0$) hypothesis $\delta\bld{\widehat{\theta}} = \bld{\widehat{\theta}}_{o}-\bld{\widehat{\theta}}_{u} \sim \mathcal{N}(0,2\bld{P}_o)$ and the quantity
\begin{align*}
    Q = (\delta\bld{\widehat{\theta}})^T \cdot \delta \bld{P} \cdot \delta\bld{\widehat{\theta}}  \qquad \text{with} \quad \delta\bld{P} = 2\bld{P}_o
\end{align*}
follows a $\chi^2$ distribution with $d = \text{dim}(\bld{\theta})$(parameter vector dimensionality) degrees of freedom. As the covariance matrix $\bld{P}_o$ corresponding to the healthy structure is unavailable, its estimated version $ \bld{\widehat{P}}_{o}$ is used. Then the following test is constructed at the $\alpha$(type I) risk level:

\begin{align}
    &Q \leqslant \chi^2_{1-\alpha}(d) \quad \Longrightarrow \qquad H_0 \quad \text{is accepted (healthy structure)}\\
    &\text{Else} \qquad \Longrightarrow \qquad H_1 \quad \text{is accepted (damaged structure)}
\end{align}
where, $\chi^2_{1-\alpha}(d)$ designates the $\chi^2$ distribution's $(1-\alpha)$ critical points.
Damage identification may be based on a multiple hypothesis testing problem comparing the parameter vector $\bld{\widehat{\theta}}_{u}$ belonging to the current state of the structure to those corresponding to different damage types $\bld{\widehat{\theta}}_{A}$, $\bld{\widehat{\theta}}_{B}$, $\cdots$.

\subsection{SVD-based Approach}

A simplified Singular Value Decomposition (SVD)-based method can also be used where eigen decomposition is performed on the diagonal matrix formed by the parameter vector. In this case, instead of projecting the parameter vector onto some lower dimensional space, the relative importance of the parameters are determined and the important $m$ parameters are kept while the rests are discarded. 

The diagonal matrix formed by the parameter vector is decomposed in the following way:

\begin{equation}
    \bld{D}(\bld{\widehat{\theta}}_o) = \bld{C} \bld{\Lambda} \bld{C}^{-1}
\end{equation}
where 

\begin{equation}
    \bld{\Lambda} = \begin{bmatrix} \lambda_1&0&\cdots&0\\
    0&\lambda_2&\cdots&0\\
    \vdots&\vdots&\ddots&\vdots\\
    0&0&\cdots&\lambda_n\end{bmatrix} 
\end{equation}

\begin{equation}
   \bld{C} = \begin{bmatrix} \bld{c_1} & \bld{c_2}& \bld{c_3}&\cdots& \bld{c_n} \end{bmatrix}
\end{equation}

The vectors $\bld{c_1},\bld{c_2},\cdots, \bld{c_n}$ are the eigen vectors of the positive definite parameter matrix $\bld{D}(\bld{\widehat{\theta}}_o)$ and the diagonal elements of $\Lambda$ are the corresponding eigenvalues $\Lambda_{ii} = \lambda_i$. The diagonal elements can be extracted into a vector $\bld{e} = \text{diag}(\Lambda)$, where,
\begin{equation}
    \bld{e} = \begin{bmatrix} \lambda_1, \lambda_2, \lambda_3, \cdots, \lambda_n \end{bmatrix}
\end{equation}
In this case, unlike the case of a singular value decomposition, the diagonal elements are not necessarily arranged in descending order. Moreover, each eigenvalue corresponds to a particular value of the parameter matrix. Knowing this, the magnitude of the eigenvalues are arranged in descending order in the vector $\bld{\widehat{e}}$, where

\begin{equation}
    \bld{\widehat{e}} = \begin{bmatrix} \widehat{\lambda}_1, \widehat{\lambda}_2, \widehat{\lambda}_3, \cdots, \widehat{\lambda}_n \end{bmatrix} \qquad \text{with} \qquad \widehat{\lambda}_1\geqslant \widehat{\lambda}_2\geqslant \cdots \widehat{\lambda}_n
\end{equation}
And the corresponding model parameters become:

\begin{align}
 \bld{\theta}_o^{eig} = \begin{bmatrix} \bld{\widehat{a}}_1^{\widehat{\lambda}_1},\bld{\widehat{a}}_2^{\widehat{\lambda}_2},\bld{\widehat{a}}_3^{\widehat{\lambda}^3}, \cdots, \bld{\widehat{a}}_n^{\widehat{\lambda}_n} \end{bmatrix}
\end{align}
Now from this parameter vector, first $m$ parameters are chosen where $m < n$. As a result, the parameter vector and the associated covariance matrix for damage detection become:

\begin{equation}
     \bld{\theta}_o^{eig} = \begin{bmatrix} \bld{\widehat{a}}_1^{\widehat{\lambda}_1},\bld{\widehat{a}}_2^{\widehat{\lambda}_2},\bld{\widehat{a}}_3^{\widehat{\lambda}^3}, \cdots, \bld{\widehat{a}}_m^{\widehat{\lambda}_m} \end{bmatrix}
\end{equation}

\begin{equation}
    \bld{D}(\bld{\widehat{\theta}}_o^{eig}) =  \bld{\widehat{C}} \bld{\widehat{\Lambda}} \bld{\widehat{C}}^{-1}
\end{equation}

where 
\begin{equation}
    \bld{\widehat{\Lambda}} = \begin{bmatrix} \widehat{\lambda}_1&0&\cdots&0\\
    0&\widehat{\lambda}_2&\cdots&0\\
    \vdots&\vdots&\ddots&\vdots\\
    0&0&\cdots&\widehat{\lambda}_m\end{bmatrix} 
\end{equation}

\subsection{PCA-based Approach}

In addition to using $d = \text{dim}(\bld{\theta}) = n$, a truncated version of the parameter vector $\bld{\theta}$ may also be used in an effort to simplify the damage detection procedure. Because all of the parameters may not be equally sensitive to the damage and may introduce artifacts in the damage detection. In order to avoid that situation, a truncation approach based on principal component analysis may be used. The idea is to project the baseline (healthy) parameter vector $\bld{\theta}_{o}$ onto a co-ordinate system where information compression is possible. A subspace of lower dimensionality, without sacrificing significant information, is subsequently selected, and the discrepancy between the baseline (healthy) and the current vector is projected onto the same subspace.

In the beginning, the covariance matrix is diagonalized via singular value decomposition (SVD) as follows:

\begin{equation}
    \bld{P}(\bld{\widehat{\theta}}_o) = \bld{U}\cdot \bld{S} \cdot \bld{U}^T 
\end{equation}
with

\begin{equation}
    \bld{S} = \begin{bmatrix} s_1&0&\cdots&0\\
    0&s_2&\cdots&0\\
    \vdots&\vdots&\ddots&\vdots\\
    0&0&\cdots&s_n\end{bmatrix} 
\end{equation}

\begin{equation}
   \bld{U} = \begin{bmatrix} \bld{u_1} & \bld{u_2}& \bld{u_3}&\cdots& \bld{u_n} \end{bmatrix}
\end{equation}
where, $\bld{U}$ is an orthonormal matrix ($\bld{U}\bld{U}^T = \bld{I}$), and the columns of which define the principal components, and form a subspace spanning the vector $\bld{\theta}_o$. The singular values $s_j \quad (j = 1,2,\cdots, n)$ are ranked in decreasing order, and represent the active energy of the associated principal components. The subspace dimensionality selection is then based on bounding the information loss, expressed in terms of active energy contribution, below a certain threshold. A measure of how well the first $m$ principal components explain the variance of $\bld{\widehat{\theta}}_o$ is given by the relative proportion

\begin{equation}
    \Psi_m = \frac{\sum_{j=1}^{m} s_j}{\sum_{j=1}^n s_j} 100(\%)
    \label{eq:energy}
\end{equation}

Selecting only the first $m$ co-ordinates $\{\bld{u_1}, \bld{u_2},\cdots,\bld{u_m} \}$ in Equation (\ref{eq:energy}) determines an $m-$dimensional subspace, and the projection of $\bld{\widehat{\theta}}_o$ on this subspace is given by:
\begin{equation}
    \bld{\widehat{\theta}}_o^{PCA} = \bld{U_m}^T\bld{\widehat{\theta}}_{mo}
\end{equation}
with
\begin{equation}
     \bld{U_m} = \begin{bmatrix} \bld{u_1} & \bld{u_2}& \bld{u_3}&\cdots& \bld{u_m} \end{bmatrix}
\end{equation}
\begin{equation}
    \bld{P}(\bld{\widehat{\theta}}_o^{PCA}) = \bld{U_m}^T \bld{P}(\bld{\widehat{\theta}}_o)\bld{U_m} = \begin{bmatrix} s_1&0&\cdots&0\\
    0&s_2&\cdots&0\\
    \vdots&\vdots&\ddots&\vdots\\
    0&0&\cdots&s_n\end{bmatrix} 
\end{equation}

\section{Damage Indices}

In this work, the following time-domain DI is employed as reference in order to compare the performance of DI-based approach and the time series model-based approach proposed herein. The DI used here was adopted from the work of Janapati et al. \cite{Janapati-etal16}, which is characterized by high sensitivity to damage size and orientation and low sensitivity to other variations such as adhesive thickness and the material properties of
the structure, sensor locations, etc. Given a baseline signal $y_o[t]$  and an unknown signal $y_u[t]$ indexed with normalized discrete time $t (t = 1, 2, 3, \cdots, N)$ where $N$ is the number of data samples considered in the calculation of the DI, the formulation of that DI is as follows:

\begin{equation}
    Y_{u}^n[t] = \frac{y_u[t]}{\sqrt{\sum_{t=1}^N y_u^2[t]}} \qquad Y_o^n[t] = \frac{y_o[t]\cdot \sum_{t=1}^N (y_o[t]\cdot Y_{u}^n[t])}{\sum_{t=1}^N y_o^2[t]}
\end{equation}

\begin{equation}
    \text{DI} = \sum_{t=1}^N (Y_{u}^n[t]-Y_o^n[t])^2
\end{equation}
In this notation, $Y_{u}^n[t]$ and $Y_o^n[t]$ are normalized unknown
(inspection) and baseline signals, respectively.

\section{Workframe of SHM}

Let $S_v$ designate the structure under consideration where $v = o$ designate the healthy state and any other $v$ from the set $v = \{ a, b, \cdots\}$ designate the damaged state of the structure. $S_a, S_b, \cdots$, etc. represent a distinct type of damage (for instance, damage in a particular region or of a particular nature). In general, each damage type may include a continuum of damages, each being characterized by its own damage magnitude.

The SHM problem may then be posed in the following way: given the structure in a currently unknown state $u$, determine whether the structure is damaged or not ($u = o$ or $u \neq o$). This is known as the \textit{damage detection sub-problem}. In case the structure is found to be damaged, determine the current damage type from the possible damage scenarios $\{a, b, \cdots \}$. This is known as the \textit{damage identification sub-problem}. The last step focuses on estimating the magnitude of the current damage, which is known as \textit{damage magnitude estimation sub-problem}. In this paper, only the first two steps are considered.

This paper considers the use of ultrasonic guided wave propagation, which are essentially non-stationary signals, as the main source of available information from the structure for solving the above-mentioned problem. As a result, the problem is classified as acousto-ultrasound-based SHM. As there is no physics-based information available, so the method is classified as the data-based method. Although guided wave signals are non-stationary in nature, in the case of weak non-stationarity, an AR model can be used for damage detection and identification.


Stochastic time series methods are commonly based on discretized excitation $x[t]$ and/or response $y[t]$ (for $t = 1,2, \cdots N$). Let the complete excitation and response signal be presented as $\bld{x}$ and $\bld{y}$, that is, $\bld{z} = [\bld{x, y}]$. In the absence of an excitation signal $\bld{z} = [\bld{y}]$. Like before, a subscript $(o, a , b, \cdots, u)$ is used for designating the corresponding state of the structure that provided the signals.

Once the raw experimental signals are collected, they need to be properly pre-processed. This may include low or band-pass filtering within the frequency range of interest, signal down-sampling (in case the originally used sampling frequency is too high), etc.

The obtained signals are subsequently analyzed by parametric stationary time series methods. In order to do so, appropriate model identification and validation is necessary. Models are identified on the basis of data $ \bld{z}_o, \bld{z}_a, \bld{z}_b, \cdots$ in the baseline phase and based on $\bld{z}_u$ in each inspection phase. From each estimated model, the corresponding estimate of a characteristic quantity $Q$ is extracted.

\begin{figure}[!t] \vspace{-8pt}
  \centering{\includegraphics[width=0.9\columnwidth]{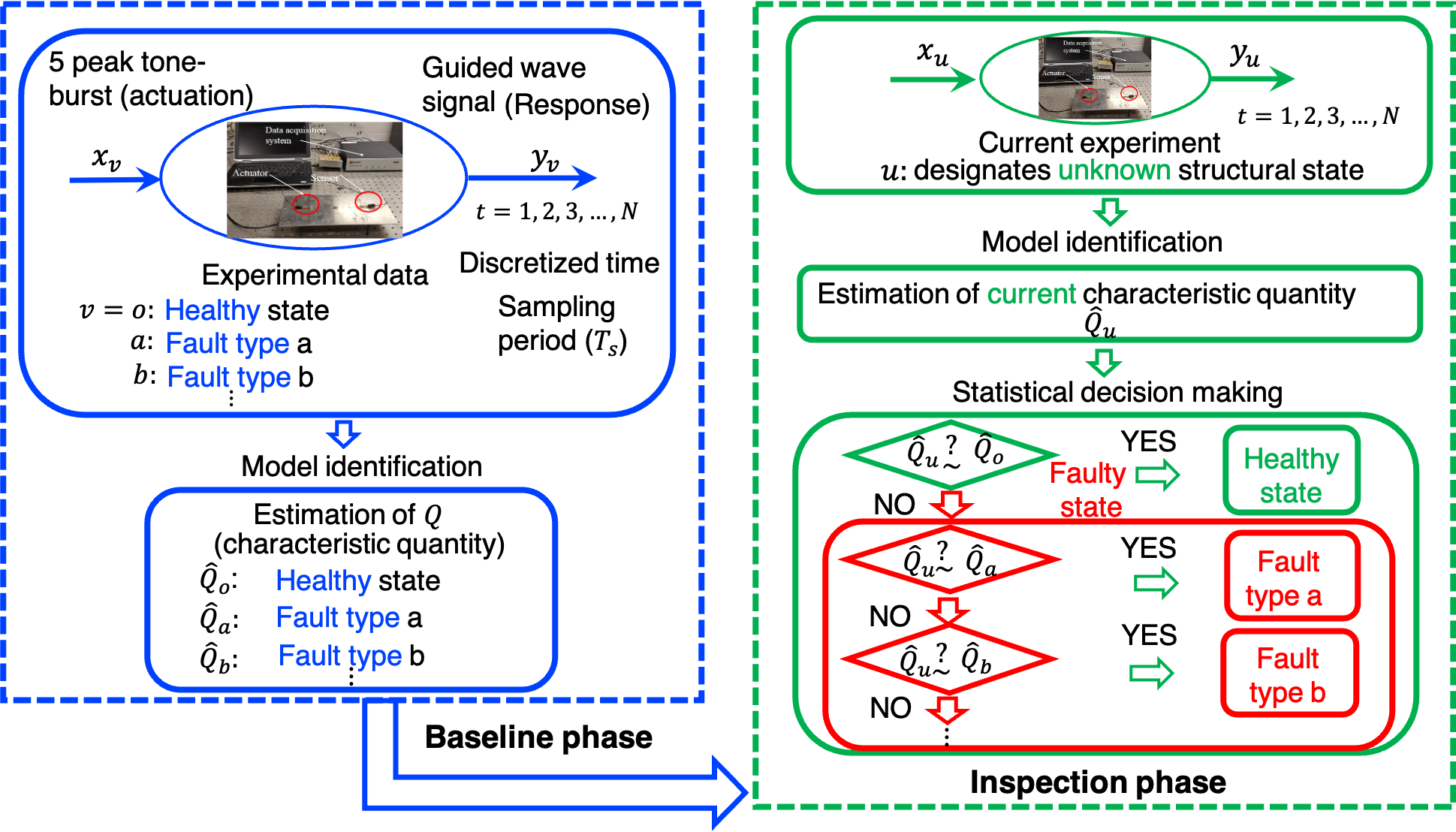}}
   \vspace{-12pt}\caption{Flow chart of the damage diagnosis algorithm}
\label{fig:flow chart}%
\end{figure}

Damage detection is then based on proper comparison of the true $Q_u$ to the true $Q_o$ via a binary statistical hypothesis test that uses the corresponding estimates. Damage identification is similarly based on the proper comparison $Q_u$ to each one of $Q_a, Q_b, \cdots$ via statistical hypothesis testing procedures that also use the corresponding estimates (see Table \ref{tab:workframe_setup} and \ref{tab:workframe_subproblems}). Damage magnitude estimation, when considered, is based on interval estimation techniques. Note that the design of a binary statistical hypothesis test is generally based on the probabilities of type I and type II error, or else the false alarm ($\alpha$) and missed damage ($\beta$) probabilities. The designs presented in this work are based on the former, but in selecting $\alpha$, it should be borne in mind that as $\alpha$ decreases (increases), $\beta$ increases (decreases). Figure \ref{fig:flow chart} concisely summarizes the work frame of the damage diagnosis process using stochastic time-series models. 

\begin{table}[t!]
\caption{Workframe setup: structural state, guided wave signals used, and the characteristic quantity (baseline and inspection phases). \label{tab:workframe_setup}}
\centering
\begin{tabular}{lcccl}
\hline
\multicolumn{5}{c}{\textbf{Baseline Phase}} \\
\hline
Structural state & $S_o$ (healthy structure) & $S_a$ (damage type a) & $S_b$ (damage type b) & \ldots \\
Guided wave signals	& $z_o=(x_o,y_o)$ & $z_a=(x_a,y_a)$ & $z_b=(x_b,y_b)$ & \ldots \\
 Characteristic quantity 	& $Q_o$ & $Q_a$ & $Q_b$ & \ldots \\
\hline
\multicolumn{5}{c}{\textbf{Inspection Phase}} \\
\hline
Structural state& \multicolumn{4}{c}{$S_u$ (current structure in unknown state)}  \\
Guided wave signals	& \multicolumn{4}{c}{$z_u=(x_u,y_u)$}  \\

Characteristic quantity  	& \multicolumn{4}{c}{$Q_u$}  \\
\hline

\end{tabular}

\end{table}

\begin{table}[b!]
\caption{Statistical hypothesis testing problems for the damage detection and identification tasks.}
\centering
\begin{tabular}{cc}
\hline
\multicolumn{2}{c}{\textbf{Damage detection}} \\
\hline
$H_o : Q_u \sim Q_o$ 	& null hypothesis -- healthy structure \\
$H_1 : Q_u \nsim Q_o$ 	& alternative hypothesis -- damaged structure \\
\hline
\multicolumn{2}{c}{\textbf{Damage identification}} \\
\hline
$H_A : Q_u \sim Q_a$ & hypothesis $a$ -- damage type a \\
$H_B : Q_u \sim Q_b$ & hypothesis $b$ -- damage type b \\
\vdots               & \vdots \\
\hline
\end{tabular}
\label{tab:workframe_subproblems}

\end{table}

In this work, the effectiveness and efficiency of the proposed AR-based damage detection and identification framework in the context of acousto-ultrasound guided wave-based SHM have been presented over the two test cases: an aluminum and a composite plate with simulated damages (weights taped on the surface to simulate damage).

\section{Test Case I: Aluminum Plate with Simulated Damage }

\subsection{Experimental Setup and Data Acquisition}

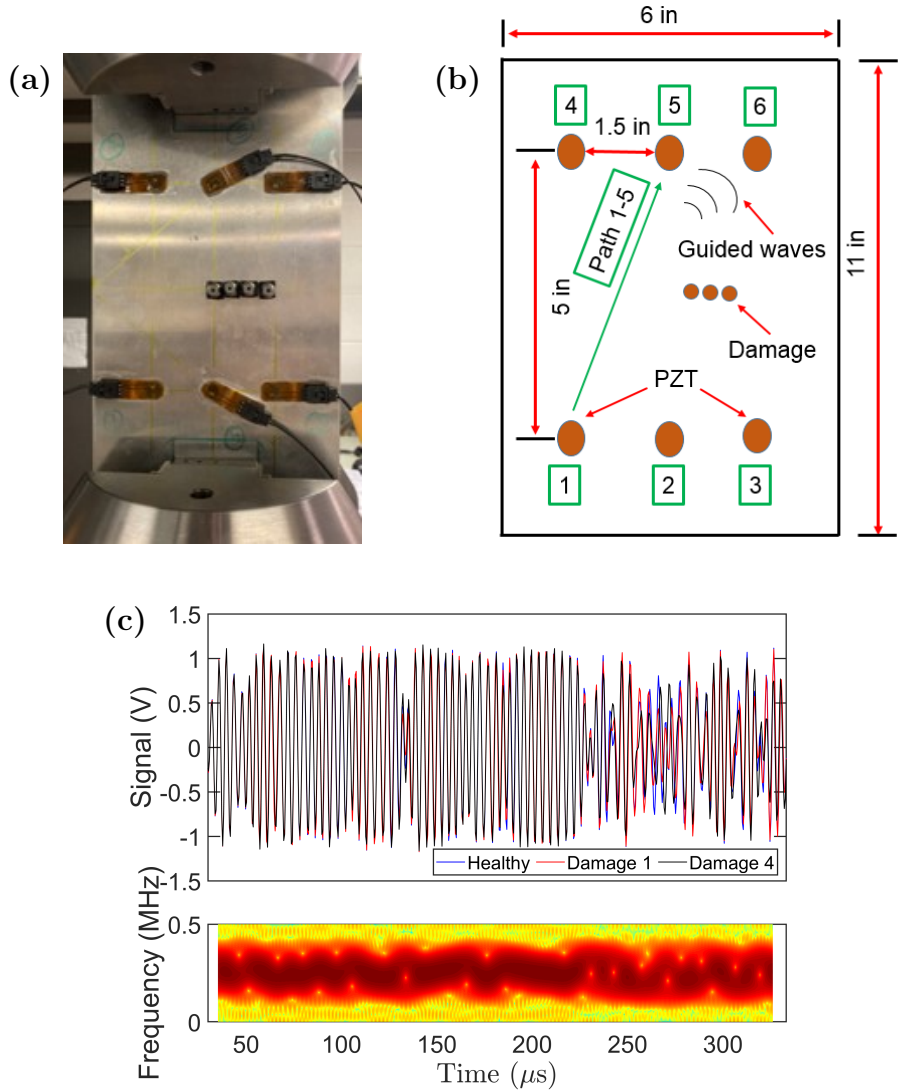
\begin{figure}[t!]
    \centering
    \begin{picture}(400,170)
    \put(50,-22){\includegraphics[width=0.24\columnwidth]{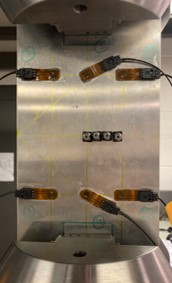}}
    \put(190,-35){\includegraphics[width=0.39\columnwidth]{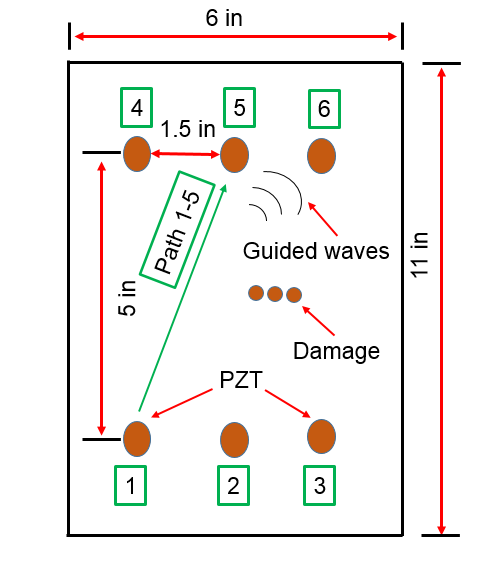}}
    \put(68,-230){\includegraphics[width=0.6\columnwidth]{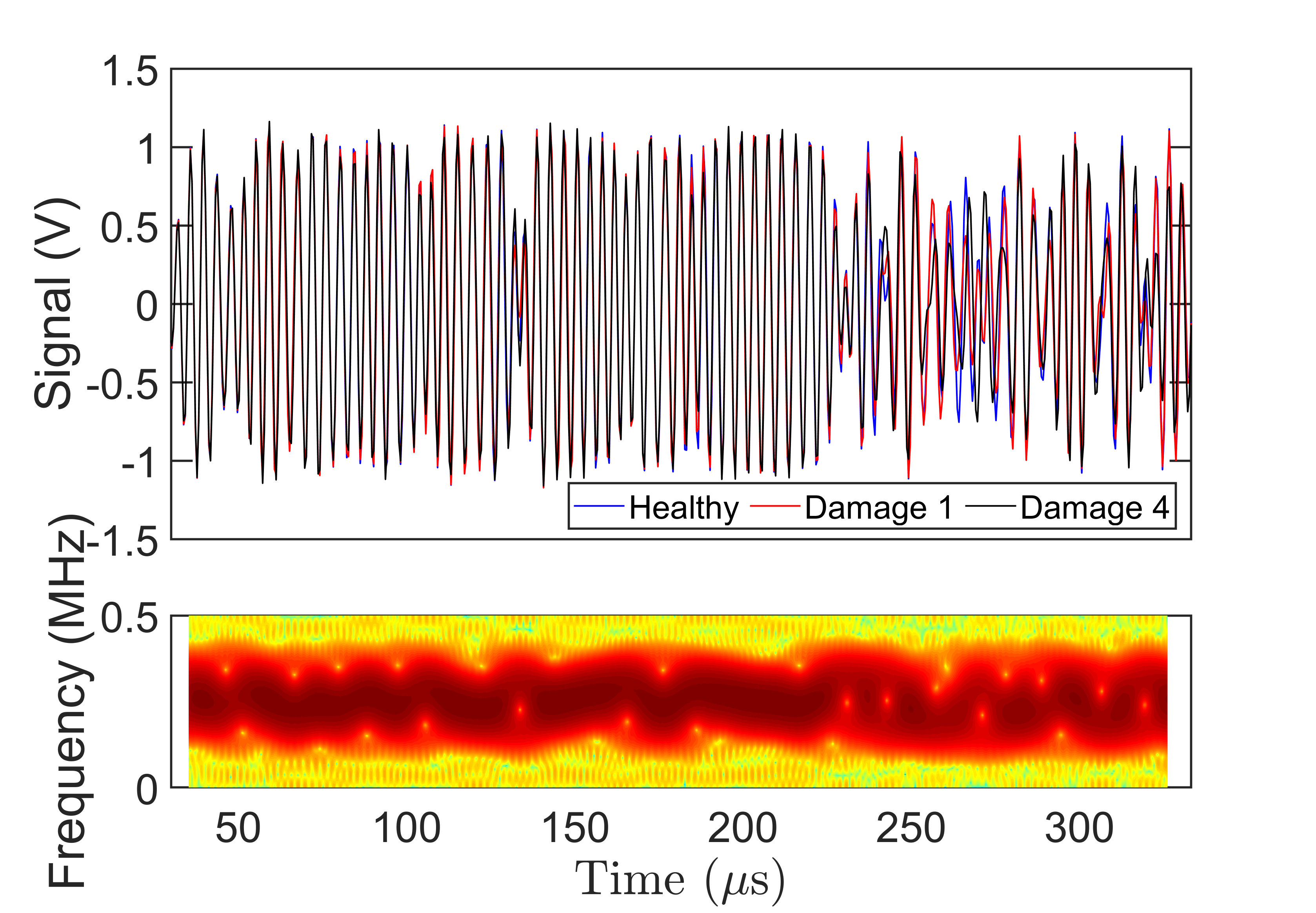}}
    \put(25,150){ \large \textbf{(a)}}
    \put(190,150){\large \textbf{(b)}}
    \put(65,-55){\large \textbf{(c)}}
    \end{picture}
    \vspace{8cm}
    \caption{(a) The aluminum plate used in this study; (b) a schematic of the plate's sensor layout and dimensions; (c) realization of the guided wave signal for healthy and damaged cases with a representative non-parametric spectrogram analysis.} 
\label{fig:plate} 
\end{figure}

In this study, a $152.4 \times 279.4$ mm ($6 \times 11$ in) 6061 aluminum coupon (2.36 mm/0.093 in thick) was used (Figure \ref{fig:plate}(a)). Using Hysol EA 9394 adhesive, six lead zirconate titanate (PZT) piezoelectric sensors (type PZT-5A, Acellent Technologies, Inc) of 6.35 mm (1/4 in) diameter and a thickness of 0.2 mm (0.0079 in) were attached to the plate and cured for 24 hours in room temperature. Figure \ref{fig:plate}(b) shows the dimensions of the plate, placement of the PZT transducers, and the path naming convention. Up to four three-gram weights were taped to the surface of the plate starting from its center-point to simulate local damage (Figure \ref{fig:plate}(b)). 

Actuation signals in the form of 5-peak tone bursts (5-cycle Hamming-filtered sine wave, 90 V peak-to-peak, 250 kHz center frequency) were generated in a pitch-catch configuration over each sensor consecutively. Data were collected using a ScanGenie III data acquisition system (Acellent Technologies, Inc) from selected sensors during each actuation cycle at a sampling frequency of 24 MHz. Twenty signals from each sensor (wave propagation path) and damage state were recorded. This led to a total of 100 data sets for each sensor. For the time-series modeling, the acquired signals were down-sampled to 2 MHz. This process resulted in 612-sample-long signals. Figure \ref{fig:plate}(c) presents indicative signal realization for different damage (health) state (top subplot) and non-parametric spectrogram of a single signal realization\footnote{window length: 30 samples; 98\% overlap; NFFT points: 30000 (zero-padding took place to obtain smooth magnitude estimates); frequency resolution $\Delta f = 666.66$ Hz.}.

\subsection{Path Selection}

\begin{figure}[t!]
    \centering
    \begin{picture}(400,130)
    \put(-40,-50){ \includegraphics[width=0.55\columnwidth]{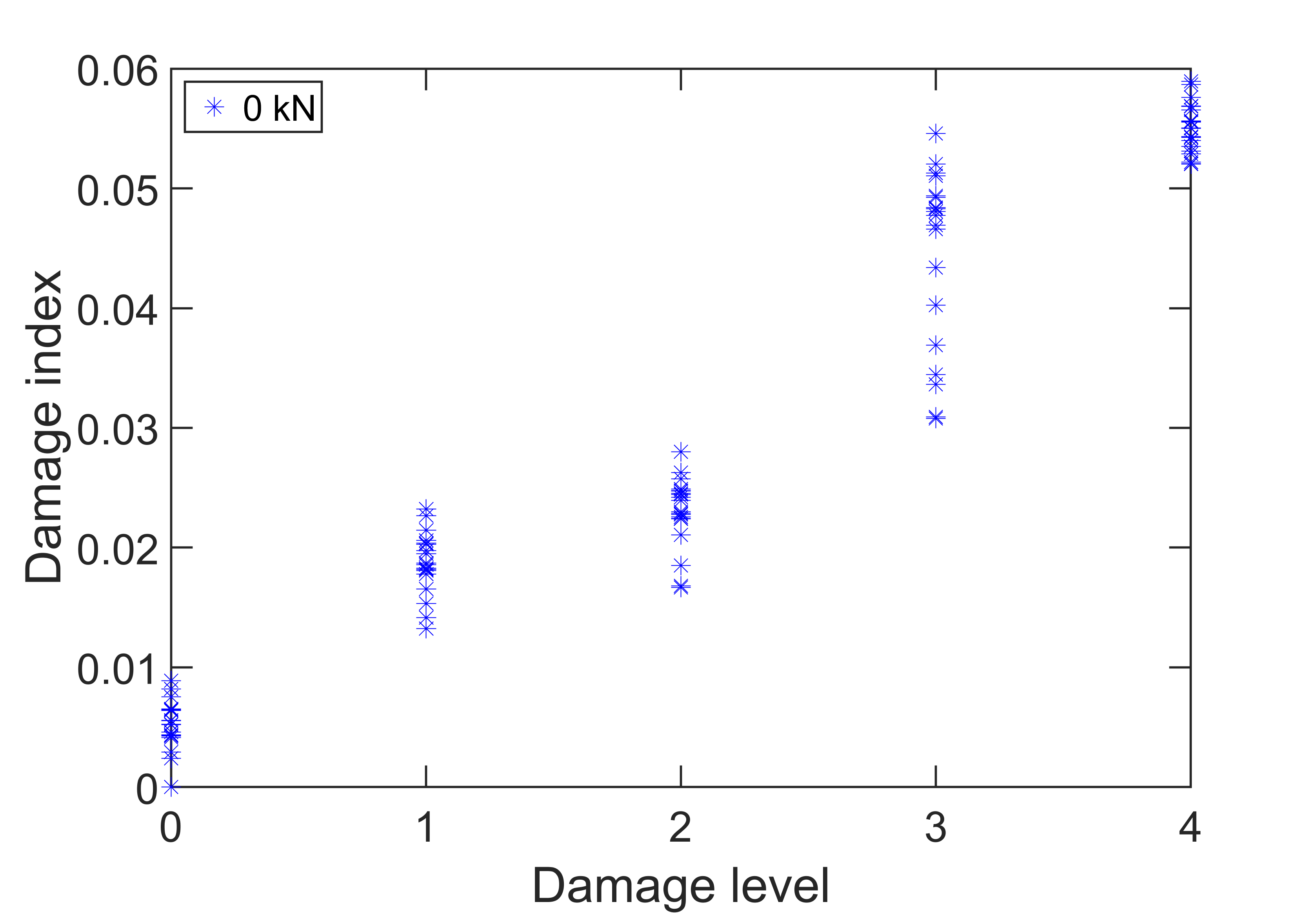}}
    \put(200,-50){\includegraphics[width=0.55\columnwidth]{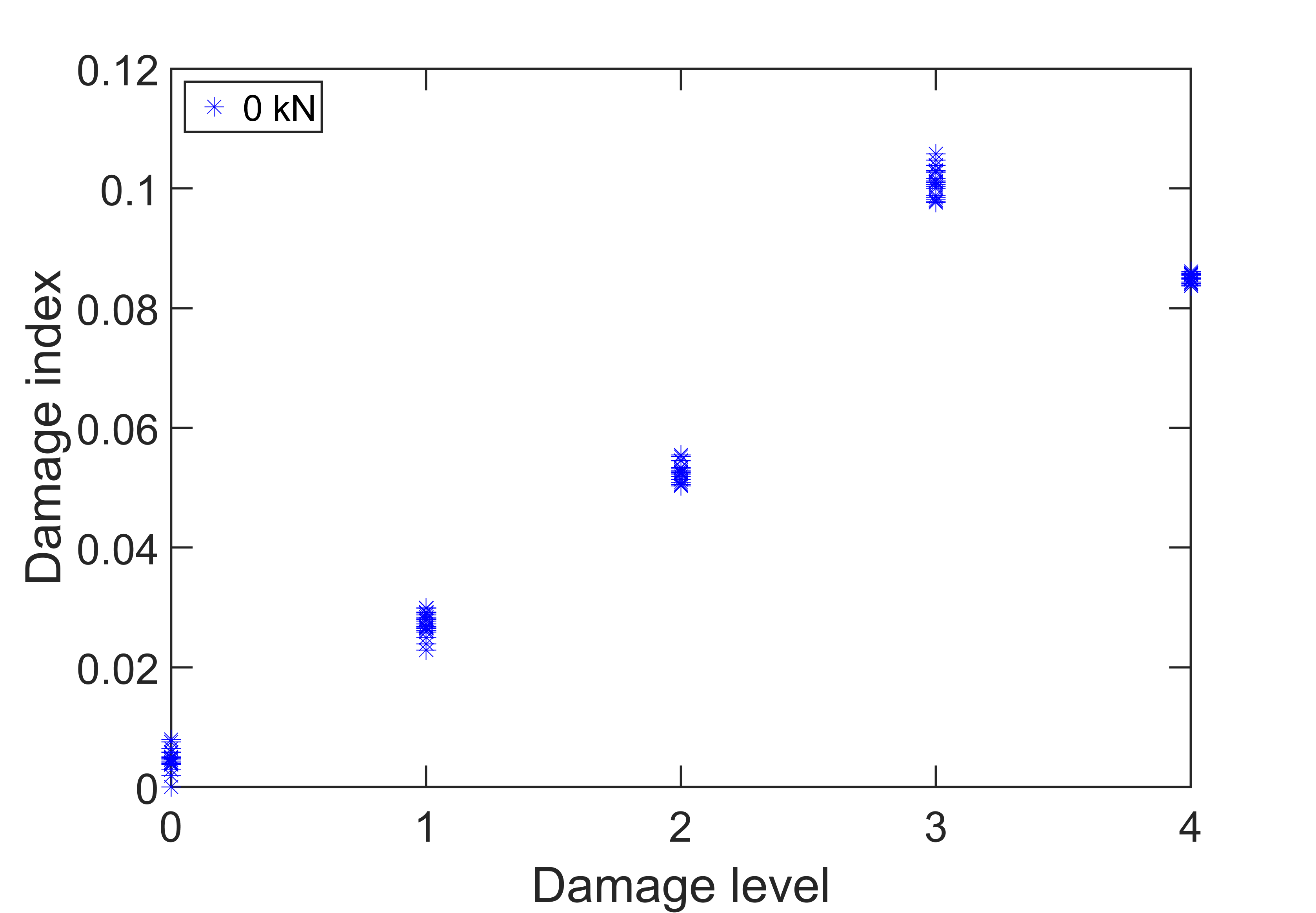}}

    \put(-15,127){\large \textbf{(a)}}
    \put(220,127){ \large \textbf{(b)}}
    \end{picture}
    \vspace{2cm}
    
    \caption{The evolution of the damage index \cite{janapati2016damage} as applied to indicative actuator-sensor paths: (a) damage non-intersecting path 1-4; (b) damage intersecting path 2-6.} 
\label{fig:DI} 
\end{figure} 

In the context of the active sensing guided wave-based method, there are often multiple sensors installed at the area being monitored, and every actuator sensor path in the network has to be examined in order to assess the integrity of the component. In the present study, Figure \ref{fig:plate}(b) shows the actuator-sensor layout, and six sensors/actuators have been used. Damage starts from the center of the plate and grows in magnitude to the right. In this study, simulated damages have been used in the form of weights mounted to the plate with tacky tapes. It has been shown that when the guided wave signal crosses the damage (known as the damage-intersecting path), a significant change can be observed in the signal with the increase in the damage size. On the other hand, for a damage non-intersecting path, one can observe that the received signals sustain significantly smaller change with the increase in damage size. Thus information from the damage non-intersecting path naturally carries less information when it comes to damage detection and identification compared to damage-intersecting paths. In order to support this point, one state-of-the-art DI from the literature was explored to see how damage intersection affects damage detection using the DI approach. Figure \ref{fig:DI}(a) and (b) show the evolution of the DI with increasing damage size for a damage non-intersecting and intersecting path, respectively. It can be observed that the magnitude of the DI for the damage non-intersecting path is much smaller than the damage intersecting path. As a result, damage detection and identification are challenging using a damage non-intersecting path. In the subsequent study, it is shown that using an AR model, perfect damage detection and identification is possible even for damage non-intersecting paths. 

%
\subsection{Parametric Identification and Damage Detection Results}

\begin{figure}[!t] 

\begin{picture}(400,130)
    \put(60,-70){ \includegraphics[width=0.65\textwidth]{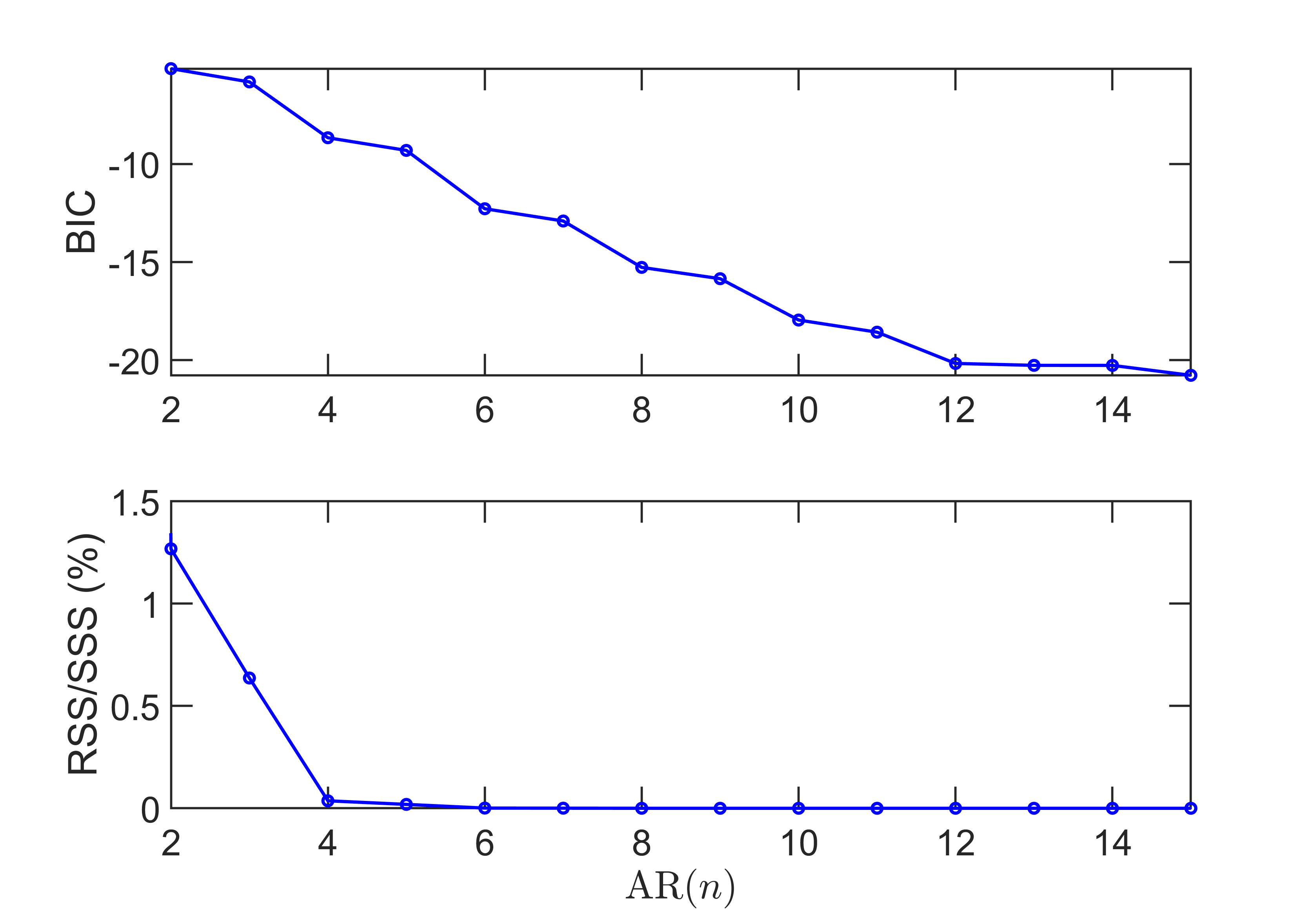}}
\end{picture}
 
   \vspace{2cm}\caption{Model order selection via the BIC (top) and RSS/SSS (bottom) criteria for damage intersecting path 2-6.}
\label{fig:AR BIC}%
\end{figure}

In order to detect and identify damage using an AR model, it is first necessary to identify the system in its healthy state for each path while the guided wave signals are being propagated. Figure \ref{fig:AR BIC} shows the AR model identification process of the aluminum plate in its healthy state for damage intersecting path 2-6. The identification of the damage non-intersecting path 1-4 provides similar results as that of Figure \ref{fig:AR BIC}. For the sake of brevity, it has not been shown here.

Model selection of AR models involves selecting the appropriate model order $na$. The RSS/SSS (Residual Sum of Squares/Signal Sum of Squares) criterion, describing the predictive ability of the model, was employed for the model selection process. AR orders from $na=2$ to $na=15$ were considered to create a pool of candidate models. Among all these models, the best model was chosen where the RSS/SSS values start to show a plateau. Following this criterion, the best model occurred for $na=4$. In addition to the RSS/SSS criterion, the Bayesian Information Criterion (BIC), which rewards the model's predictive capability while penalizing model complexity for increasing model order, was also taken into account (Figure \ref{fig:AR BIC}). Model validation took place via examination of the whiteness, or uncorrelatedness, normality hypothesis of the model residuals. It should be mentioned here that as stationary AR models are being used to model a non-stationary response signal, perfect white residuals are not expected with an arbitrarily large model order keeping a reasonable sample per parameter value (SPP).

It is to be mentioned here that for the estimation of the model parameters, an asymptotically efficient weighted least squares (WLS) estimator was employed in this study. The use of a different estimator may provide a slightly different parameter estimate, which may subsequently affect the damage diagnosis process. 

\begin{figure}[!t] 

\begin{picture}(400,130)
    \put(-30,-50){ \includegraphics[width=0.57\columnwidth]{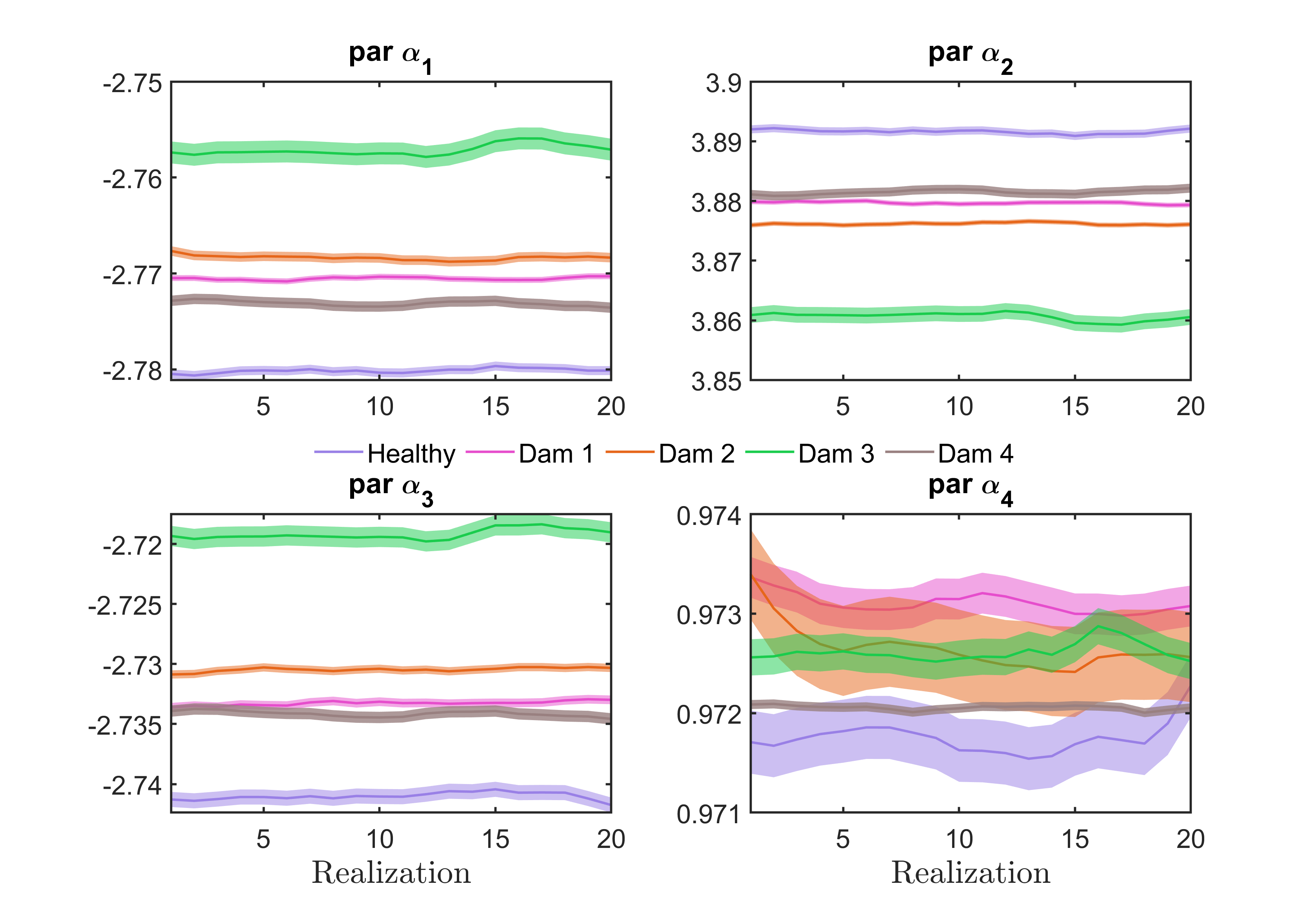}}
    \put(220,-50){\includegraphics[width=0.57\columnwidth]{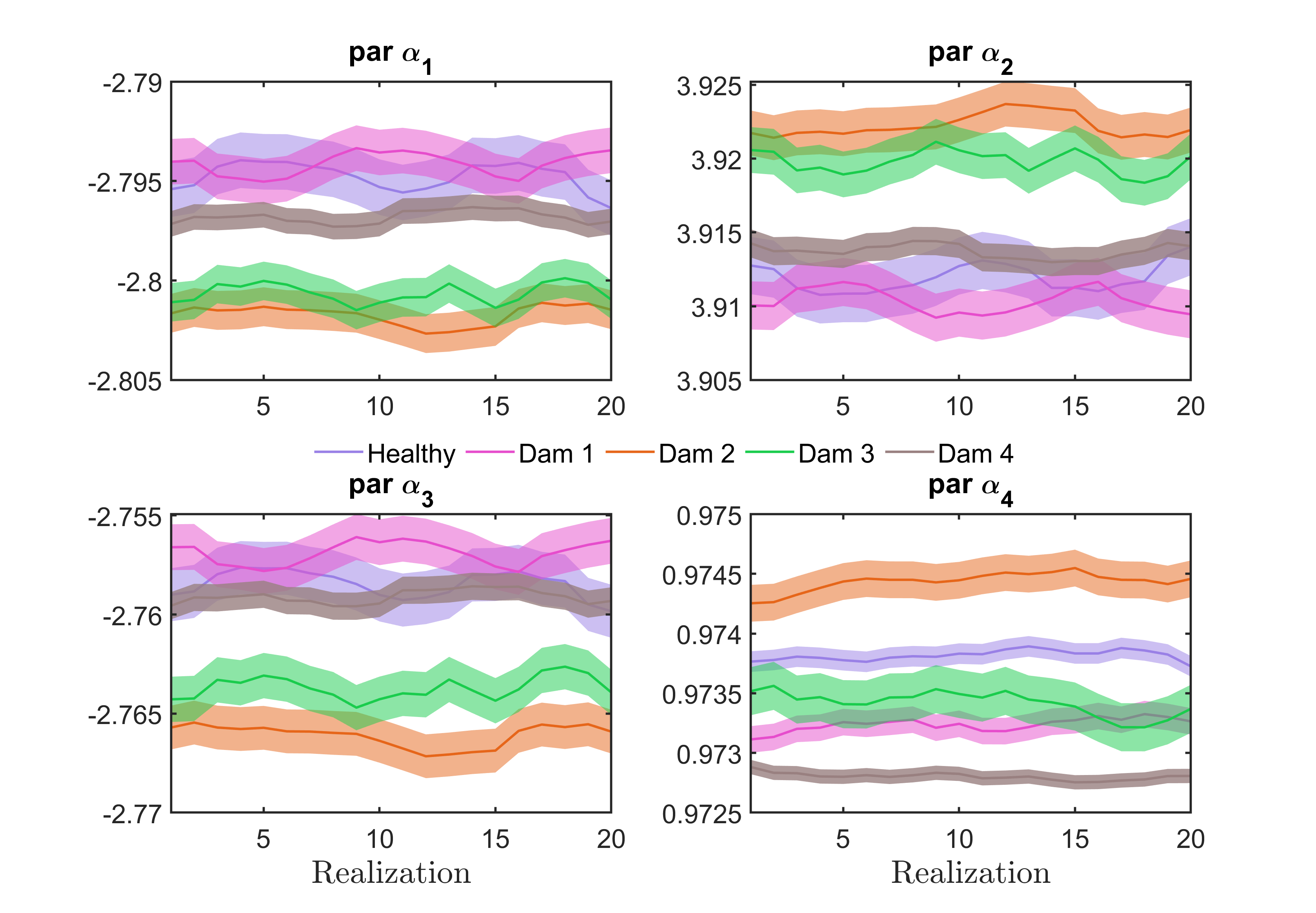}}

    \put(-15,127){\large \textbf{(a)}}
    \put(220,127){ \large \textbf{(b)}}
    \end{picture}
 
   \vspace{1.5cm}\caption{AR($4$)-based model parameters for different structural states: the parameter mean is shown as solid lines and the associated $\pm 2$ standard deviation is shown as shaded regions; (a) damage intersecting path 2-6; (b) damage non-intersecting path 1-4.}
\label{fig:AR par}%
\end{figure}

\begin{figure}[t!]
    \centering
    \begin{picture}(400,130)
    \put(-40,-50){ \includegraphics[width=0.55\columnwidth]{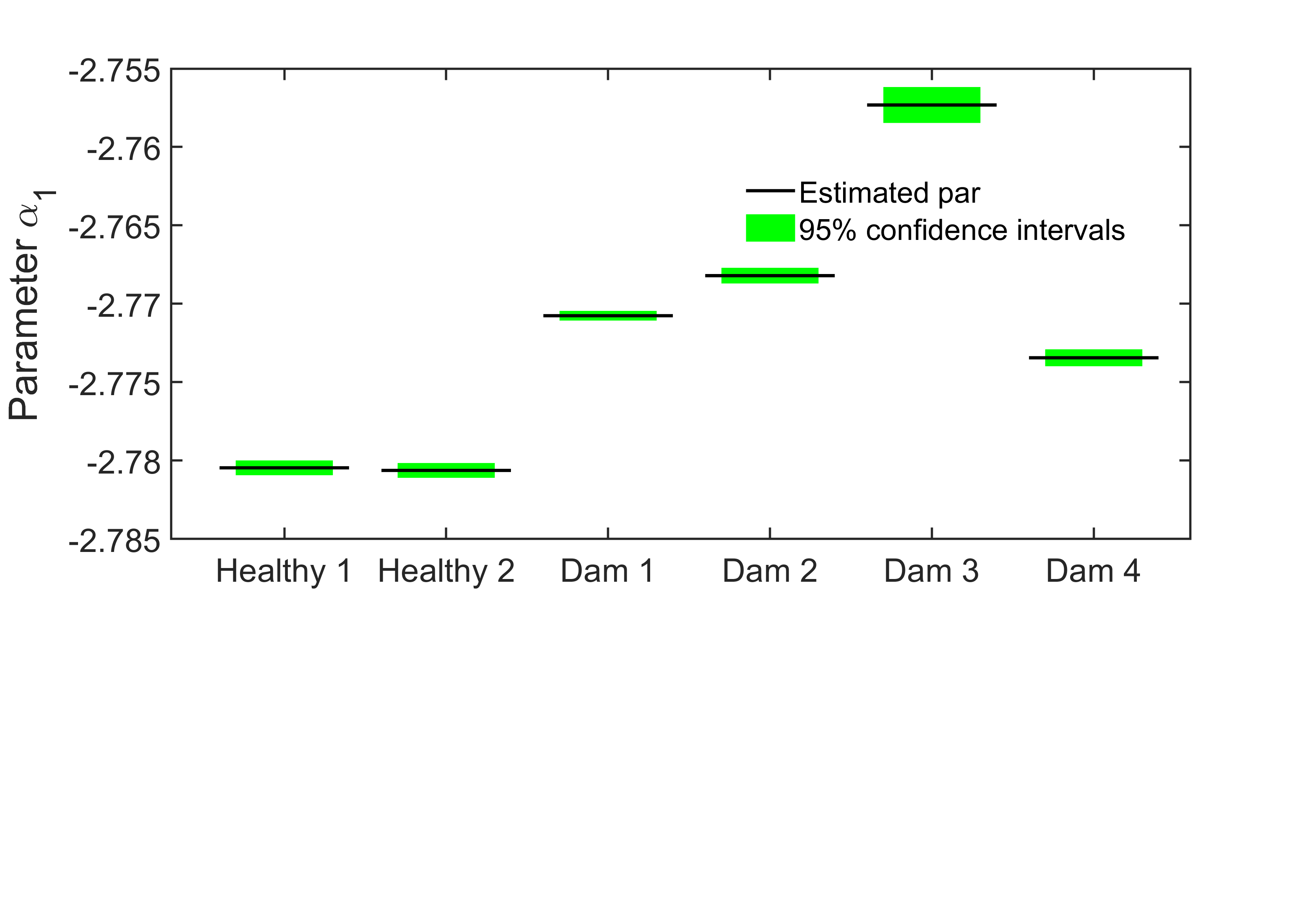}}
    \put(200,-50){\includegraphics[width=0.55\columnwidth]{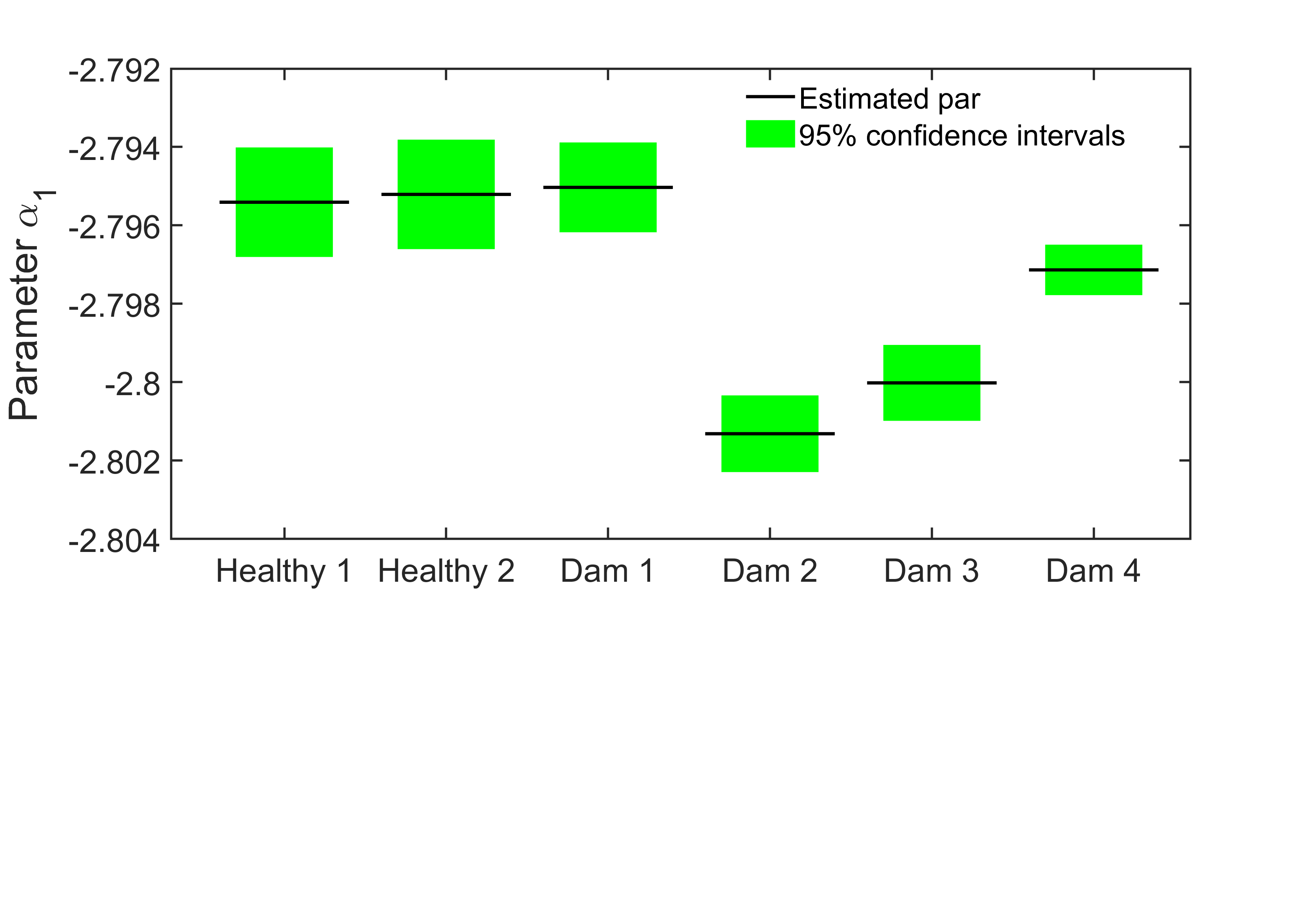}}

    \put(-15,127){\large \textbf{(a)}}
    \put(220,127){ \large \textbf{(b)}}
    \end{picture}
    \vspace{-0.5cm}
    
    \caption{Estimated model parameters for healthy and damaged states and the corresponding 95\% confidence intervals: (a) parameter $\alpha_1$ for damage intersecting path 2-6; (b) parameter $\alpha_1$ for damage non-intersecting path 1-4.} 
\label{fig:par damage} 
\end{figure} 

Figure \ref{fig:AR par}(a) and (b) depict the AR model parameters for damage intersecting path 2-6 and damage non-intersecting path 1-4, respectively, for all different structural states, namely: healthy, damage level 1, damage level 2, damage level 3, and damage level 4. For each state, 20 realizations are shown. The solid lines represent the mean parameter values, and the shaded regions represent the $\pm 2$ standard deviation confidence intervals. As the model order $na = 4$, the number of estimated parameters is also four. Note that the parameters of the damage intersecting path 2-6 for different realizations and structural states are well separated and the confidence intervals are also narrower compared to the damage non-intersecting path 1-4. However, an exception occurred for parameter $\alpha_4$, where different structural states got overlapped on each other. Figure \ref{fig:par damage}(a) and (b) show the evolution of parameter $\alpha_1$ for different structural states for damage intersecting path 2-6 and damage non-intersecting path 1-4, respectively. The black lines represent the mean parameter values and the green regions represent the 95\% confidence intervals. It can be observed that for damage intersecting path 2-6, different states are well separated and the confidence intervals also do not overlap on each other. However, for damage non-intersecting path 1-4, the confidence bound for damage level 1 overlaps with the healthy case, and the confidence bound for damage level 2 and 3 overlaps on each other.

\begin{figure}[t!]
    \centering
    \begin{picture}(400,130)
    \put(-40,-50){ \includegraphics[width=0.55\columnwidth]{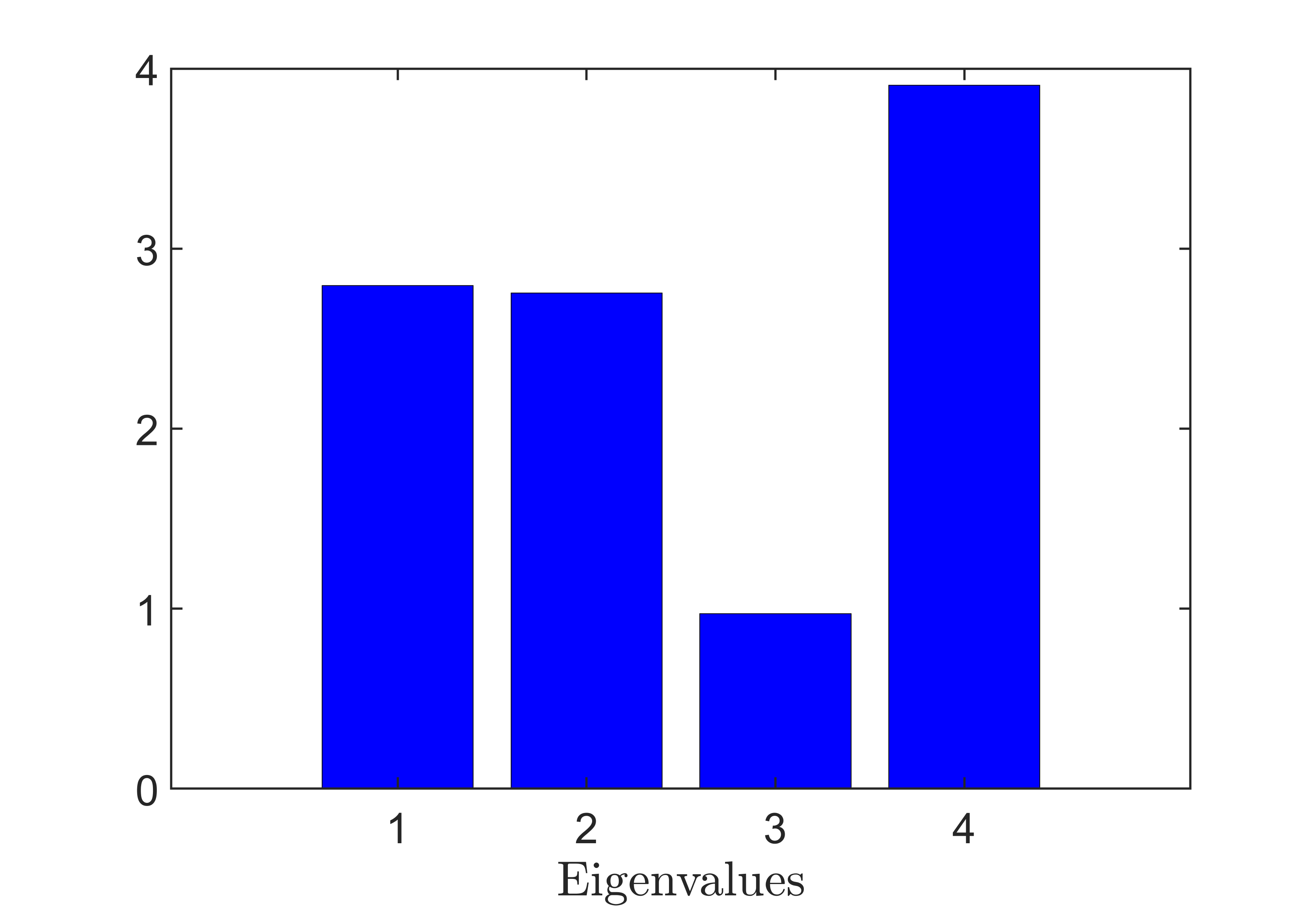}}
    \put(200,-50){\includegraphics[width=0.55\columnwidth]{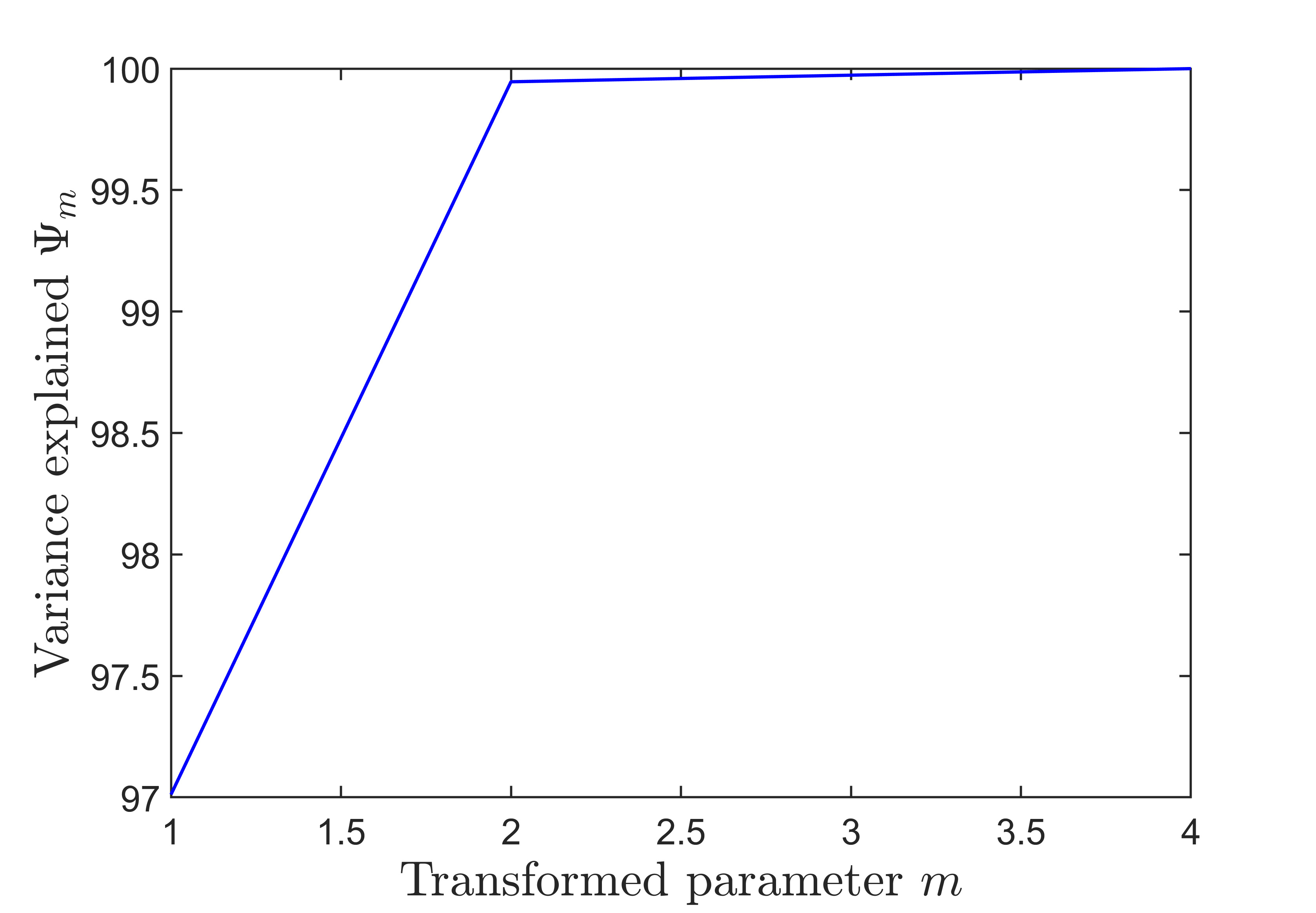}}

    \put(-15,127){\large \textbf{(a)}}
    \put(220,127){ \large \textbf{(b)}}
    \end{picture}
    \vspace{1.5cm}
    
    \caption{AR parameter selection process for damage detection and identification of the damage intersecting path 2-6: (a) SVD-based approach; (b) PCA-based approach.} 
\label{fig:par select} 
\end{figure} 

For damage detection, all four parameters may not be needed or may be redundant in some cases. The use of all the parameters may reduce the damage detection performance when model parameters are correlated. As described in the theoretical section, three approaches have been taken in this regard. The first one is the standard one, where all the model parameters are used. The second one is referred to as the SVD-based approach, and the third one is known as the PCA-based approach. Figure \ref{fig:par select} shows the parameter selection process for the purpose of damage detection and identification in an aluminum plate for damage intersecting path 2-6. From Figure \ref{fig:par select}(a), it can be observed that the eigenvalue $\lambda_1$ and $\lambda_4$ have the highest magnitude. As a result, the corresponding parameter $a_1$ and $a_4$ were used for the subsequent damage detection using the SVD-based approach. From Figure \ref{fig:par select}(b), it can be observed that after projecting the parameters onto some lower dimensional space, only two parameters are needed for explaining the total variance. As a result, the truncated parameters and the associated covariance matrix were used for the subsequent damage detection and identification using the PCA-based approach. Similar results were obtained for the damage non-intersecting path 1-4, and for the sake of brevity, those are not shown.

\begin{figure}[t!]
    \centering
    \begin{picture}(400,130)
    \put(-40,-50){ \includegraphics[width=0.55\columnwidth]{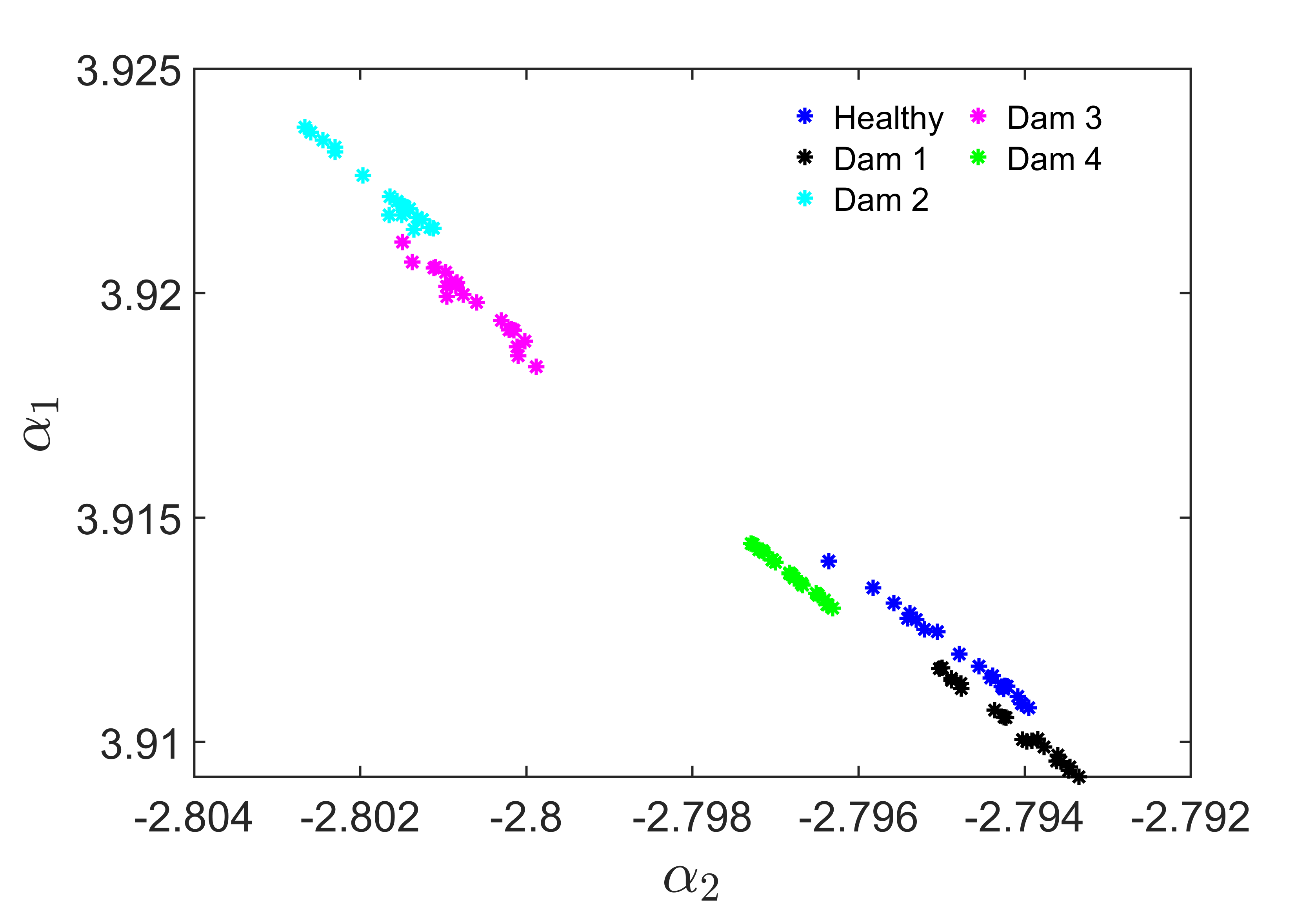}}
    \put(200,-50){\includegraphics[width=0.55\columnwidth]{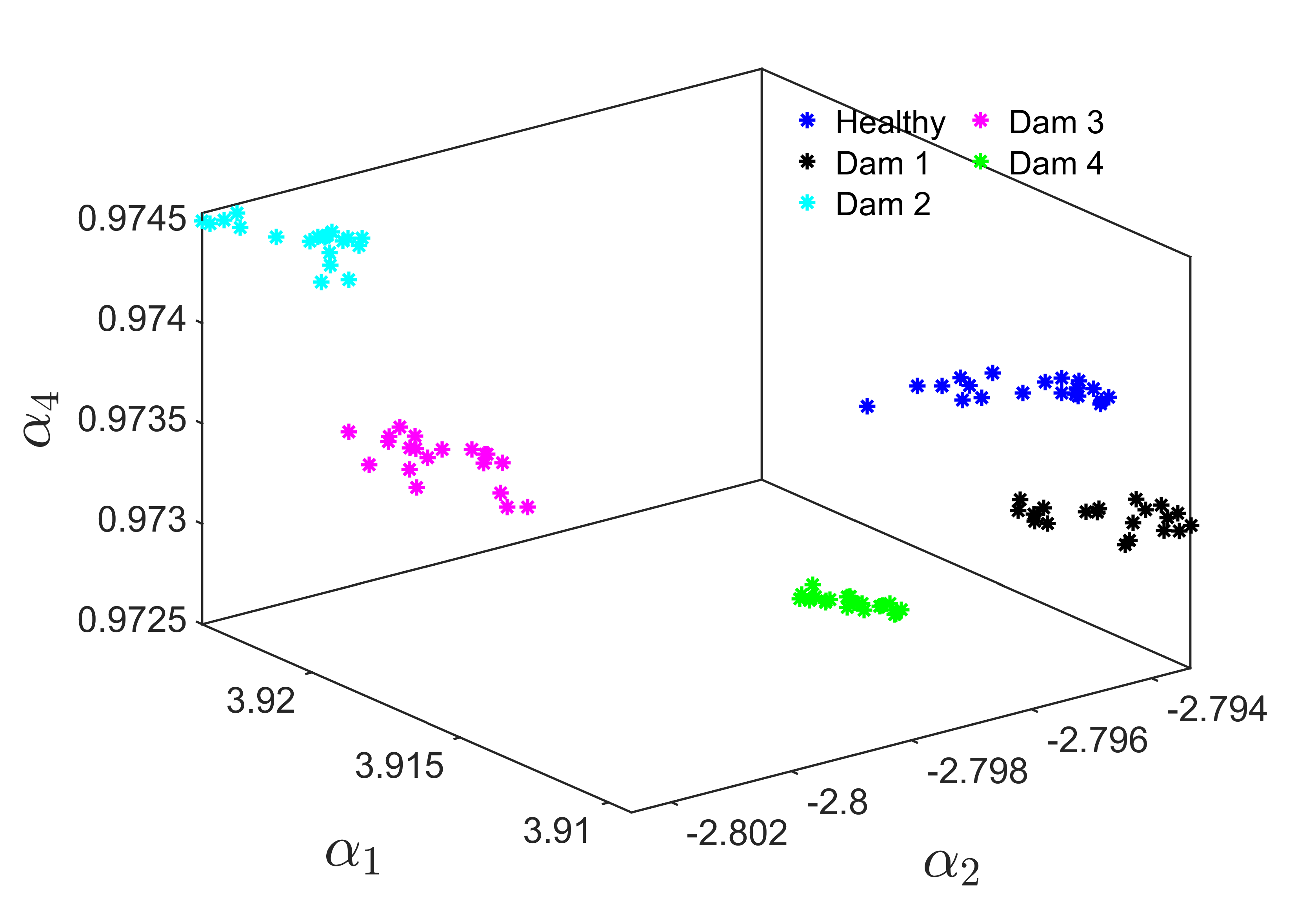}}
    \put(-40,-230){ \includegraphics[width=0.55\columnwidth]{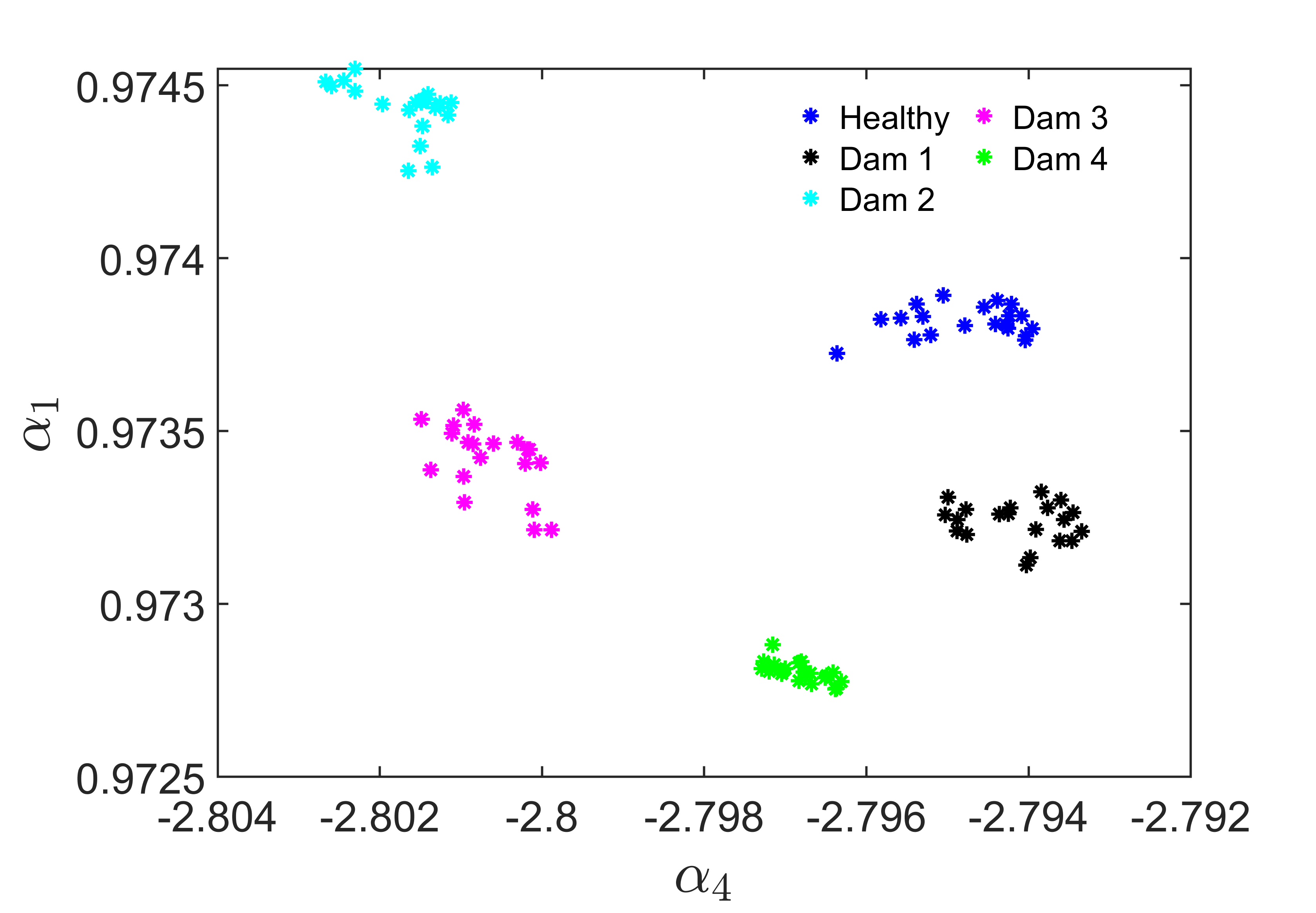}}
    \put(200,-230){\includegraphics[width=0.55\columnwidth]{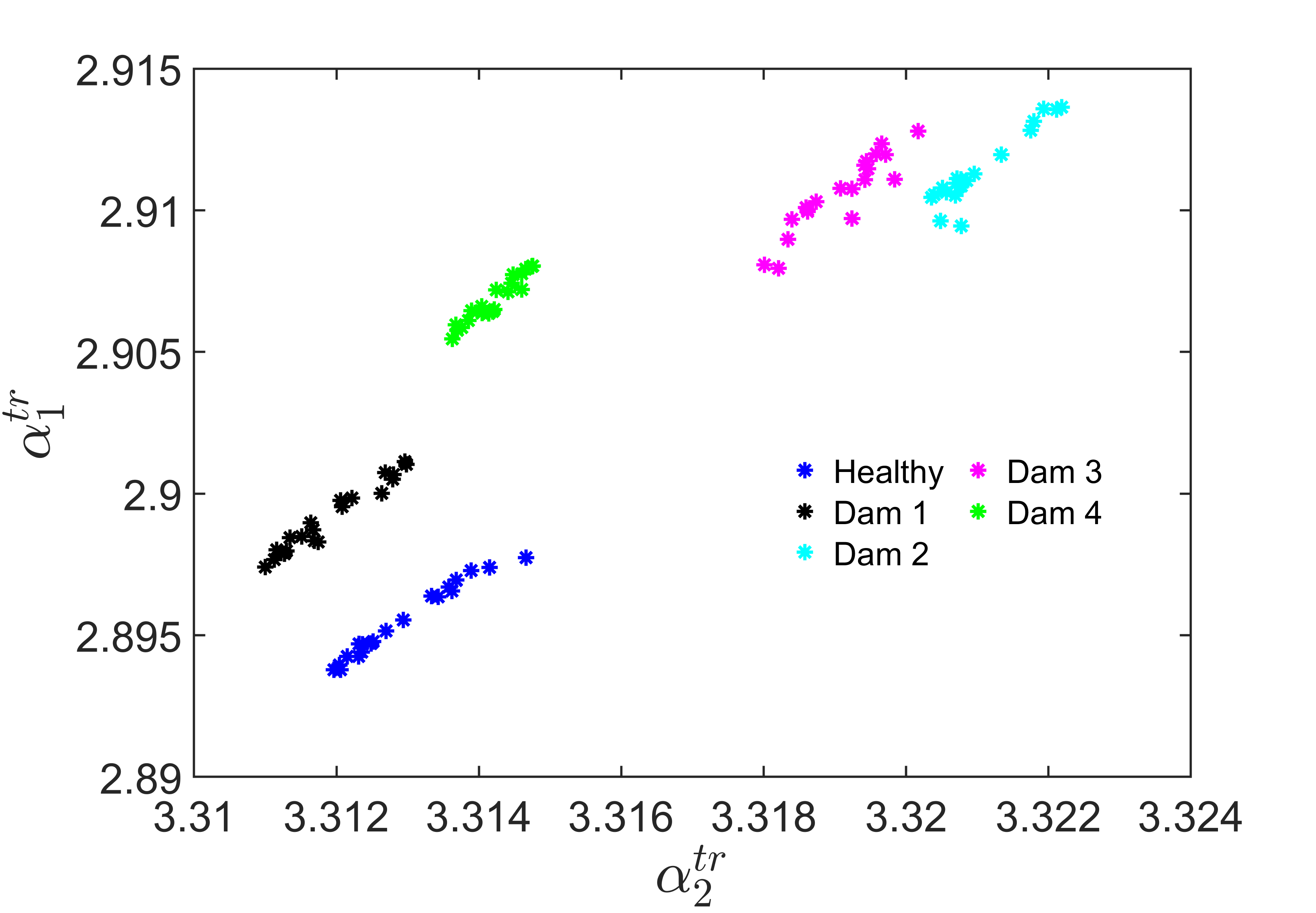}}

    \put(-15,127){\large \textbf{(a)}}
    \put(220,110){ \large \textbf{(b)}}
    \put(-15,-55){\large \textbf{(c)}}
    \put(220,-55){ \large \textbf{(d)}}
    \end{picture}
    \vspace{7.5cm}
    
    \caption{Indicative AR($4$) model parameters are shown for damage non-intersecting path 1-4: (a) model parameter $\alpha_1$ and $\alpha_2$; (b) model parameter $\alpha_1$, $\alpha_2$ and $\alpha_4$; (c) model parameter to be used in damage detection indicated by SVD approach ($\alpha_1$ and $\alpha_4$); (d) truncated model parameters from PCA transformation } 
\label{fig:par par} 
\end{figure}
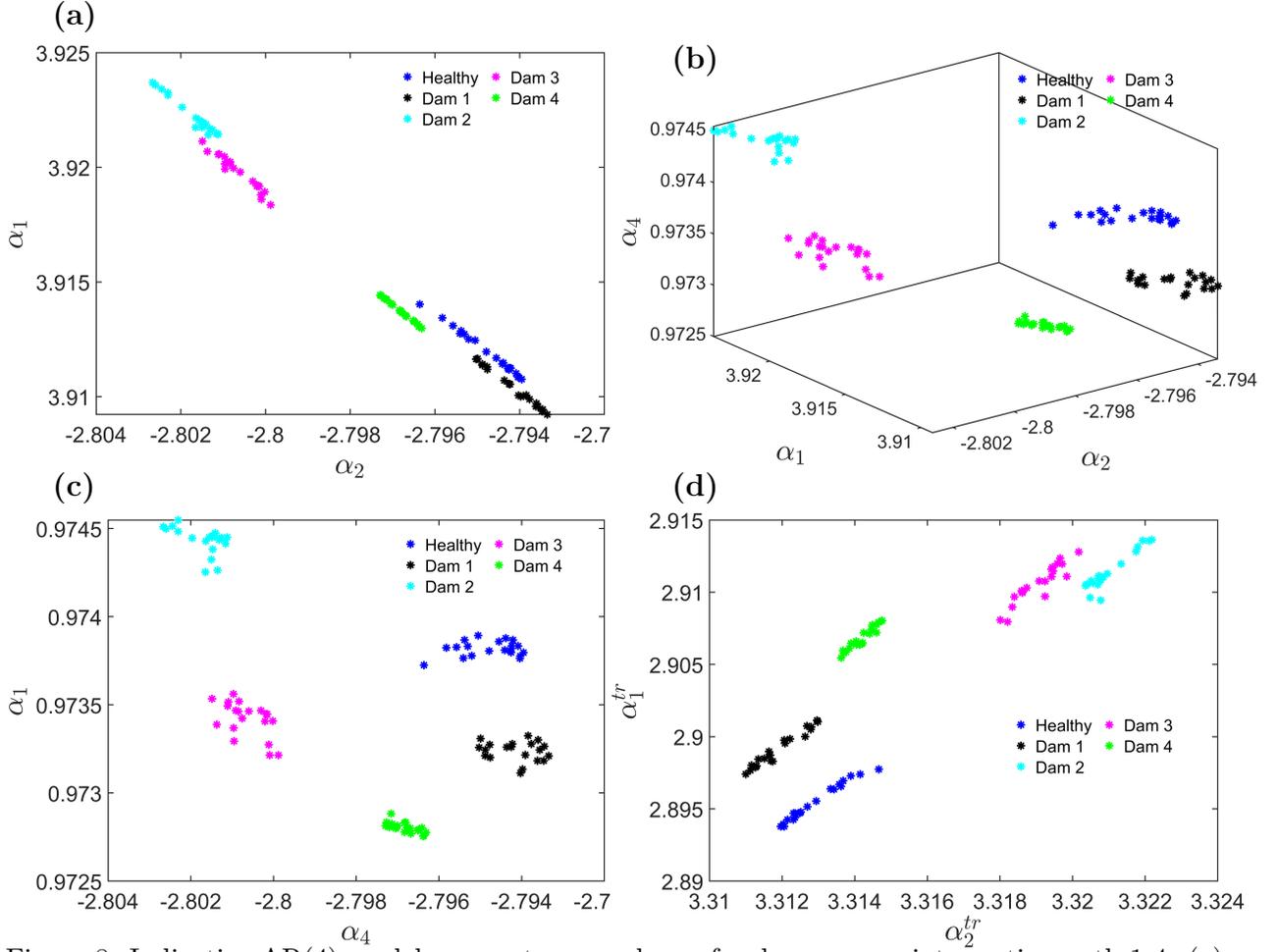 

Figure \ref{fig:par par} shows the correlation between the model parameters. Figure \ref{fig:par par}(a) plots parameter $\alpha_1$ and $\alpha_2$ and they are highly correlated. Including highly correlated parameters in the damage detection algorithm reduces the performance of the damage detection process. Figure \ref{fig:par par}(b) shows a 3-dimensional view of parameter $\alpha_1$, $\alpha_2$, and $\alpha_4$. Figure \ref{fig:par par}(c) shows the correlation between $\alpha_1$ and $\alpha_4$. It can be observed that the parameters are not correlated and different damage states are separated. Also note from Figure \ref{fig:par select}(a) that the eigenvalues of the parameter $\alpha_4$ and $\alpha_1$ has the highest magnitude which suggests the use of parameter $\alpha_4$ and $\alpha_1$ in the damage diagnosis process. Figure \ref{fig:par par}(d) shows that after performing PCA, the model parameters become more uncorrelated and separated for different structural states.

\begin{figure}[t!]
    \centering
    \begin{picture}(400,130)
     \put(-46,10){\includegraphics[width=0.36\columnwidth]{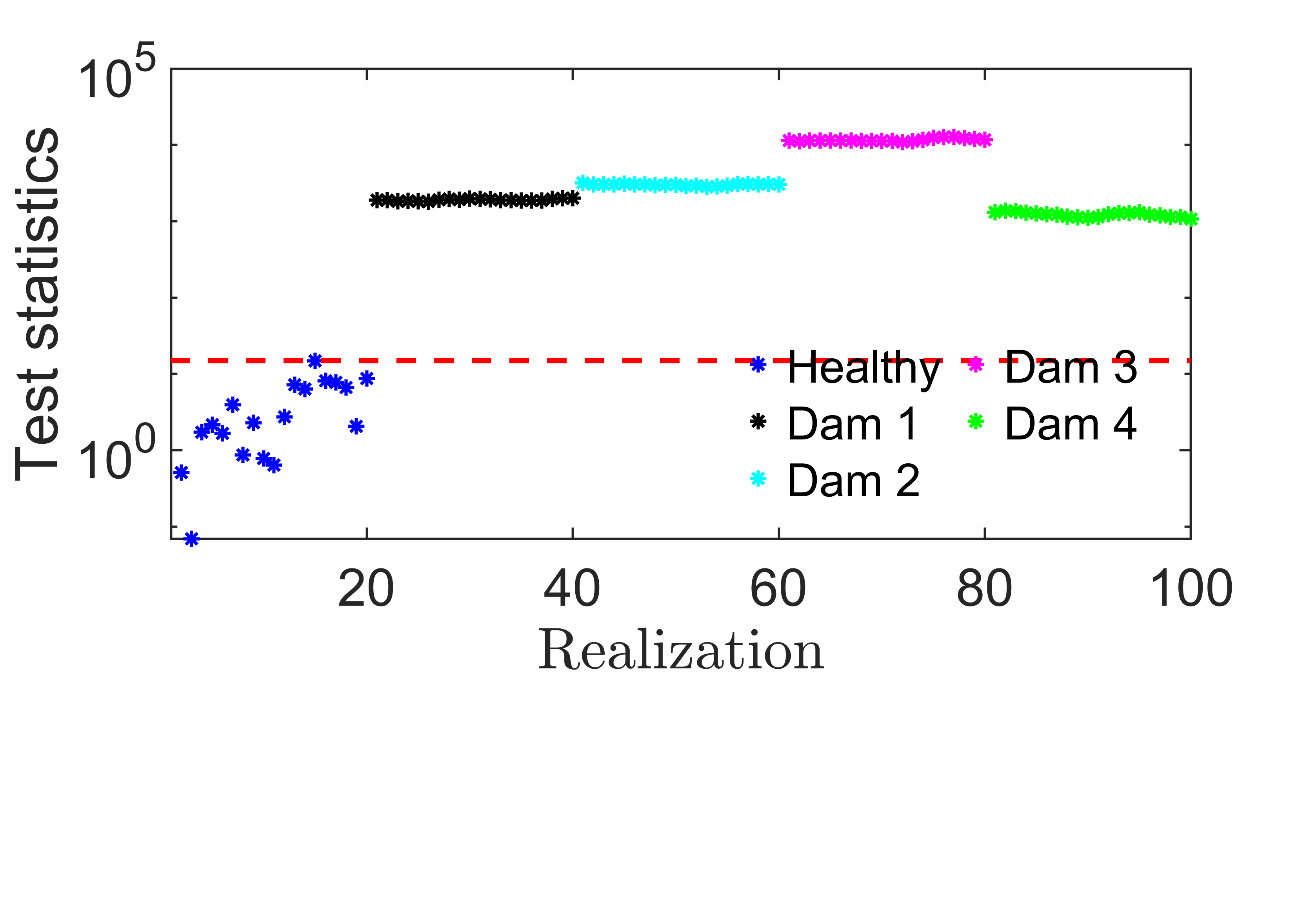}}
    \put(114,10){\includegraphics[width=0.36\columnwidth]{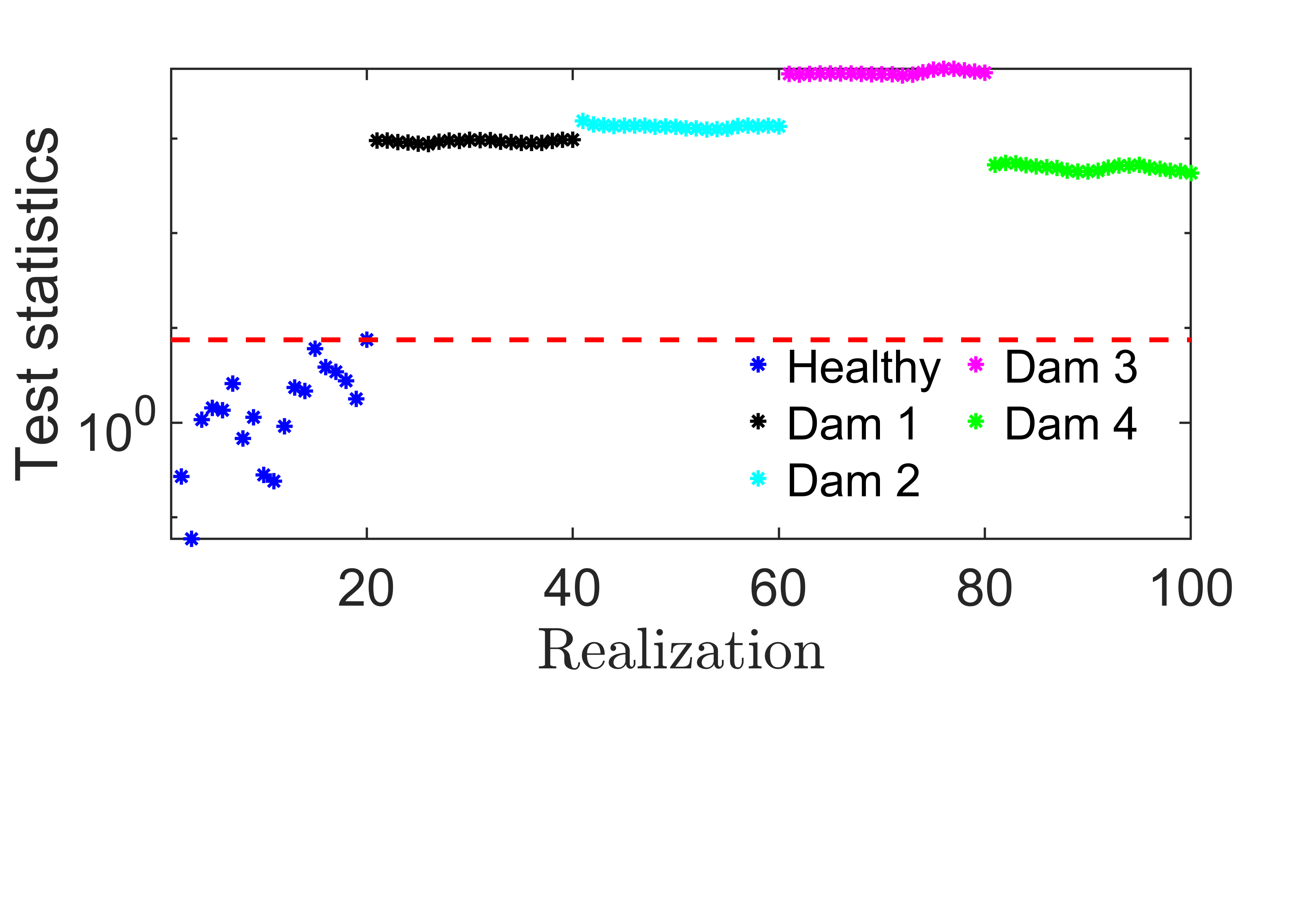}}
    \put(275,10){\includegraphics[width=0.36\columnwidth]{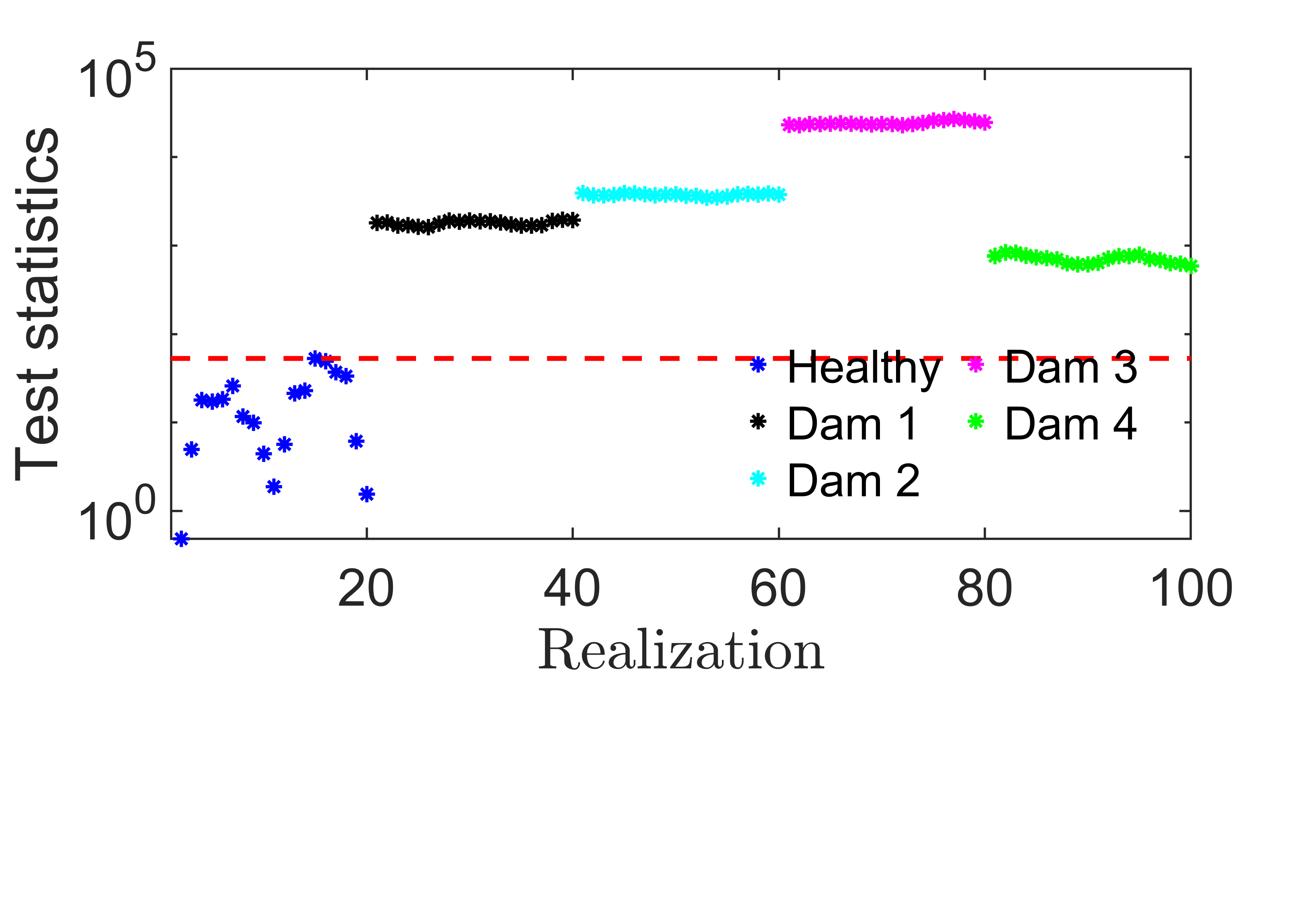}}

    \put(-20,125){\large \textbf{(a)}}
    \put(140,125){\large \textbf{(b)}}
    \put(290,125){ \large \textbf{(c)}}
    \end{picture}
    \vspace{-1.5cm}
    
    \caption{Damage detection performance comparison for damage intersecting path 2-6 using the covariance matrix derived from 20 experimental healthy signals: (a) standard AR approach; (b) SVD-based approach; (c) PCA-based approach.} 
\label{fig:dam intersect exp} 
\end{figure}
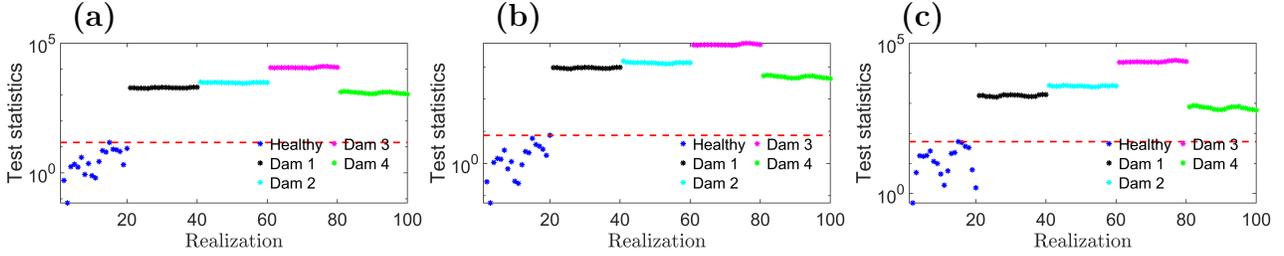 

\begin{figure}[t!]
    \centering
    \begin{picture}(400,130)
     \put(-43,10){\includegraphics[width=0.36\columnwidth]{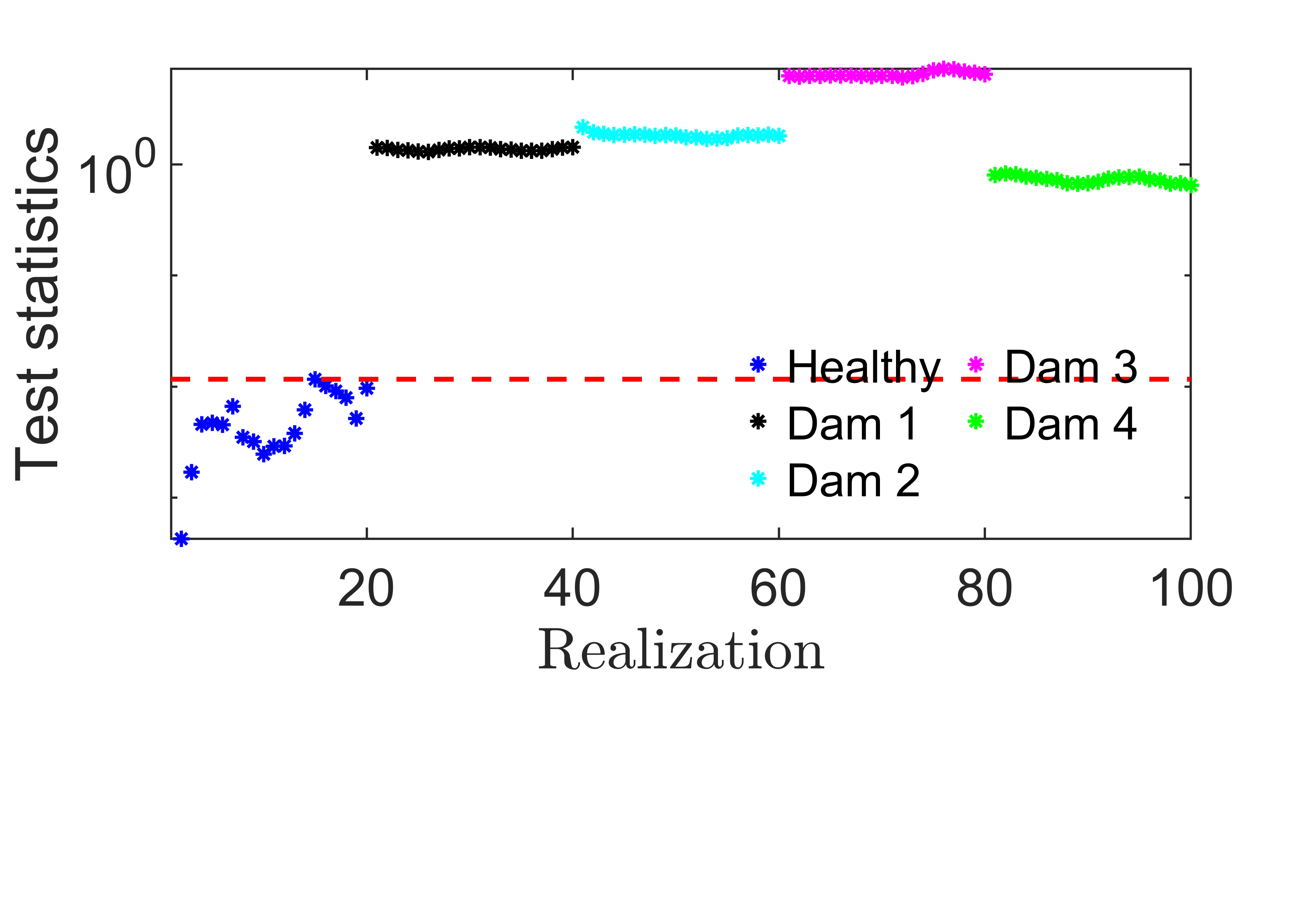}}
    \put(114,10){ \includegraphics[width=0.36\columnwidth]{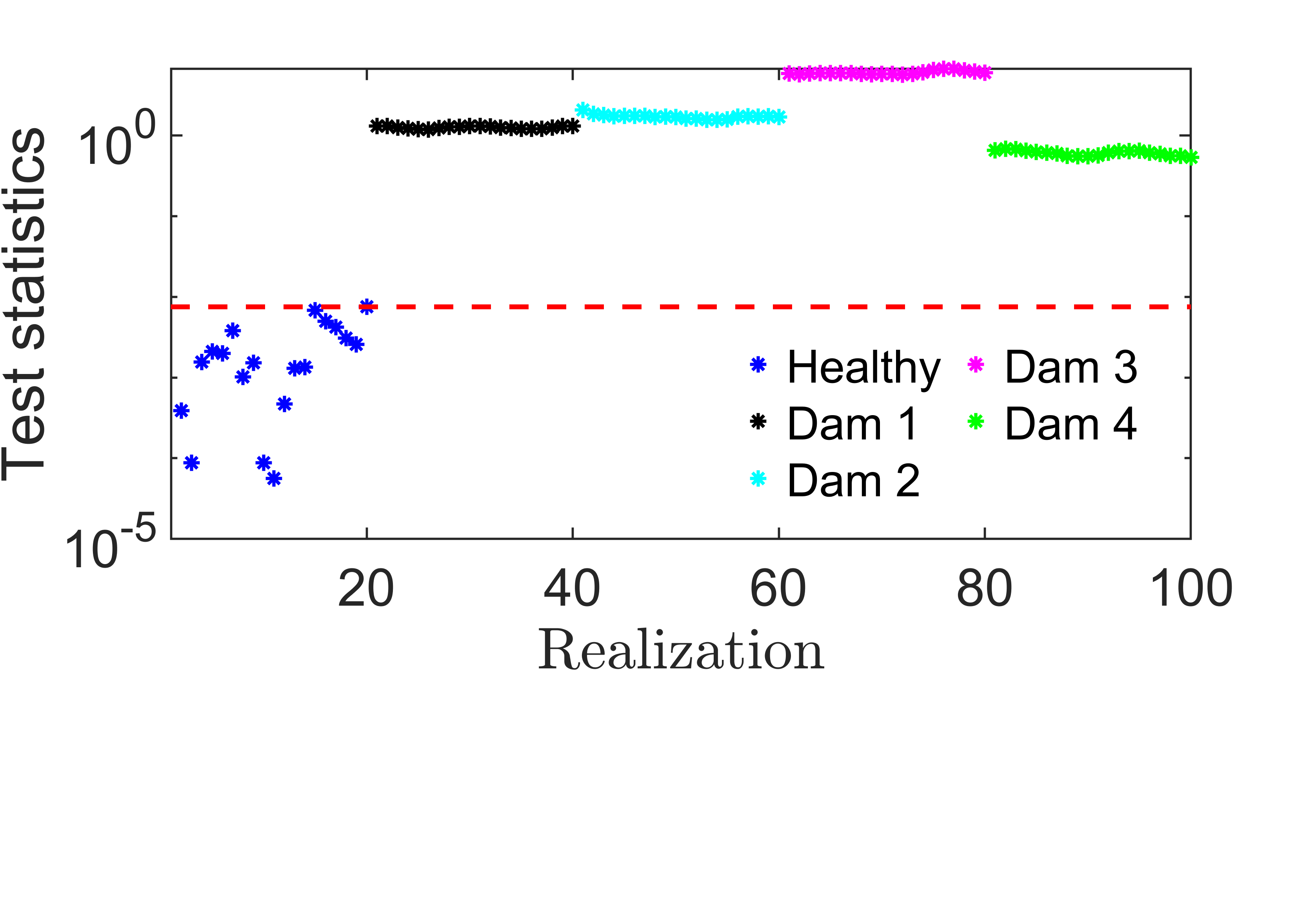}}
    \put(280,10){\includegraphics[width=0.36\columnwidth]{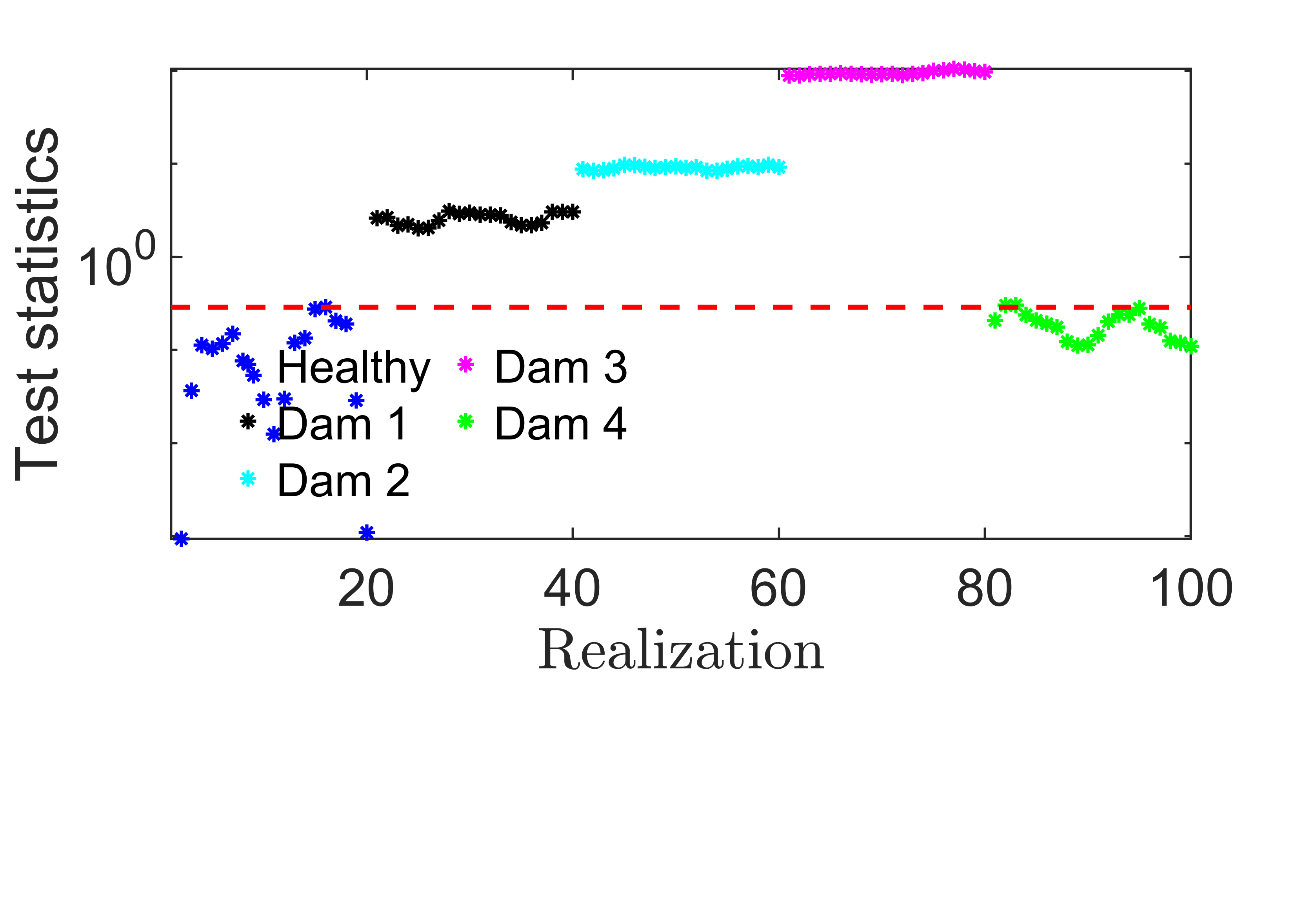}}

    \put(-20,125){\large \textbf{(a)}}
    \put(140,125){\large \textbf{(b)}}
    \put(290,125){ \large \textbf{(c)}}
    \end{picture}
    \vspace{-1.5cm}
    
    \caption{Damage detection performance comparison for damage intersecting path 2-6 using the AR($4$)-based covariance matrix: (a) standard AR approach; (b) SVD-based approach; (c) PCA-based approach.} 
\label{fig:dam intersect theory} 
\end{figure}
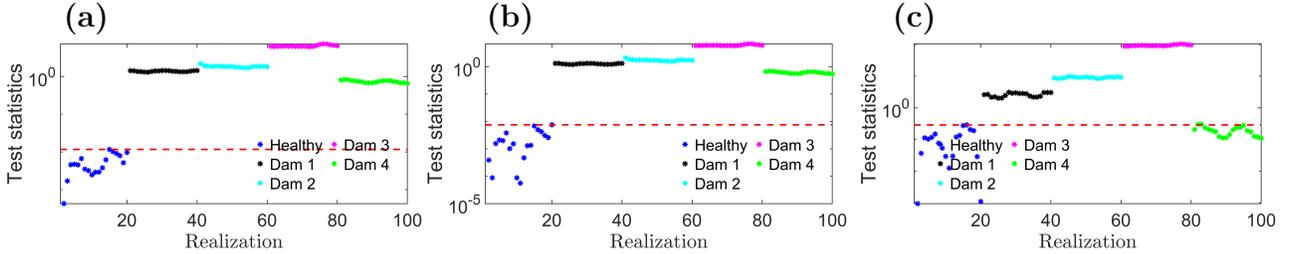

Figure \ref{fig:dam intersect exp} shows the damage detection performance of the damage intersecting path 2-6 using the standard AR, SVD, and PCA-based approach. In this case, the covariance matrix was derived from the 20 experimental healthy signals. It can be observed that for all three cases, perfect damage detection was achieved. The $\alpha$ level used for the standard AR, SVD and PCA-based approach was $1 \times 10^{-3}$, $0.1$ and $1 \times 10^{-11}$, respectively. Similarly, Figure \ref{fig:dam intersect theory} shows the damage detection using the above-mentioned three methods using the covariance matrix derived from the AR($4$) model. It can be observed that for the standard AR and SVD-based method, perfect damage detection was achieved. However, for the PCA-based approach, damage level 4 goes inside the threshold. For all three cases, the thresholds were manually adjusted as $\alpha$ level exceeds the numerical limit when using AR($4$)-based covariance.

\begin{figure}[t!]
    \centering
    \begin{picture}(400,130)
    \put(-43,10){\includegraphics[width=0.36\columnwidth]{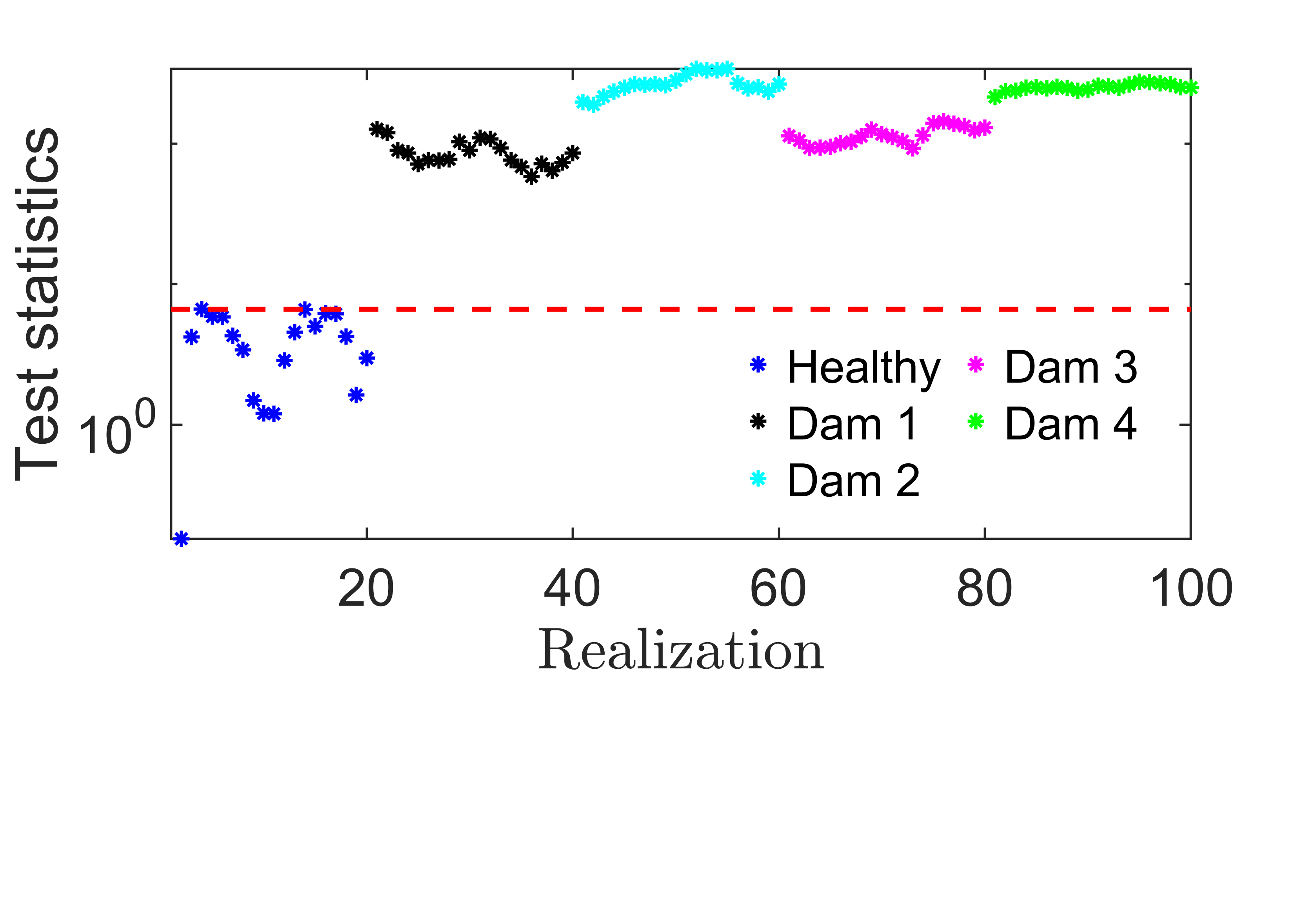}}
    \put(114,10){ \includegraphics[width=0.36\columnwidth]{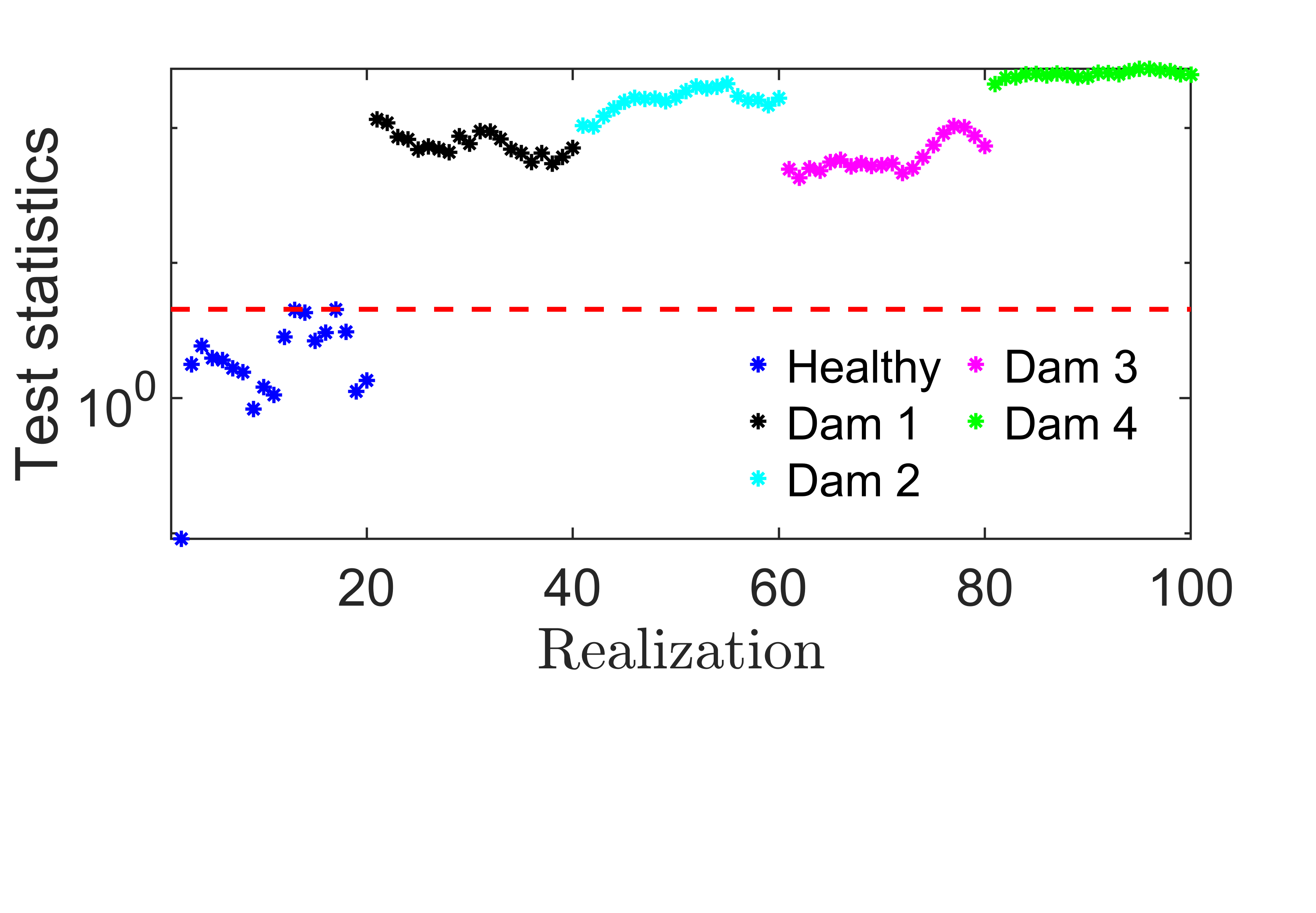}}
    \put(280,10){\includegraphics[width=0.36\columnwidth]{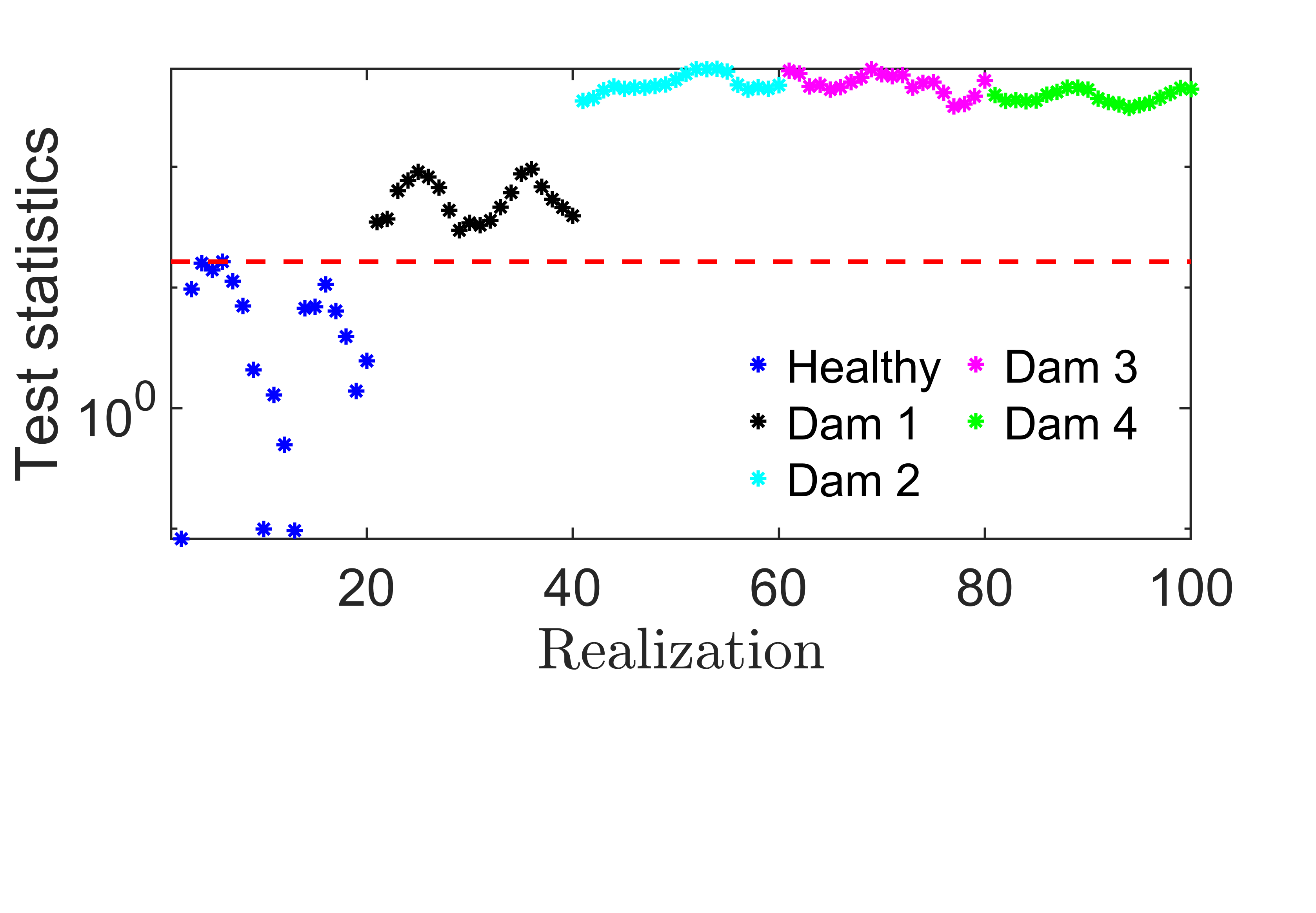}}

    \put(-20,125){\large \textbf{(a)}}
    \put(140,125){\large \textbf{(b)}}
    \put(290,125){ \large \textbf{(c)}}
    \end{picture}
    \vspace{-1.5cm}
    
    \caption{Damage detection performance comparison for damage non-intersecting path 1-4 using the covariance matrix derived from 20 experimental healthy signals: (a) standard AR approach; (b) SVD-based approach; (c) PCA-based approach.} 
\label{fig:dam nonintersect exp} 
\end{figure}

Figure \ref{fig:dam nonintersect exp} shows the damage detection performance of the damage non-intersecting path 1-4 using the standard AR, SVD, and PCA-based approach. In this case, the covariance matrix was derived from the 20 experimental healthy signals. It can be observed that perfect damage detection was achieved for all three cases when experimental covariance was used. The $\alpha$ level used for the standard AR, SVD and PCA-based approach was $0.15$, $0.99995$, and $1 \times 10^{-3}$, respectively.

\begin{figure}[t!]
    \centering
    \begin{picture}(400,130)
    \put(-43,10){\includegraphics[width=0.36\columnwidth]{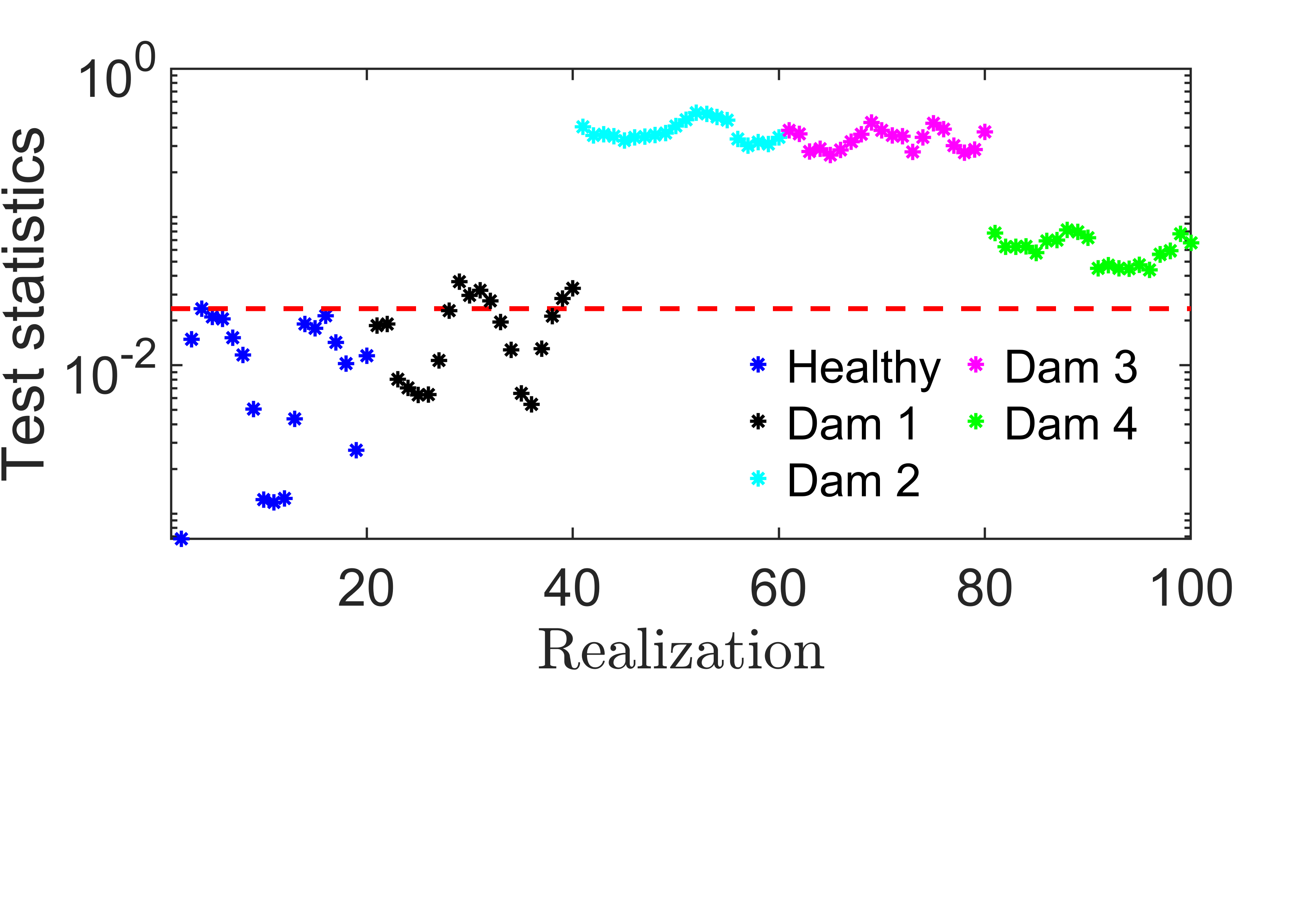}}
    \put(114,10){ \includegraphics[width=0.36\columnwidth]{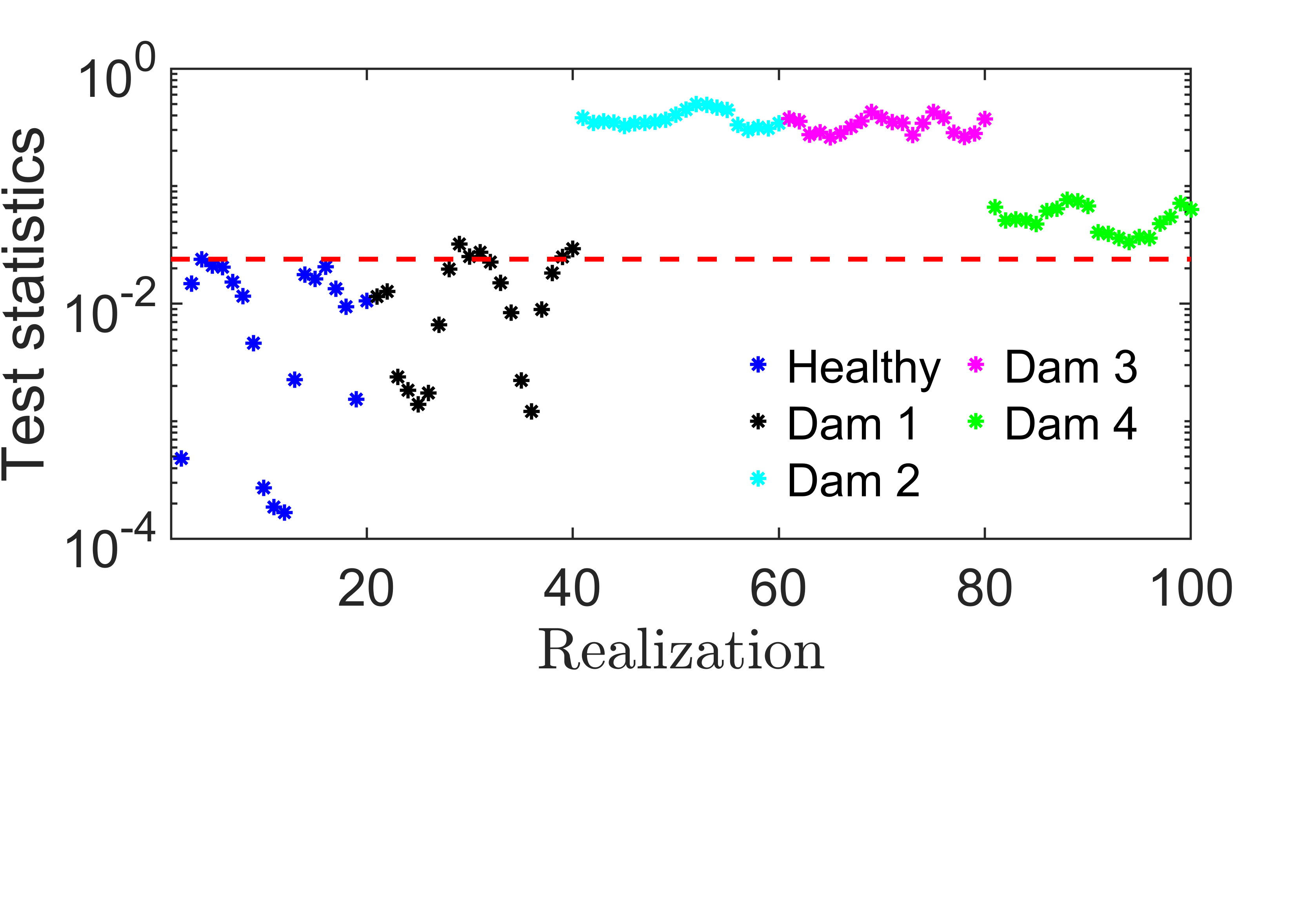}}
    \put(280,10){\includegraphics[width=0.36\columnwidth]{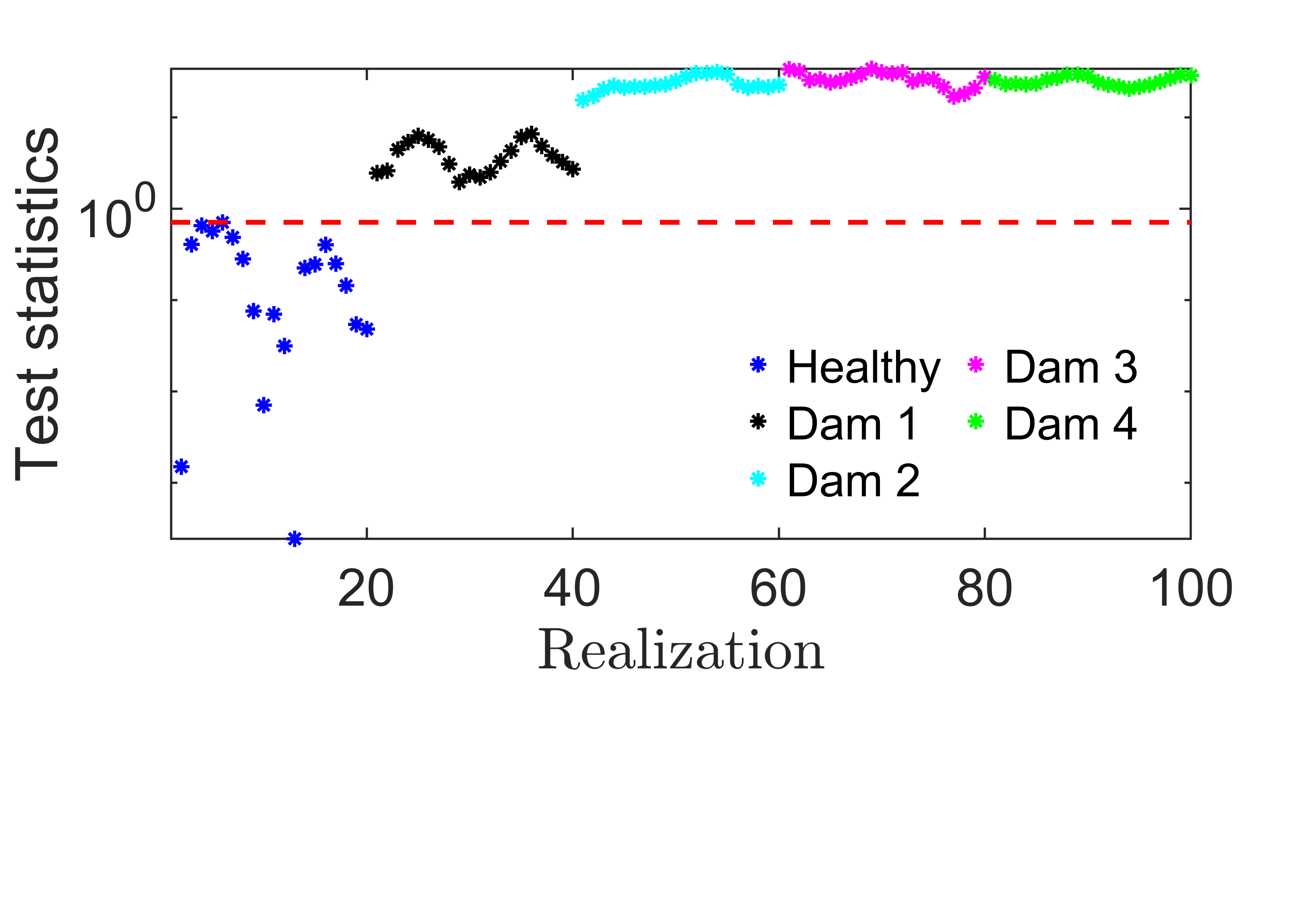}}

    \put(-20,125){\large \textbf{(a)}}
    \put(140,125){\large \textbf{(b)}}
    \put(290,125){ \large \textbf{(c)}}
    \end{picture}
    \vspace{-1.5cm}
    
    \caption{Damage detection performance comparison for damage non-intersecting path 1-4 using the AR($4$)-based covariance matrix: (a) standard AR approach; (b) SVD-based approach; (c) PCA-based approach.} 
\label{fig:dam nonintersect theory} 
\end{figure}

When using the AR($4$)-based covariance matrix, the performance of standard AR and SVD-based method was bad as the damage level 1 was missed(Figure \ref{fig:dam nonintersect theory}(a) and (b)). However, the PCA-based approach perfectly detects damage when the AR($4$)-based covariance was used (Figure \ref{fig:dam nonintersect theory}(c)). For all three cases, the thresholds were manually adjusted as $\alpha$ level exceeds the numerical limit when using AR($4$)-based covariance. Table \ref{tab: alpha al} compactly summarizes the alpha level used for the aluminum plate. 

\begin{table}[t!]
\caption{$\alpha$-level chart for aluminum plate}\label{tab: alpha al}
\centering
{\footnotesize
\centering
\begin{tabular}{lllc}
\hline
  
Method & Path & Covariance & $\alpha$-level  \\

 \hline
 Standard AR & 1-4 & Experiment &  $0.15$ \\
 & 1-4 & Theory & manual \\
  & 2-6 & Experiment & $1\times 10^{-3}$  \\
  & 2-6 & Theory & manual \\
\hline
SVD-based & 1-4 & Experiment &0.99995 \\
 & 1-4 & Theory & manual \\
  & 2-6 & Experiment & $0.1$\\
  & 2-6 & Theory & manual \\
\hline
PCA-based & 1-4 & Experiment & $1\times 10^{-3}$ \\

 & 1-4 & Theory & manual\\
 
 & 2-6 & Experiment & $1\times 10^{-11}$  \\
 & 2-6 & Theory & manual \\

\hline
\end{tabular} }
\end{table}

\begin{figure}[t!]
    \centering
    \begin{picture}(400,130)
    \put(-40,-50){ \includegraphics[width=0.55\columnwidth]{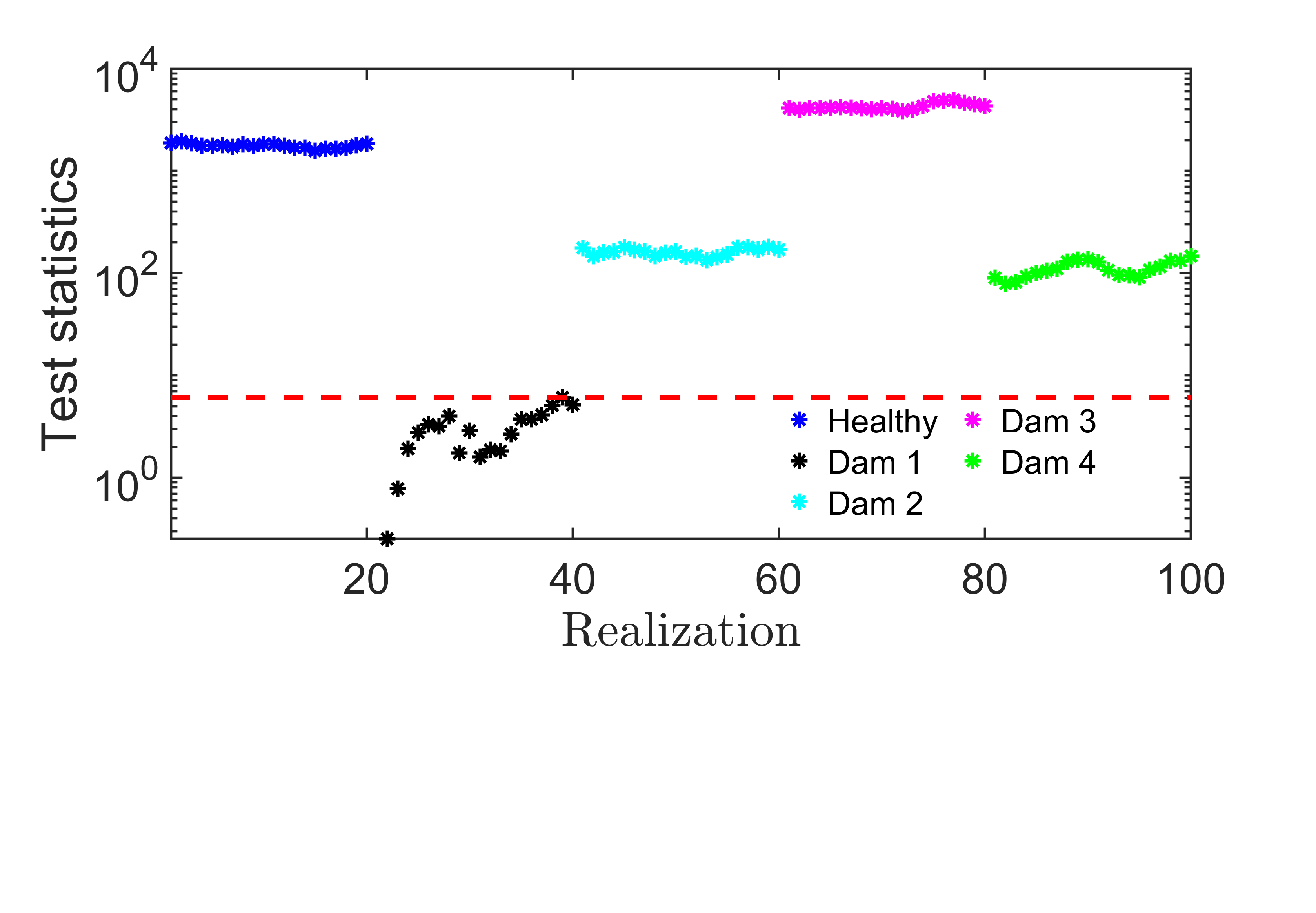}}
    \put(200,-50){\includegraphics[width=0.55\columnwidth]{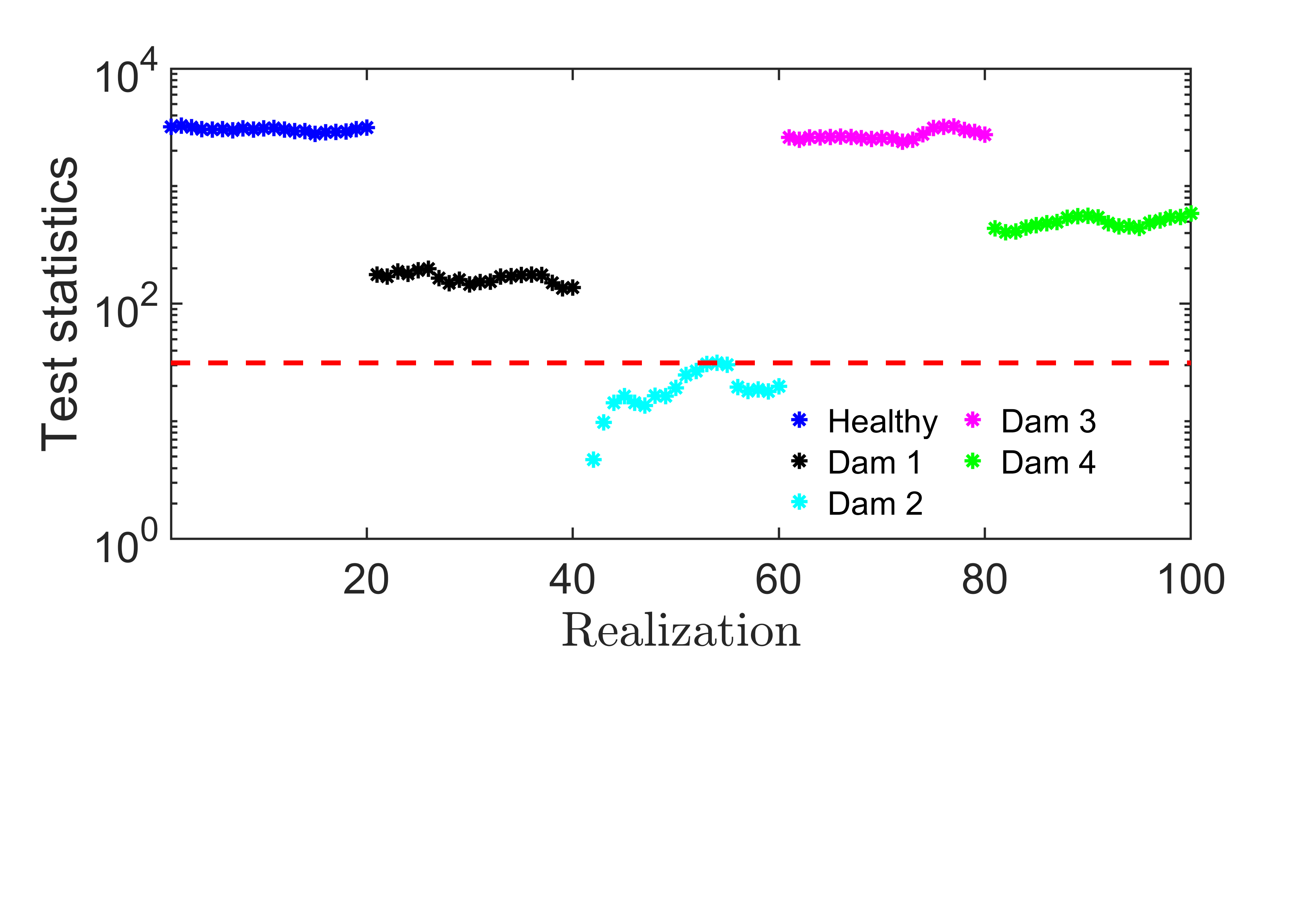}}
    
    \put(-40,-179){ \includegraphics[width=0.55\columnwidth]{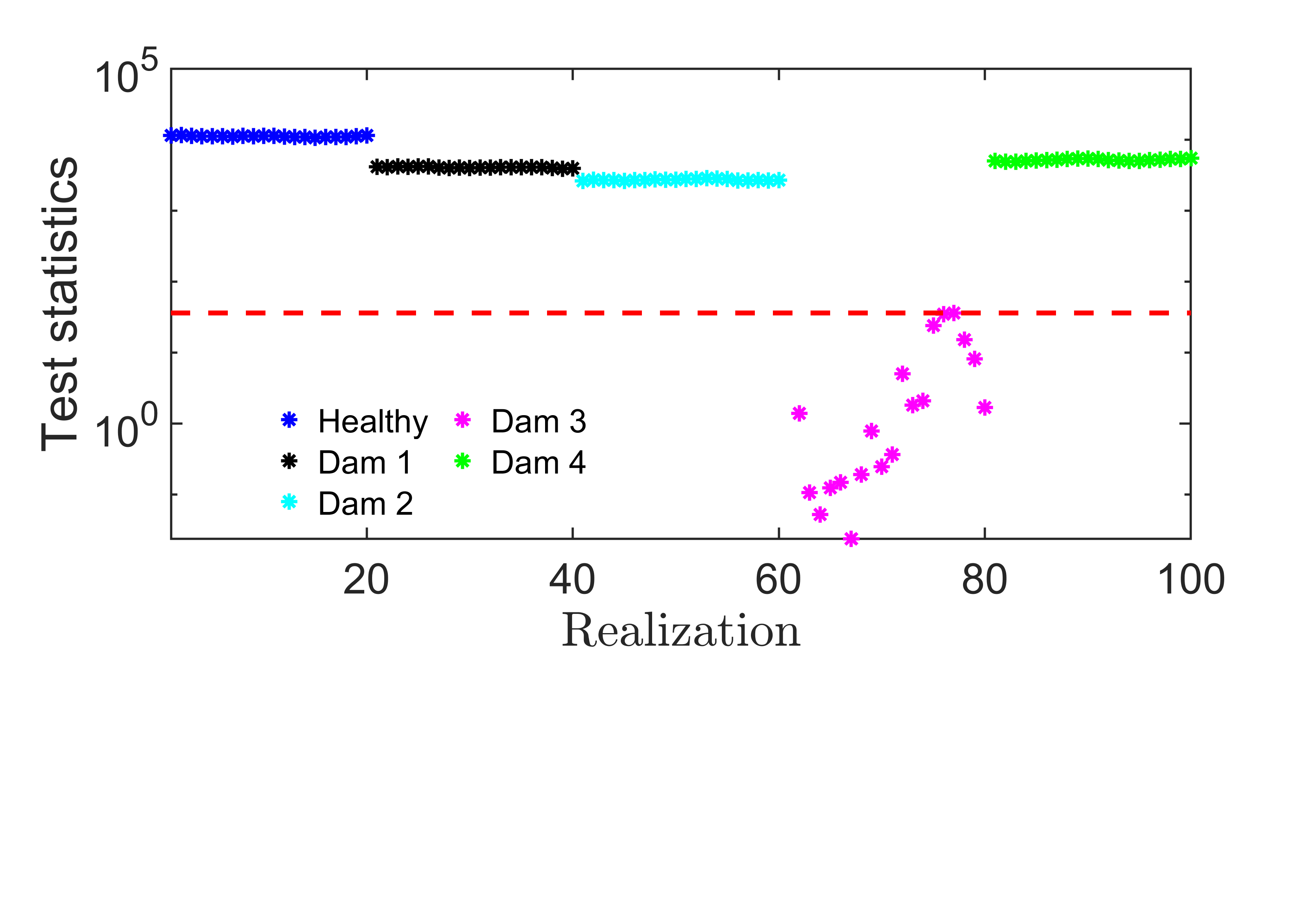}}
    \put(200,-179){\includegraphics[width=0.55\columnwidth]{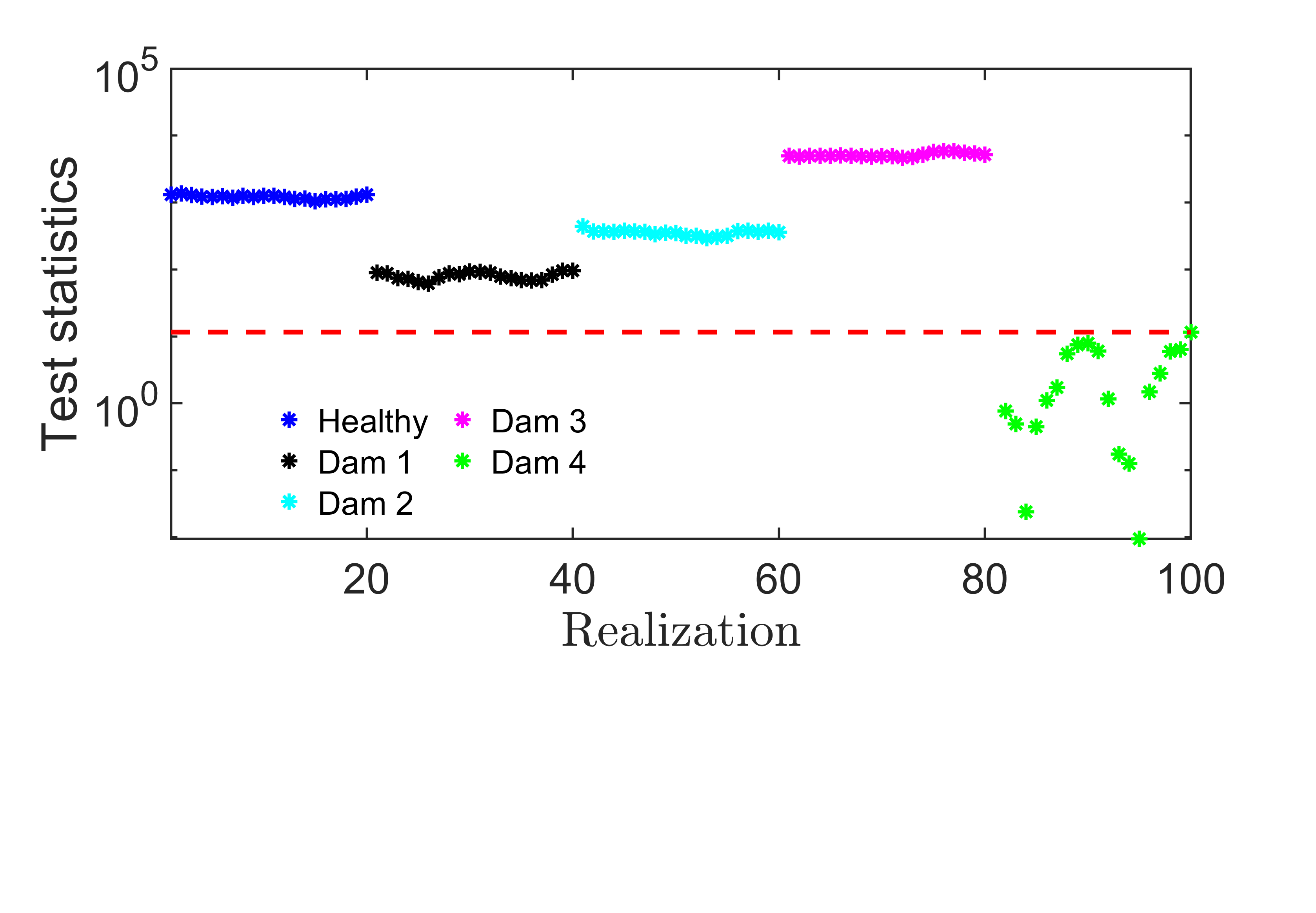}}

    \put(-15,127){\large \textbf{(a)}}
    \put(220,127){ \large \textbf{(b)}}
     \put(-15,-3){\large \textbf{(c)}}
    \put(220,-3){ \large \textbf{(d)}}
    \end{picture}
    \vspace{4.5cm}
    \caption{Damage identification results for the damage intersecting path 2-6 using the standard AR approach and using experimental covariance: (a) damage level 1; (b) damage level 2; (c) damage level 3; (d) damage level 4.} 
\label{fig:dam iden theory} 
\end{figure}
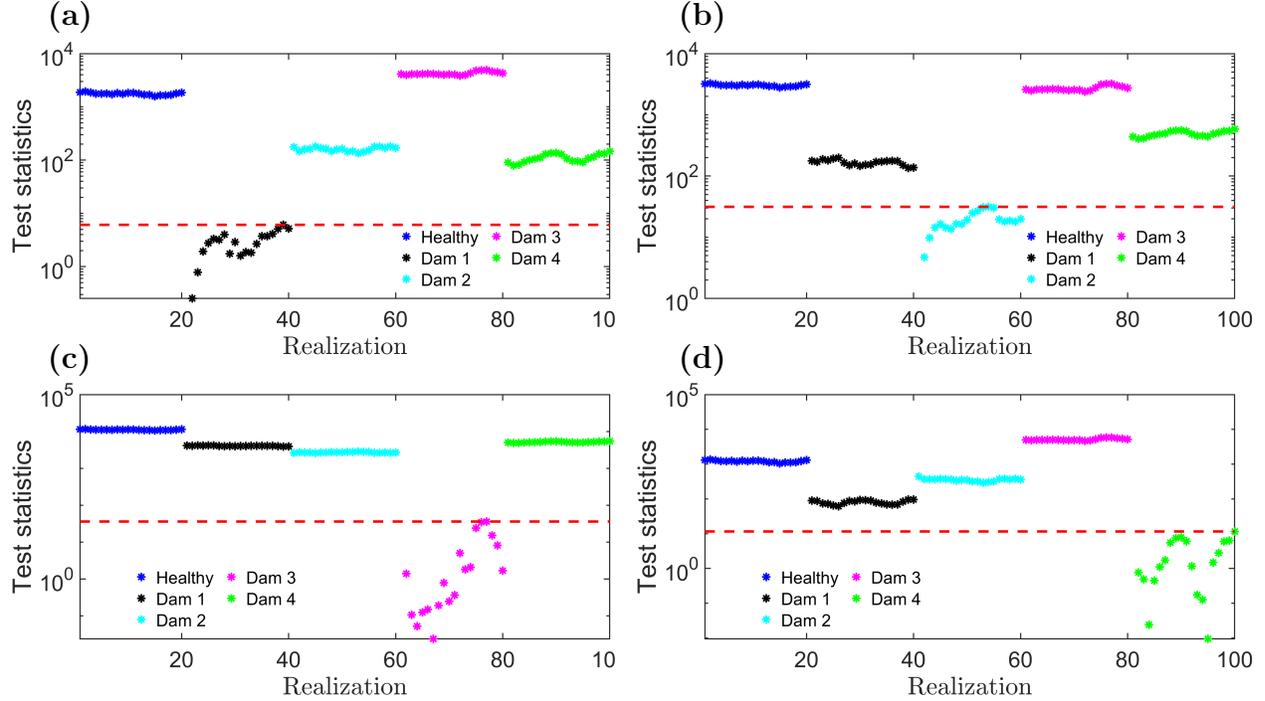 

Figure \ref{fig:dam iden theory} shows the damage identification results for the damage intersecting path 2-6 for the aluminum plate using the experimental covariance matrix. It can be observed that perfect damage identification was achieved with no missed classification. Table \ref{tab:sum_res} compactly summarizes the damage detection and identification results for the standard AR, SVD and PCA-based methods for the two indicative paths and the associated covariance matrix used. It can be observed that, for the aluminum plate, perfect damage detection and identification is possible both for damage intersecting and non-intersecting paths when the experimental covariance matrix is used.

\begin{table}[t!]
\caption{Damage detection and identification summary results for the aluminum plate.}\label{tab:sum_res}
\centering
\renewcommand{\arraystretch}{1.2}
{\footnotesize
\centering
\begin{tabular}{lllcccccc}
\hline
   &  &  &  \multicolumn{5}{|c}{\bf Damage Detection} \\  
 \cline{4-9}
Method & Path & Covariance& \multicolumn{1}{|c}{False alarms} & \multicolumn{5}{|c}{Missed damage} \\
\cline{4-9}
 \multicolumn{3}{c|}{}  &Healthy & Damage 1 & Damage 2 & Damage 3 & Damage 4 & \\
 \hline
 Standard AR & 1-4 & Experiment & 0/20&0/20 &0/20 &0/20 &0/20 & \\
 & 1-4 & Theory &0/20 &16/20 &0/20 &0/20 &0/20 & \\
  & 2-6 & Experiment &0/20 &0/20 &0/20 &0/20 &0/20 & \\
  & 2-6 & Theory &0/20 &0/20 &0/20 &0/20 &0/20 & \\
\hline
SVD-based & 1-4 & Experiment & 0/20&0/20 &0/20 &0/20 &0/20 & \\
 & 1-4 & Theory &0/20 & 15/20&0/20 &0/20 &0/20 & \\
  & 2-6 & Experiment &0/20 &0/20 &0/20 &0/20 &0/20 & \\
  & 2-6 & Theory &0/20 &0/20 &0/20 &0/20 &0/20 & \\
\hline
PCA-based & 1-4 & Experiment &0/20 &0/20 &0/20 &0/20 &0/20 & \\

 & 1-4 & Theory &0/20 &0/20 &0/20 &0/20 &0/20 & \\
 
 & 2-6 & Experiment &0/20 &0/20 &0/20 &0/20 &0/20 & \\
 & 2-6 & Theory &0/20 &0/20 &0/20 &0/20 &20/20 & \\

\hline
\end{tabular} 

\begin{tabular}{lllccccc}

   &  &  &  \multicolumn{4}{|c}{\bf Damage Identification} \\  
 \cline{4-8}
Method & Path & Covariance& \multicolumn{5}{|c}{Missed damage} \\
\cline{4-8}
 \multicolumn{3}{c|}{}  & Damage 1 & Damage 2 & Damage 3 & Damage 4 & \\
\hline
Standard AR & 1-4 & Experiment &(--,0,0,0) &(0,--,0,0) &(0,0,--,0) &(0,0,0--) & \\
 & 1-4 & Theory &(--,0,0,0) &(0,--,13,0) &(0,20,--,0) &(0,0,0,--) & \\
  & 2-6 & Experiment  &(--,0,0,0) &(0,--,0,0) &(0,0,--,0) &(0,0,0,--) & \\
  & 2-6 & Theory & (--,0,0,0)&(0,--,0,0) &(0,0,--,0) &(0,0,0,--) & \\
\hline
SVD-based & 1-4 & Experiment &(--,0,0,0) &(0,--,0,0) &(0,0,--,0) &(0,0,0,--) & \\
 & 1-4 & Theory &(--,0,0,0) &(0,--,12,0) &(0,18,--,0) &(0,0,0,--) & \\
  & 2-6 & Experiment  &(--,0,0,0) &(0,--,0,0) &(0,0,--,0) &(0,0,0,--) & \\
  & 2-6 & Theory & (--,0,0,0)&(0,--,0,0) &(0,0,--,0) &(0,0,0,--) & \\
\hline
PCA-based & 1-4 & Experiment & (--,0,0,0)&(0,--,18,0) &(0,20,--,8) & (0,0,0,--) & \\

 & 1-4 & Theory &(--,0,0,0) &(5,--,17,20) &(0,19,--,20) &(0,14,13,--) & \\
 
 & 2-6 & Experiment &(--,0,0,0) & (0,--,0,0)&(0,0,--,0) &(0,0,0,--) & \\
 & 2-6 & Theory &(--,0,0,0) &(0,--,0,0) &(0,0,--,0) &(0,0,0,--) & \\

\hline
\end{tabular} }

\end{table}

\begin{figure}[t!]
    \centering
    \begin{picture}(400,130)
    \put(-40,-50){ \includegraphics[width=0.5\columnwidth]{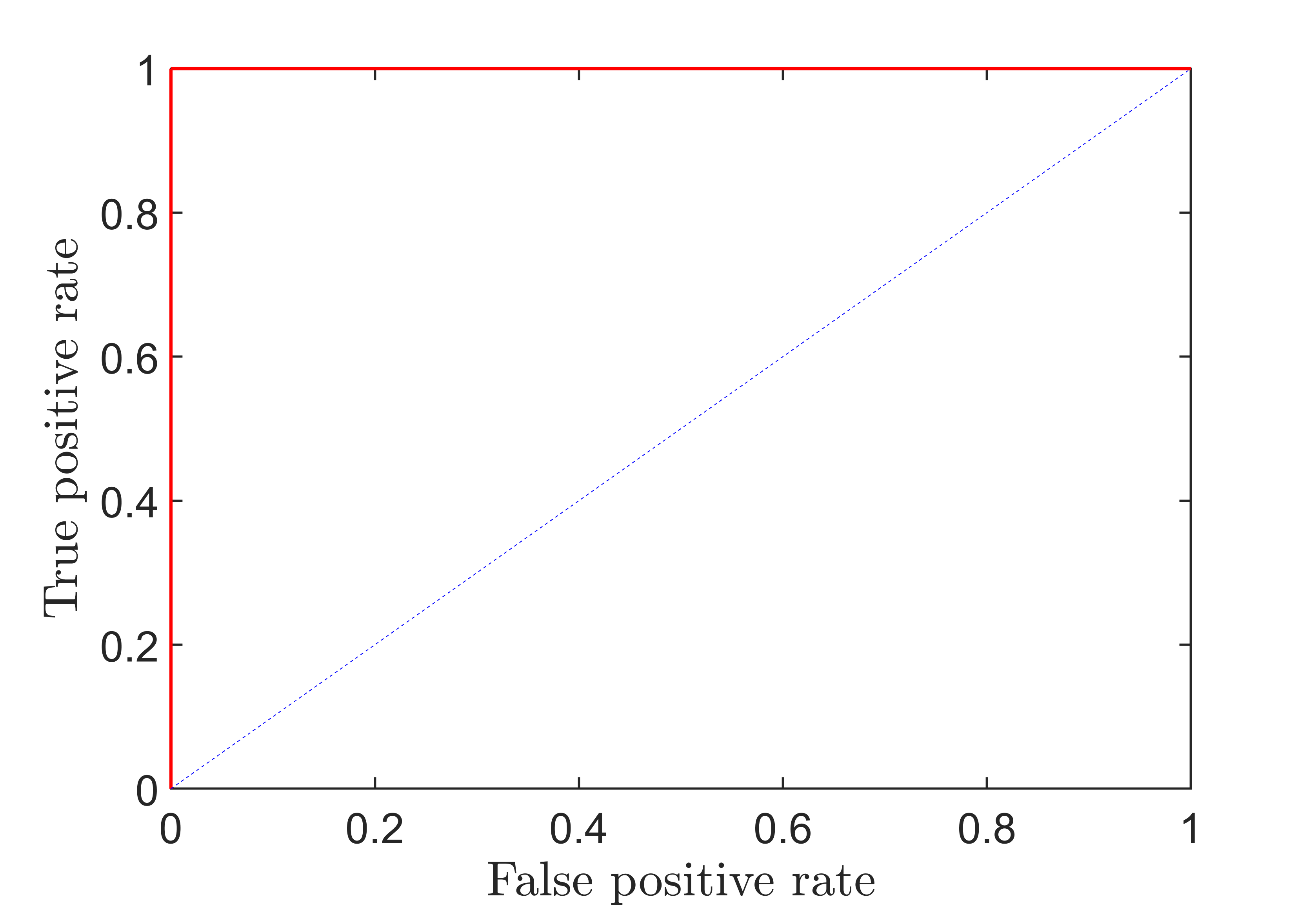}}
    \put(200,-50){\includegraphics[width=0.5\columnwidth]{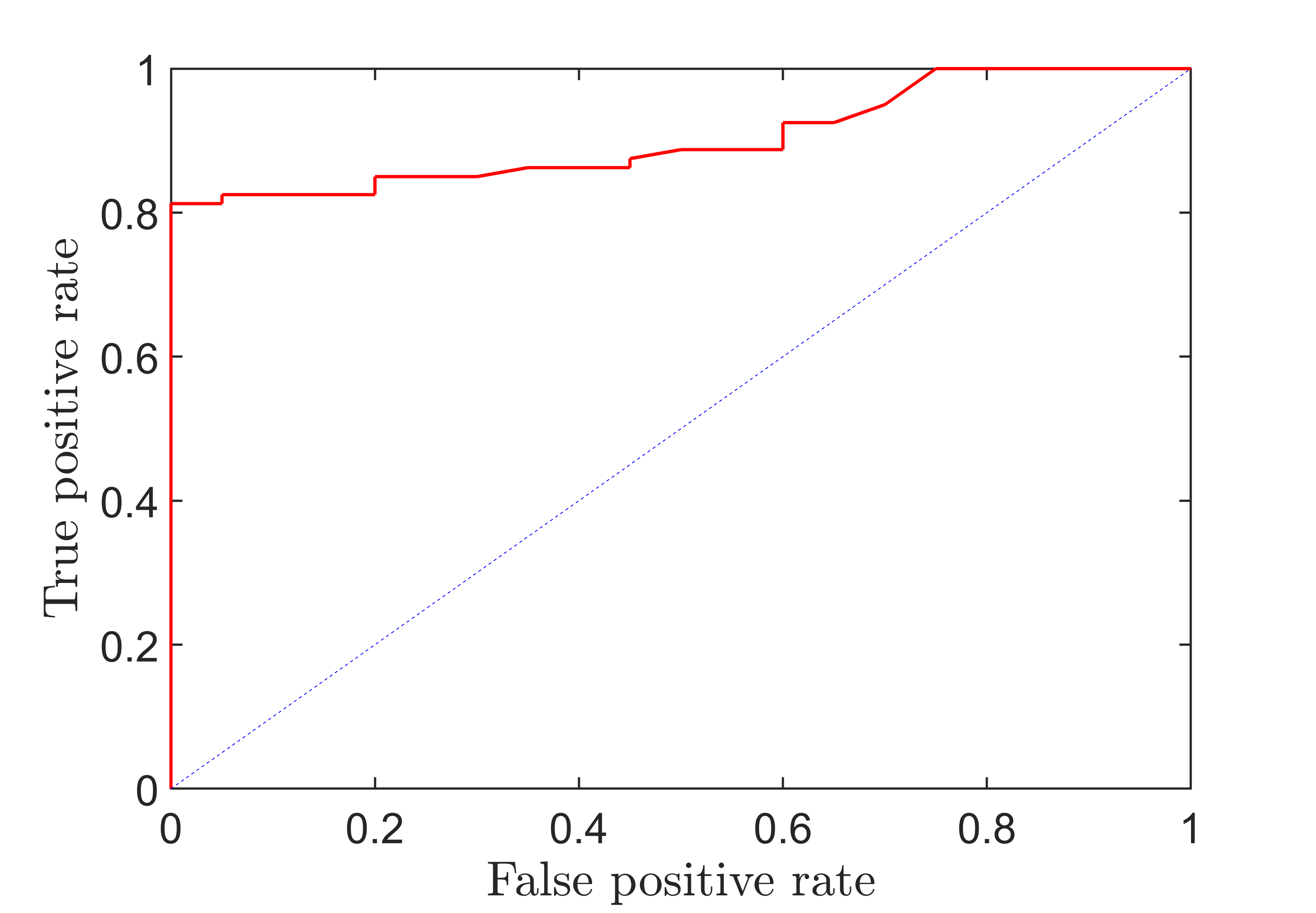}}
    
    \put(-40,-230){ \includegraphics[width=0.5\columnwidth]{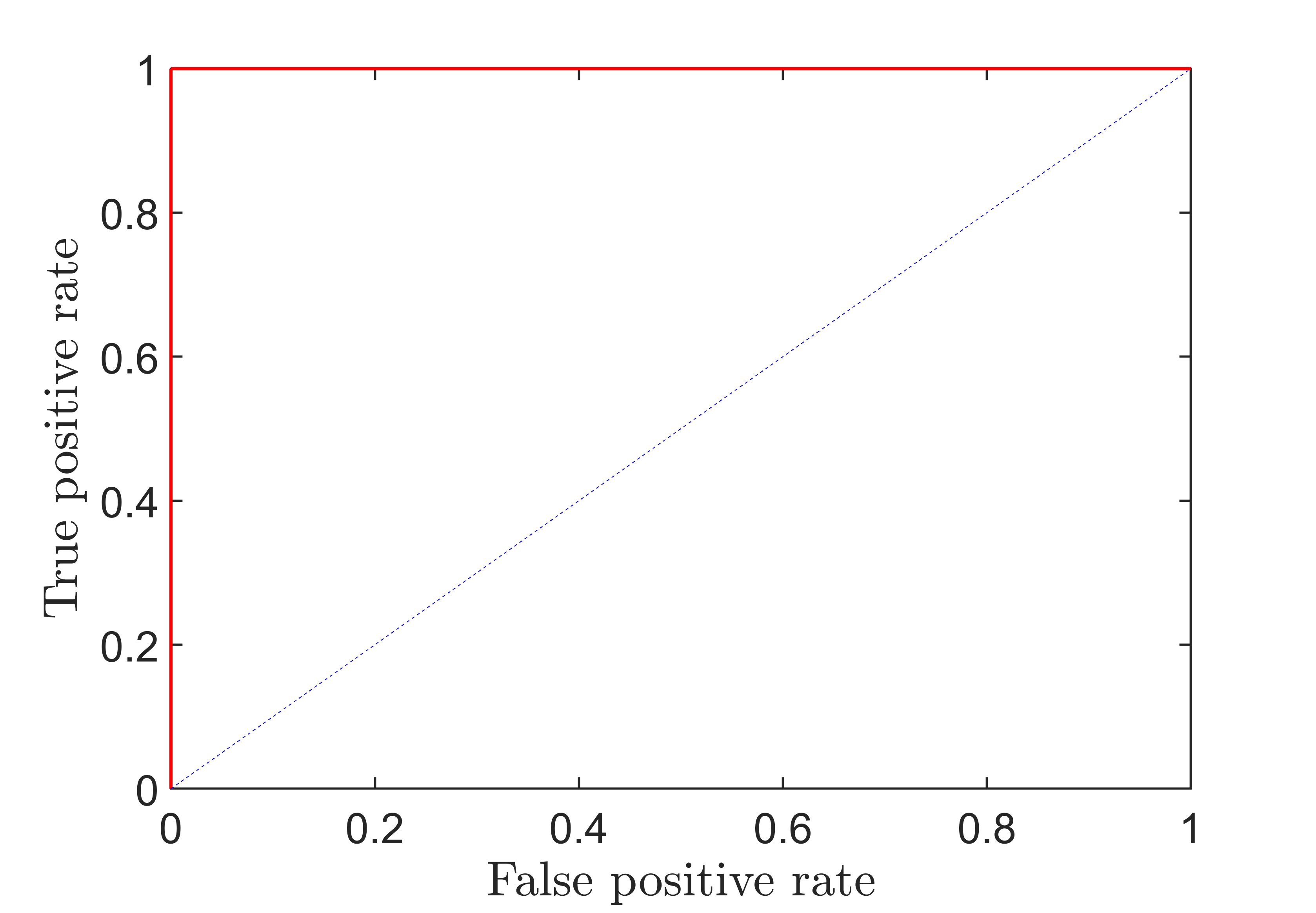}}
    \put(200,-230){\includegraphics[width=0.5\columnwidth]{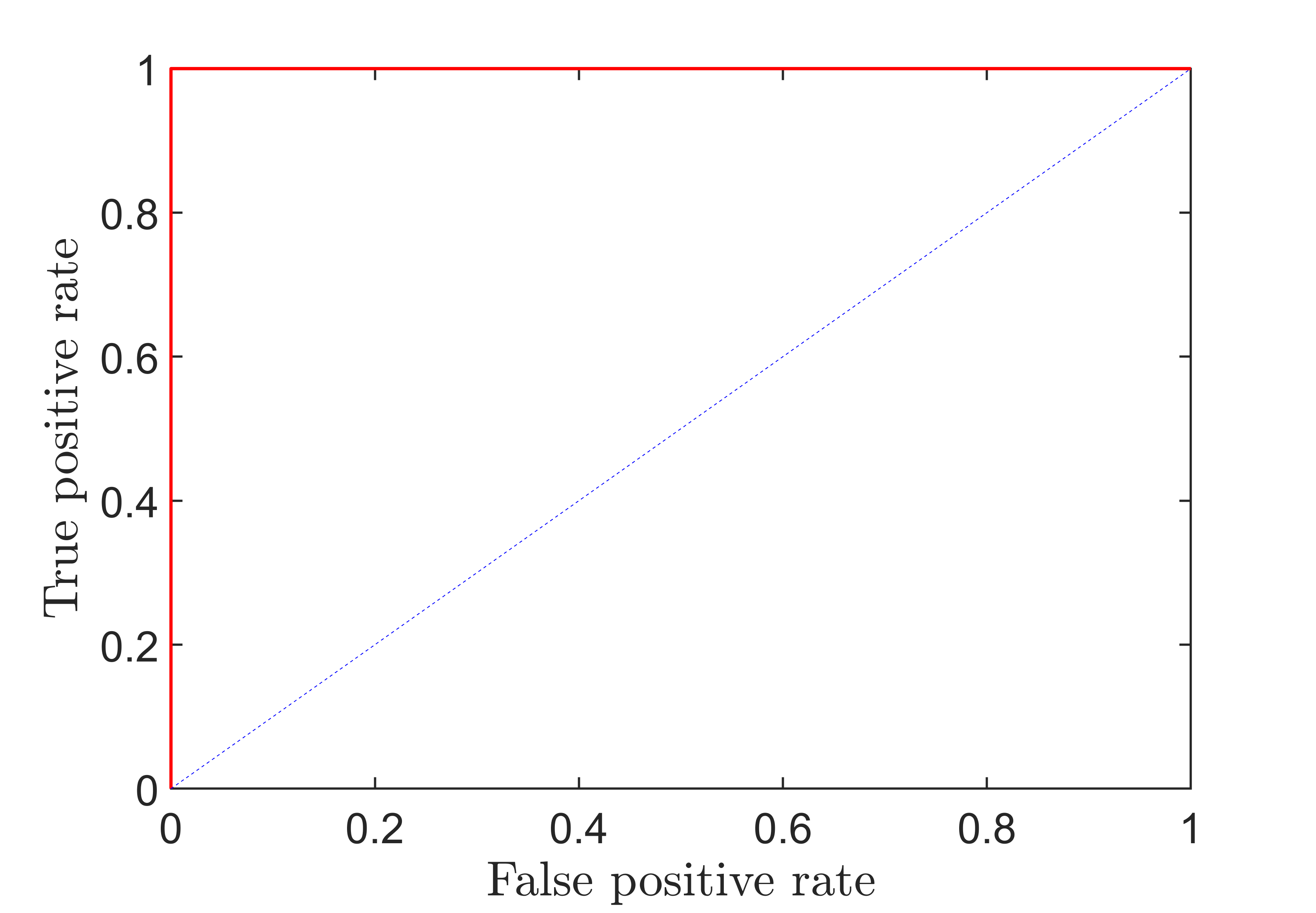}}

    \put(-15,110){\large \textbf{(a)}}
    \put(220,110){ \large \textbf{(b)}}
     \put(-15,-70){\large \textbf{(c)}}
    \put(220,-70){ \large \textbf{(d)}}
    \end{picture}
    \vspace{8cm}
    \caption{Receiver operating characteristic (ROC) plots comparing the SVD-based damage detection methods for the aluminum plate for different paths and covariance used: (a) path 1–4 with the experimental covariance; (b) path 1–4 with the AR($4$)-based covariance; (c) path 2–6 with the experimental covariance; and (d) path 2–6 with the AR($4$)-based covariance.} 
\label{fig:ROC al} 
\end{figure} 

The proper selection of the risk level $\alpha$ (type I error) is of crucial importance as it is associated with the methods’ robustness and effectiveness. If the $\alpha$-level is not properly adjusted, damage diagnosis will be ineffective, as false alarm, missed damage, and damage misclassification cases may occur. In order to take into account this issue, receiver operating characteristics (ROC) curves can be plotted, which investigates the relationship between the false positive rate and true positive rate for different $\alpha$-levels. ROC plots may provide insight into the methods’ robustness and effectiveness.

Figure \ref{fig:ROC al} shows the ROC plots of the SVD-based method for the damage non-intersecting path 1-4 and the damage intersecting path 2-6 using the experimental as well as the theoretical covariance matrix. In constructing each plot, the threshold of the Q-statistics was varied from -100 to $10^5$ with an increment of 1, that is, covering all possible values. It can be observed that for the aluminum plate, in terms of damage detection, the damage intersecting paths perform better than the damage non-intersecting paths.

\section{Test Case II: CFRP Plate with Simulated Damage }

\subsection{Experimental Setup and Data Acquisition}

\begin{figure}[t!]
    \centering
    \begin{picture}(400,170)
    \put(0,-30){\includegraphics[width=0.25\columnwidth]{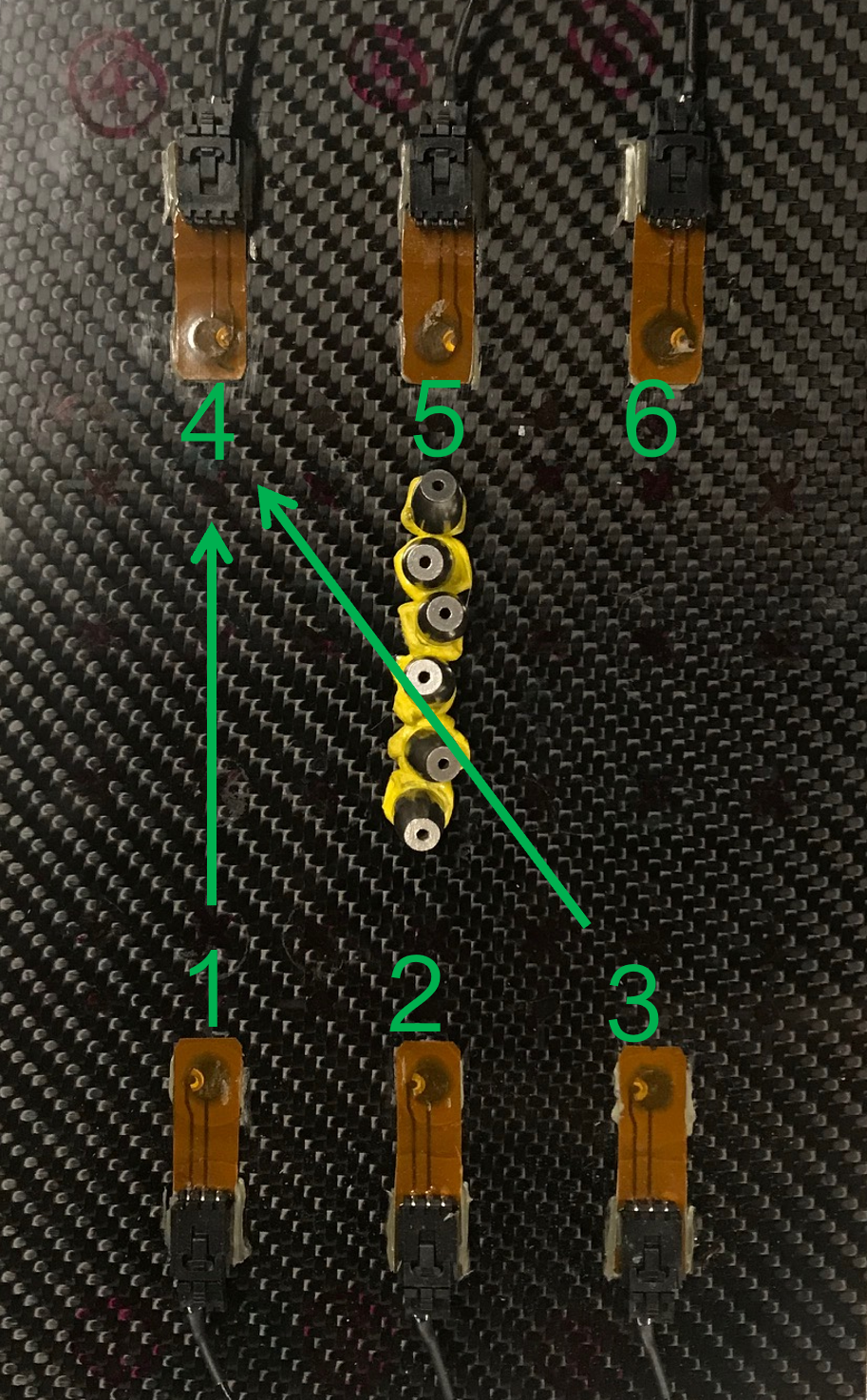}}
    
    \put(140,-50){\includegraphics[width=0.7\columnwidth]{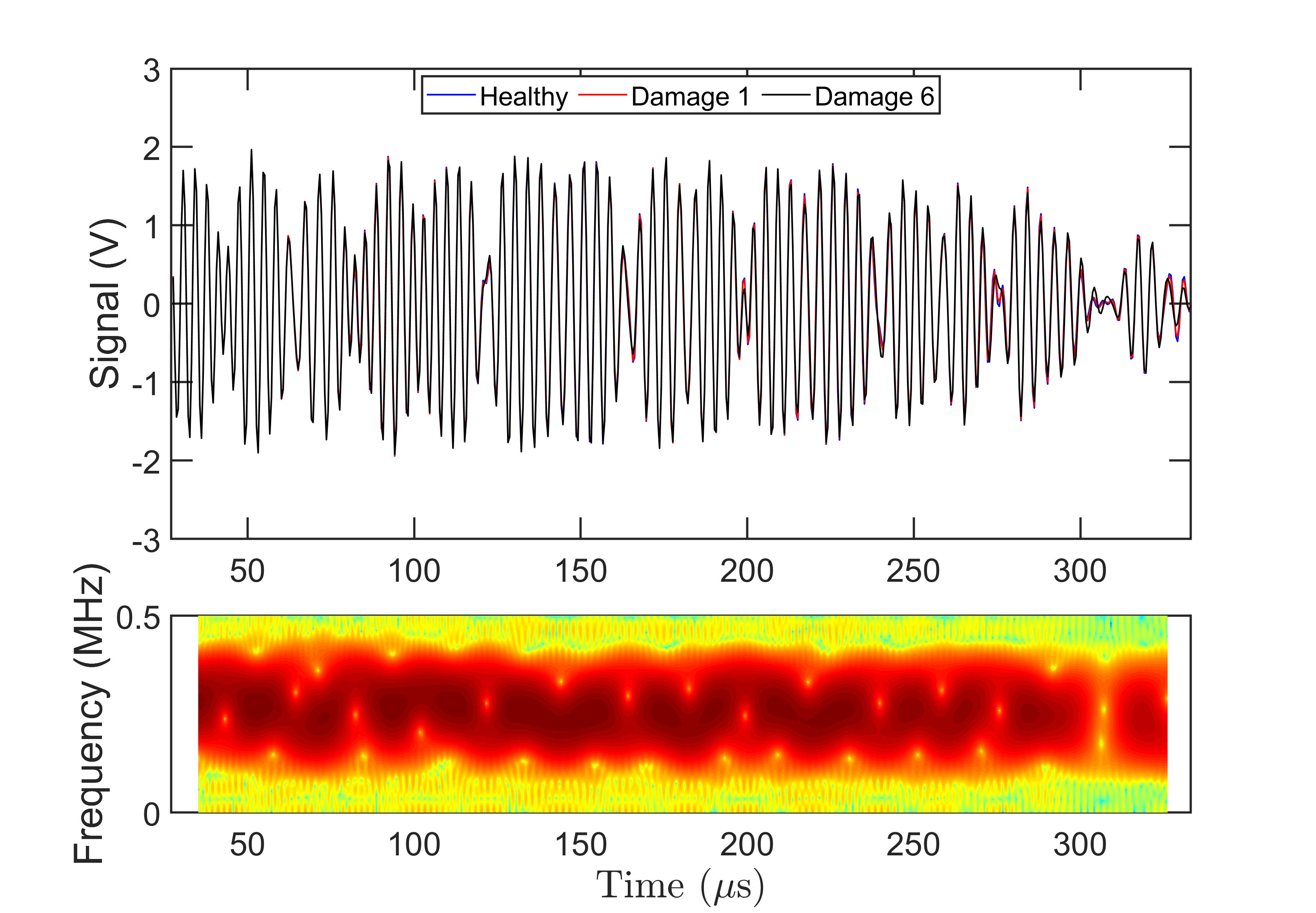}}
    \put(-22,150){ \large \textbf{(a)}}
    
    \put(150,150){\large \textbf{(b)}}    
    \end{picture}
    \vspace{1.5cm}
    \caption{(a) The CFRP plate used in this study fitted with the PZT sensors; (b) realization of the guided wave signal for healthy and damaged cases with a representative non-parametric spectrogram analysis.} 
\label{fig:composite plate} \vspace{-10pt}
\end{figure}

The second experimental setup consists of carbon fiber reinforced plastic (CFRP) coupon with a dimension of $152.4 \times 254 \times 2.36$ mm ($6\times 10 \times 0.093$ in). The coupons were acquired from ACP Composites which had $0/90^0$ unidirectional layup with Carbon fiber prepreg. Similar to the case of the aluminum coupon, the CFRP coupon was also fitted with six PZT sensors (PZT-5A) using Hysol EA 9394 adhesive. In order to simulate damage, up to six three-gram weights were sequentially attached to the surface of the plate using tacky tape as shown in Figure \ref{fig:composite plate}. Data acquisition and analysis were done in the same manner as in the case of the aluminum coupon.

\subsection{Path Selection}

\begin{figure}[t!]
    \centering
    \begin{picture}(400,135)
    \put(-40,-50){ \includegraphics[width=0.55\columnwidth]{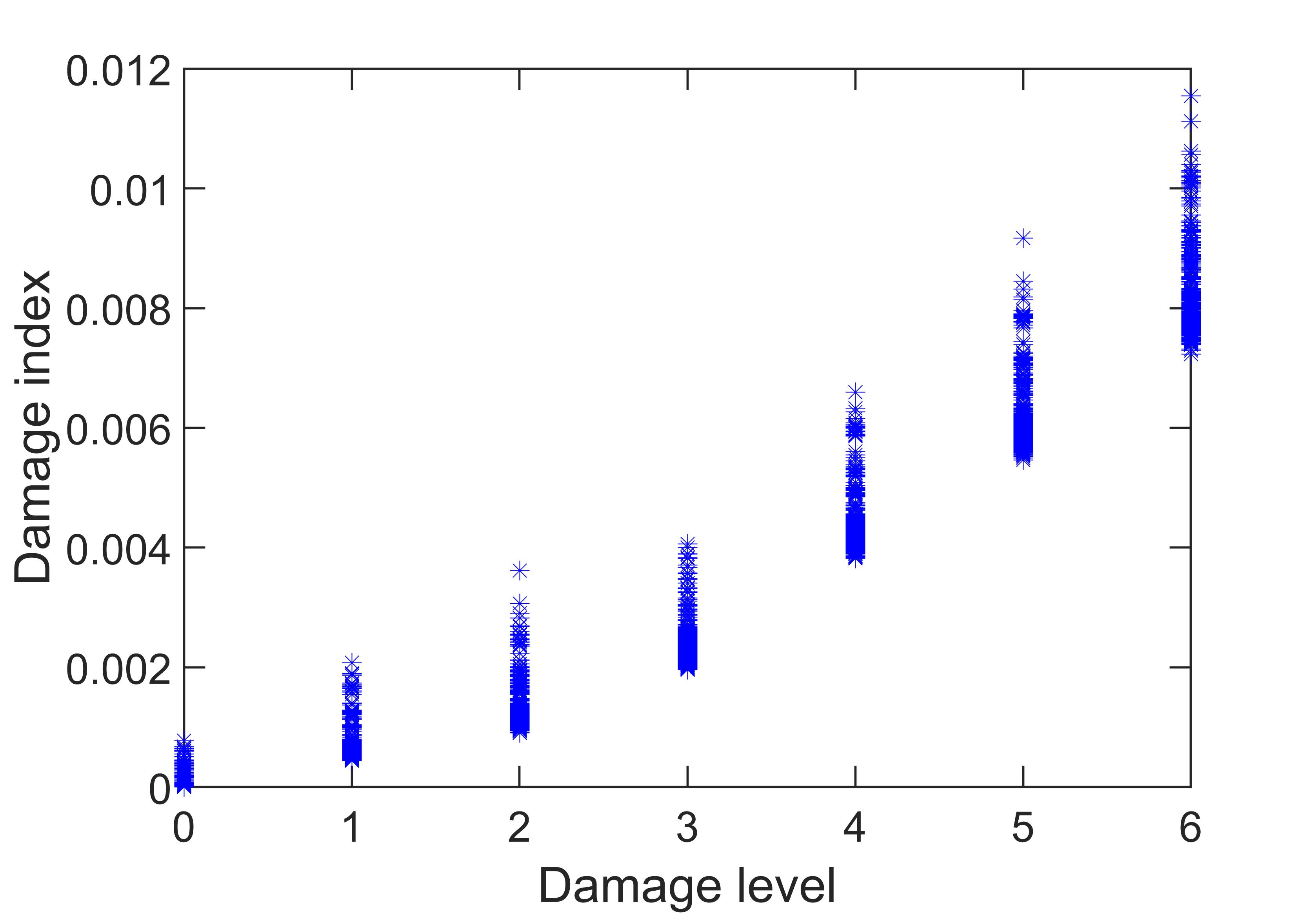}}
    \put(200,-50){\includegraphics[width=0.55\columnwidth]{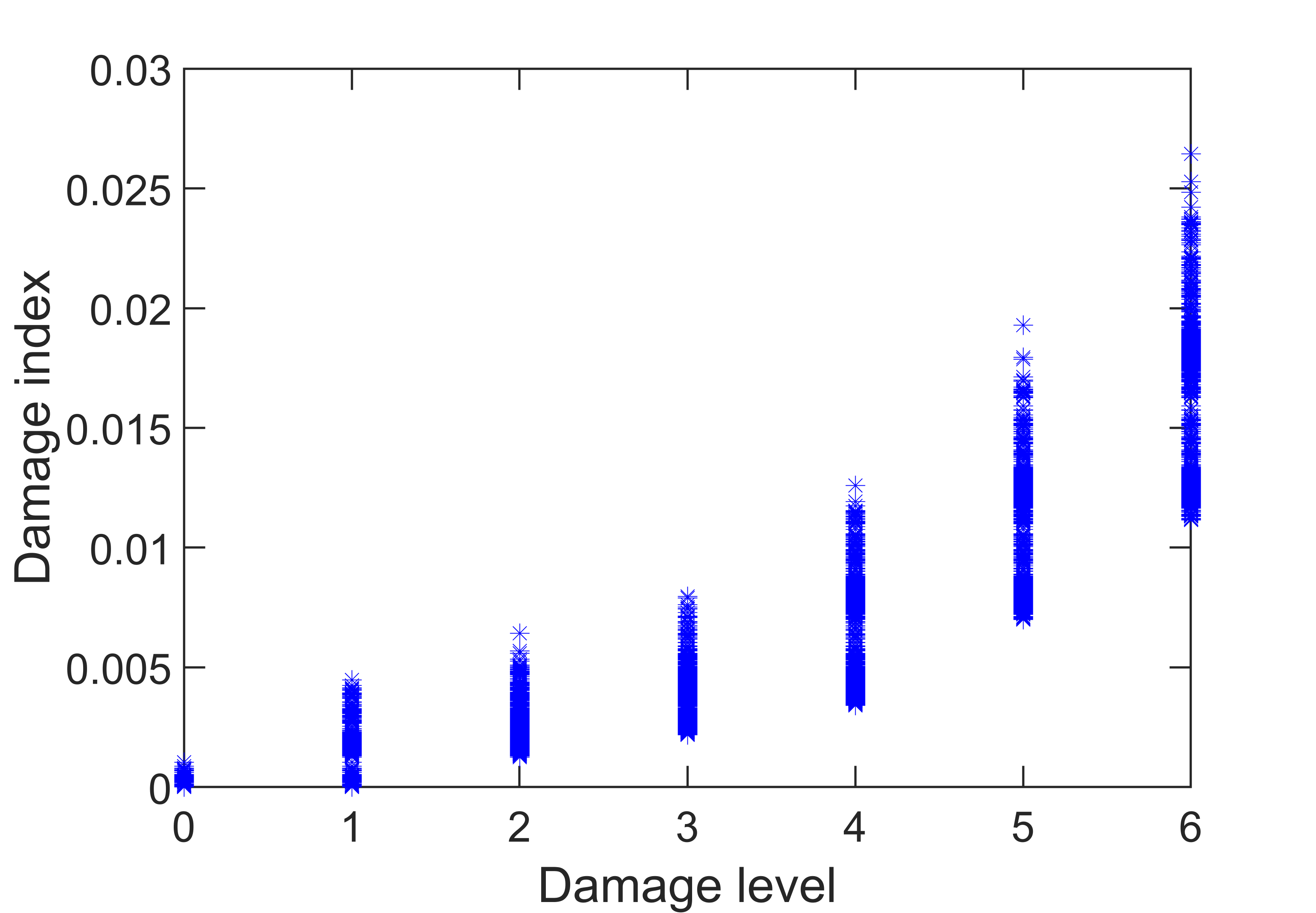}}

    \put(-15,127){\large \textbf{(a)}}
    \put(220,127){ \large \textbf{(b)}}
    \end{picture}
    \vspace{1.5cm}
    
    \caption{The evolution of the damage index \cite{janapati2016damage} as applied to indicative actuator-sensor paths: (a) damage non-intersecting path 1-4; (b) damage intersecting path 3-4.} 
\label{fig:DI composite} 
\end{figure}

In the present study, Figure \ref{fig:composite plate}(a) shows the six sensors/actuators layout in the CFRP plate. Simulated damage starts from just below sensor 5 and increases in magnitude towards sensor 2. Figure \ref{fig:DI composite}(a) and (b) show the evolution of the DI with increasing damage size for damage non-intersecting path 1-4 and damage intersecting path 3-4, respectively. It can be observed that the magnitude of the DI for the damage non-intersecting path is smaller than the damage intersecting path. However, a higher degree of overlap of the DIs among different damage levels for different realizations exists for the damage intersecting path 3-4 than the damage non-intersecting path 1-4. As a result, damage detection and identification are challenging using the damage intersecting path 3-4 for the CFRP coupon, even though the magnitude of the DIs are higher for damage intersecting path 3-4. In the subsequent study, it is shown that using an AR model, perfect damage detection was achieved for the CFRP coupon, however, perfect damage identification remains challenging using the stationary time series models. 

\subsection{Parametric Identification and Damage Detection Results}

\begin{figure}[!t] 
  \centering
   \begin{picture}(400,135)
    \put(40,-50){ \includegraphics[width=0.6\columnwidth]{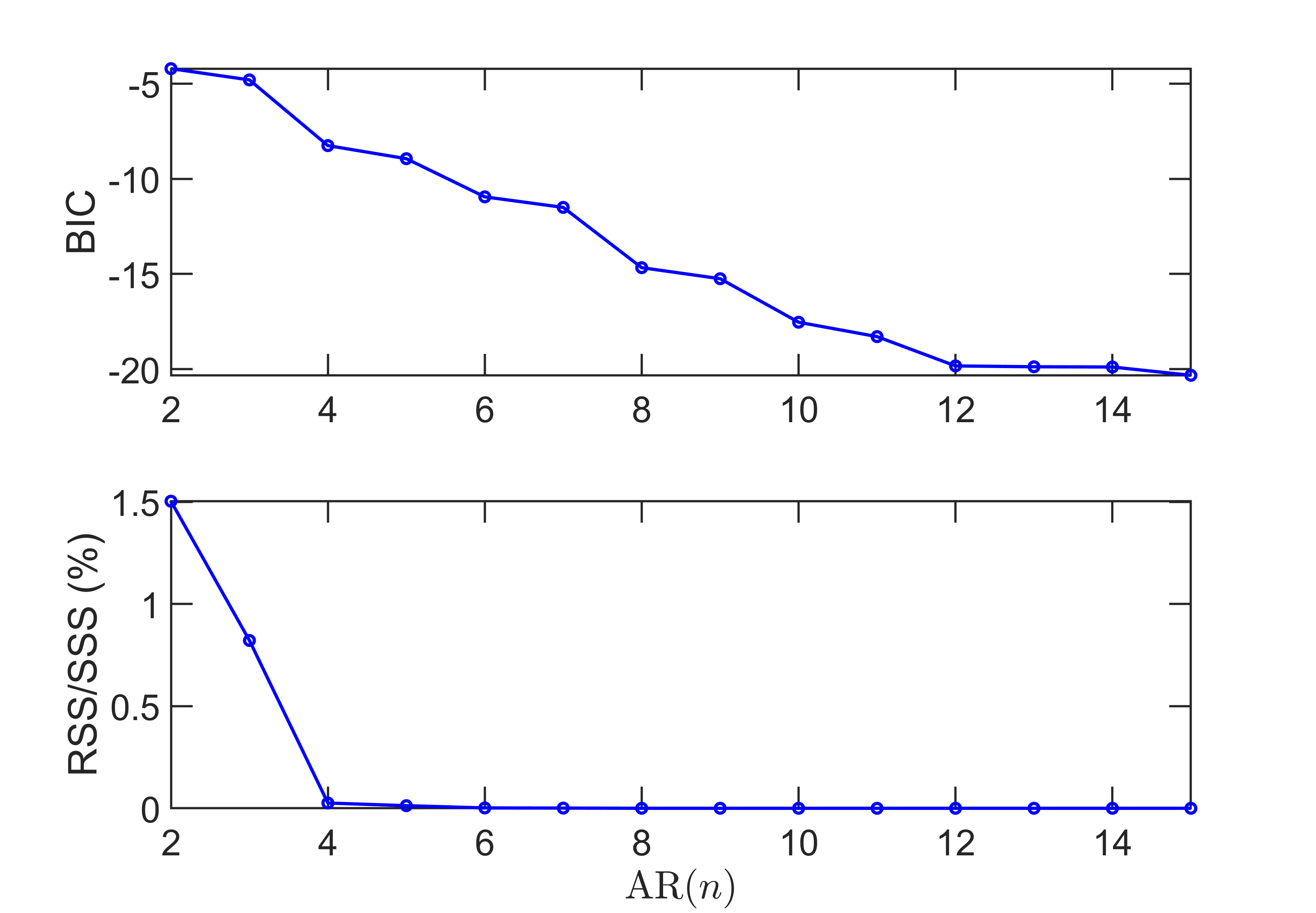}}
       
    \end{picture}
    
    \vspace{1.5cm}
    \caption{Model order selection via the BIC (top) and RSS/SSS (bottom) criteria for the CFRP plate for damage non-intersecting path 1-4.}
\label{fig:AR BIC composite} 
\end{figure}

Similar to the case of the aluminum plate, for composites, the RSS/SSS (Residual Sum of Squares/Signal Sum of Squares) criterion, describing the predictive ability of the model, was employed for the model selection process. AR orders from $na=2$ to $na=15$ were considered to create a pool of candidate models. Among all these models, the best model can be chosen where the RSS/SSS values start to show a plateau. Following this criterion, from Figure \ref{fig:AR BIC composite}, it can be observed that for $na=4$, the plateau starts to occur (both for damage intersecting and non-intersecting path). However, selecting the model order $na=4$ results in highly correlated model parameters, and the model parameters of different damage states get overlapped on each other. As a result, $na=6$ was selected as the final model order for the purpose of damage detection and identification in the CFRP plate.

\begin{figure}[!t] 

 \begin{picture}(400,145)
    \put(-10,-50){ \includegraphics[width=0.55\columnwidth]{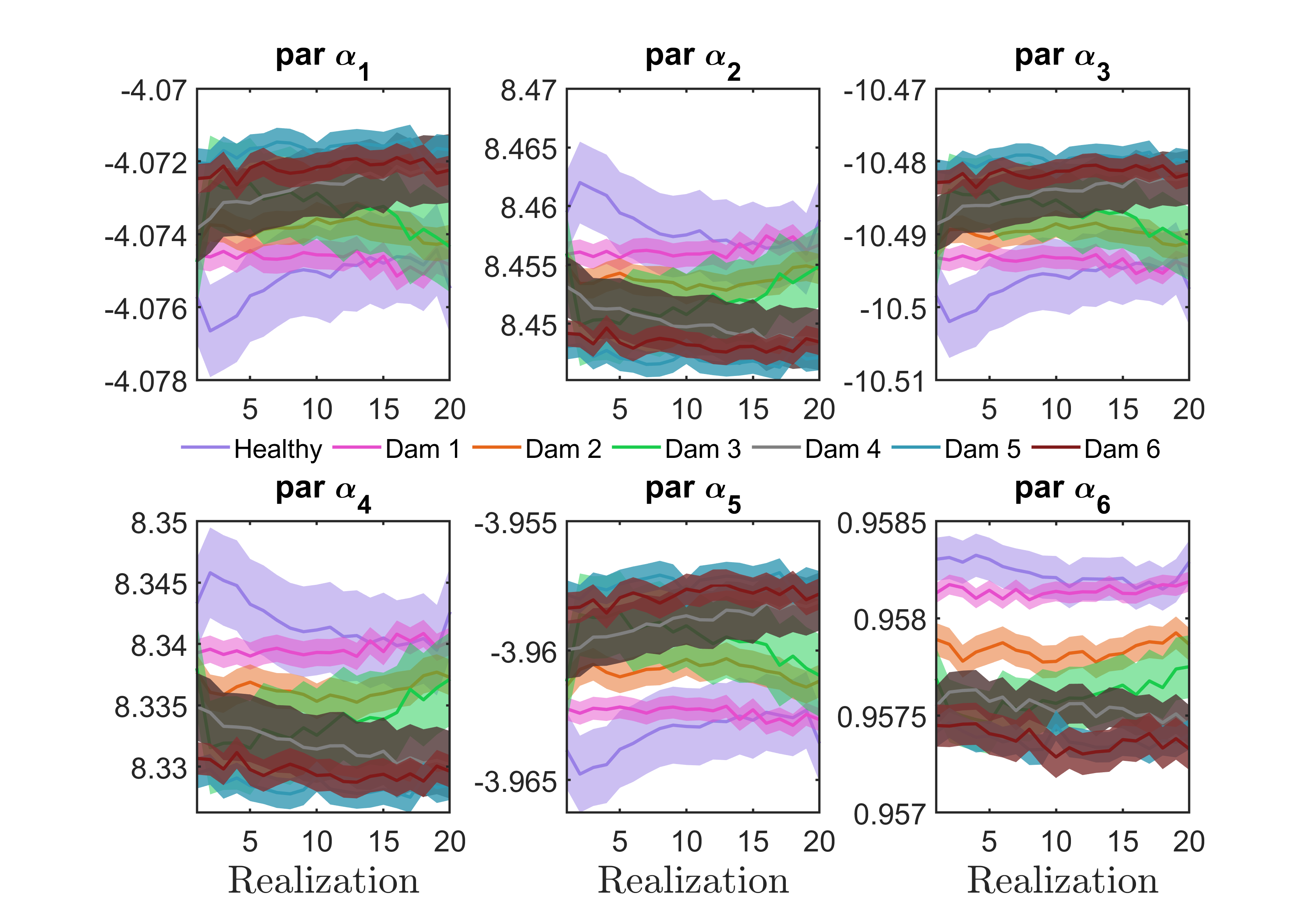}}
    \put(240,-50){\includegraphics[width=0.55\columnwidth]{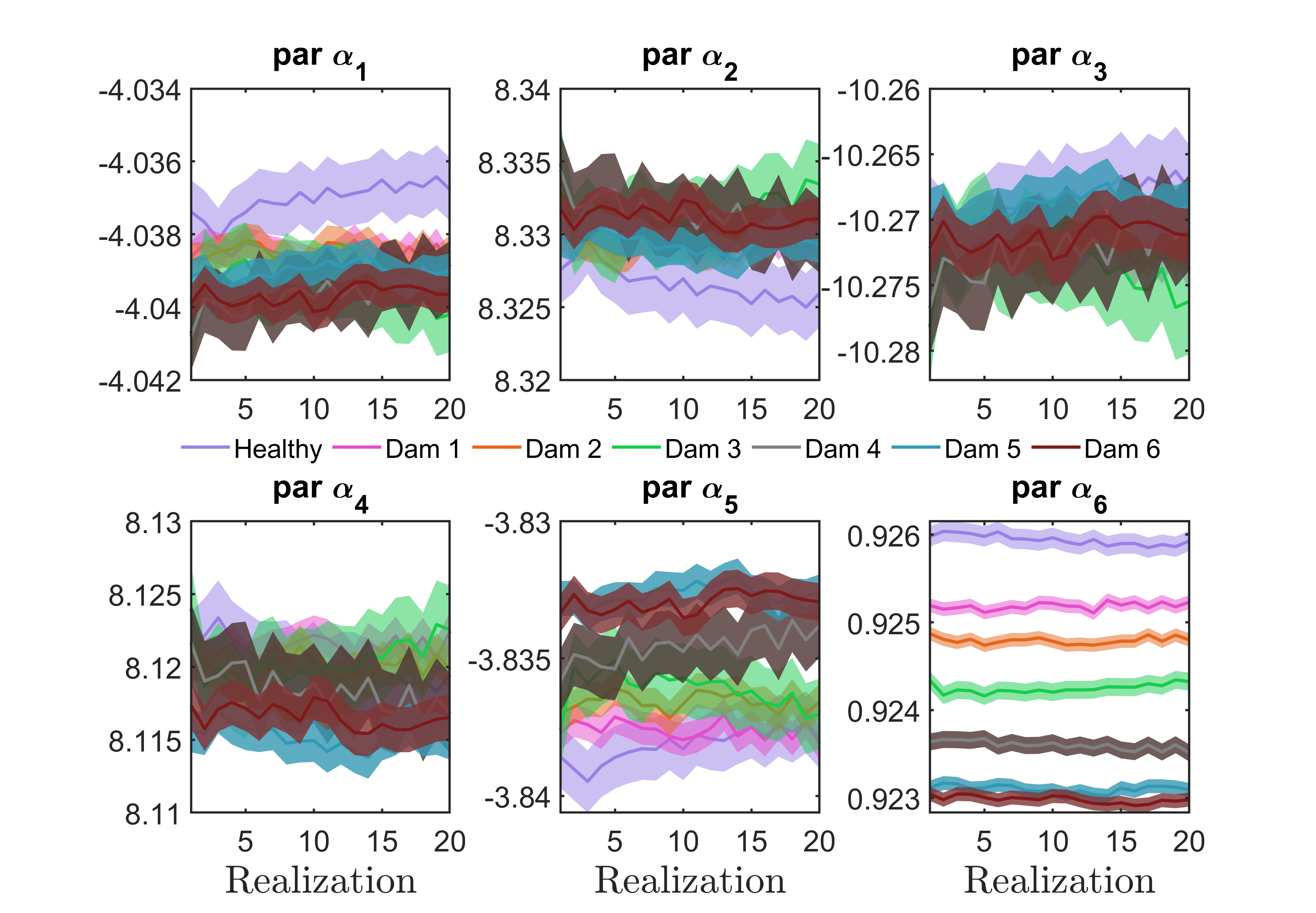}}

   \put(-15,127){\large \textbf{(a)}}
    \put(230,127){ \large \textbf{(b)}}
    \end{picture}

   \vspace{1.5cm}\caption{AR($6$) model parameters for different structural states for the CFRP plate: the parameter mean is shown as solid lines and the associated $\pm 2$ standard deviations as shaded regions; (a) damage intersecting path 3-4; (b) damage non-intersecting path 1-4.}
\label{fig:AR par composite}%
\end{figure}

Figure \ref{fig:AR par composite}(a) and (b) depict the AR model parameters for all realizations of the guided wave signals acquired from the damage intersecting path 3-4 and the damage non-intersecting path 1-4 of the CFRP plate for all different structural states, namely: healthy, damage level 1, damage level 2, damage level 3, damage level 4, damage level 5, and damage level 6. For each state, 20 realizations are shown. The solid lines represent the mean parameter values, and the shaded regions represent the $\pm 2$ standard deviation confidence intervals. As the model order $na = 6$, the number of estimated parameters is also six. Note that the model parameters of different states intersect on each other at different realizations both for damage intersecting and non-intersecting paths. For parameter $\alpha_6$, the degree of overlapping is higher for the damage intersecting path 3-4 than the damage non-intersecting path 1-4 for different structural states.

\begin{figure}[t!]
    \centering
    \begin{picture}(400,130)
    \put(-40,-50){ \includegraphics[width=0.55\columnwidth]{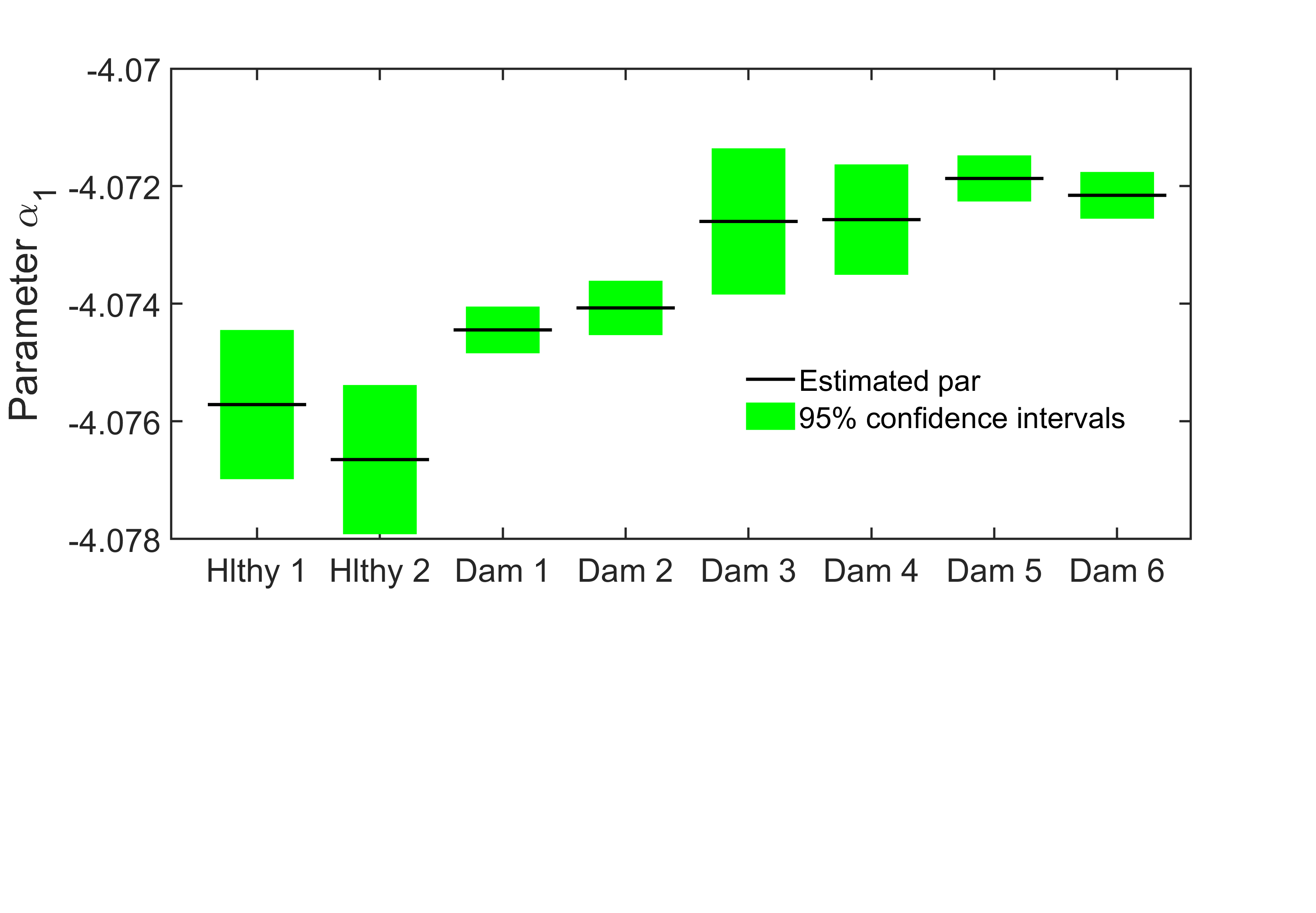}}
    \put(200,-50){\includegraphics[width=0.55\columnwidth]{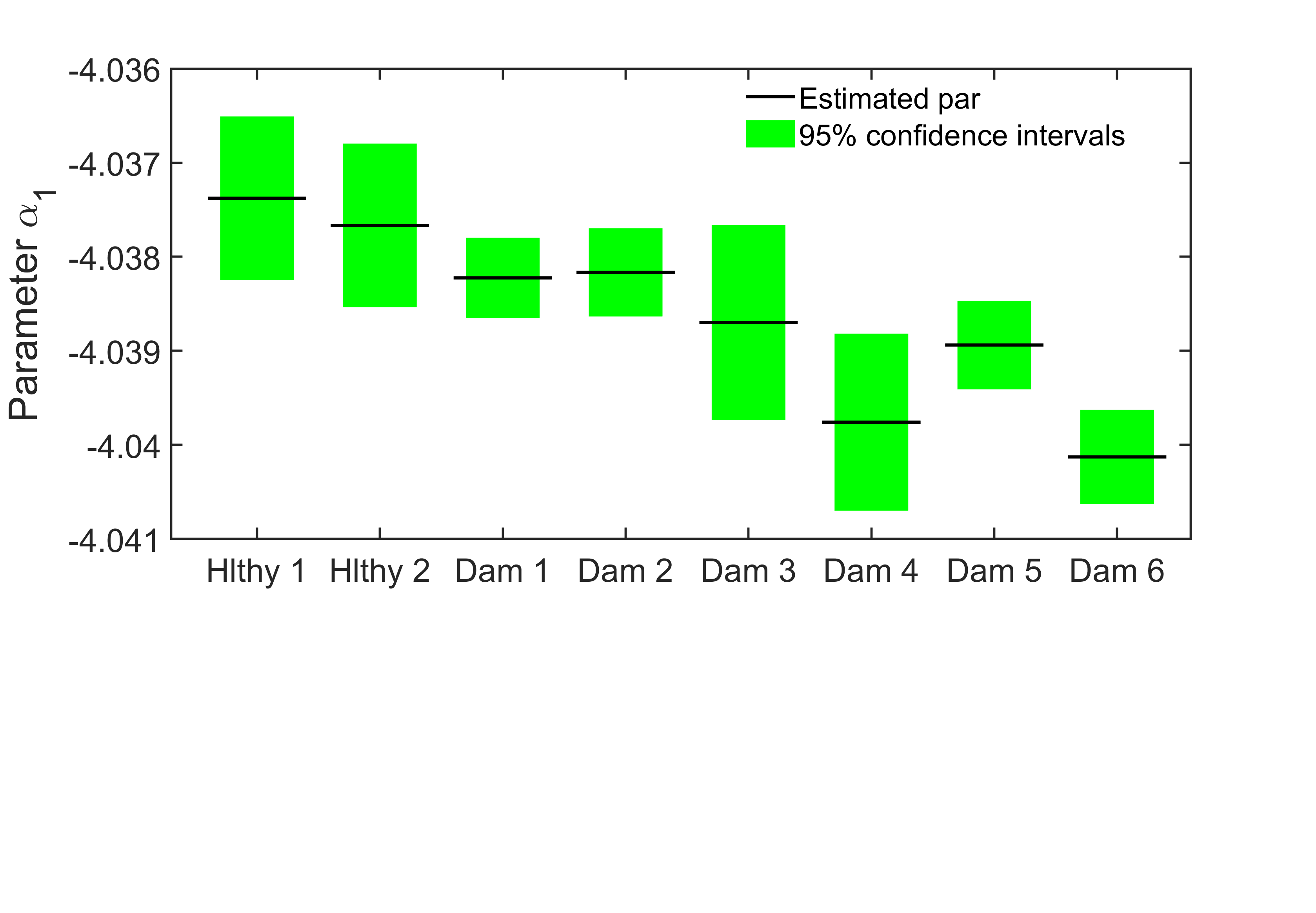}}

    \put(-15,127){\large \textbf{(a)}}
    \put(220,127){ \large \textbf{(b)}}
    \end{picture}
    \vspace{-0.5cm}
    
    \caption{Estimated model parameters for healthy and damaged states and the corresponding 95\% confidence intervals of the CFRP coupon : (a) parameter $\alpha_1$ for damage intersecting path 3-4; (b) parameter $\alpha_1$ for damage non-intersecting path 1-4.} 
\label{fig:par damage composite} 
\end{figure} 

Figure \ref{fig:par damage composite}(a) and (b) shows the evolution of parameter $\alpha_1$ for different structural states for damage intersecting path 3-4 and damage non-intersecting path 1-4, respectively. The black lines represent the mean parameter values and the green regions represent the 95\% confidence intervals. It can be observed that both for damage intersecting path 3-4 and damage non-intersecting path 1-4, a certain degree of overlap exists among different structural states of the CFRP plate. This is one reason that makes damage detection and identification in the CFRP plate more challenging.

\begin{figure}[t!]
    \centering
    \begin{picture}(400,130)
    \put(-40,-50){ \includegraphics[width=0.55\columnwidth]{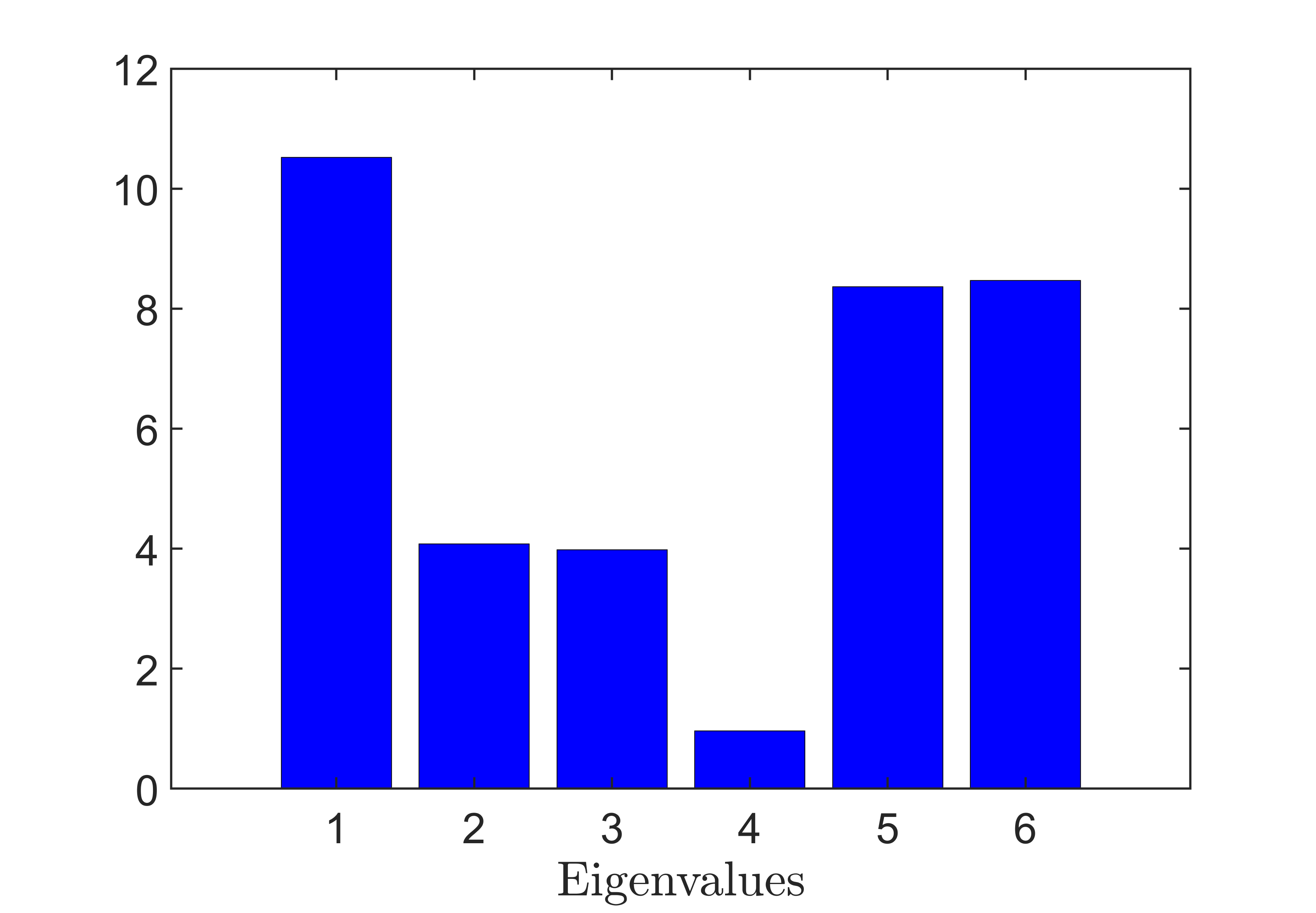}}
    \put(200,-50){\includegraphics[width=0.55\columnwidth]{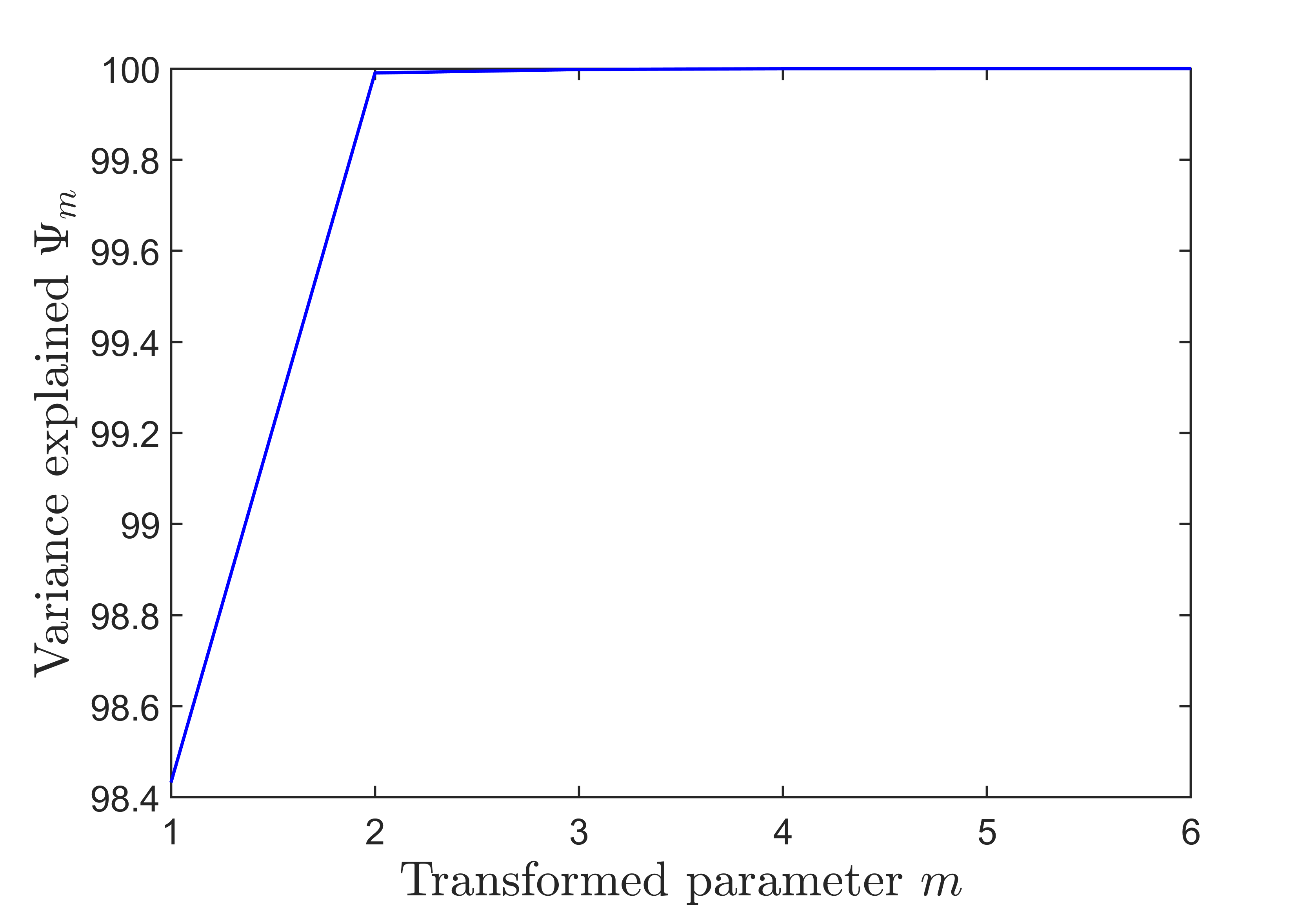}}

    \put(-15,127){\large \textbf{(a)}}
    \put(220,127){ \large \textbf{(b)}}
    \end{picture}
    \vspace{2cm}
    
    \caption{AR parameter selection process for damage detection of the CFRP plate: (a) SVD-based approach; (b) PCA-based approach.} 
\label{fig:par select composite} 
\end{figure} 

Figure \ref{fig:par select composite} shows the parameter selection process for the damage detection and identification in CFRP plate. From Figure \ref{fig:par select composite}(a), it can be observed that the eigenvalue $\lambda_1$ and $\lambda_6$ have the highest magnitude. As a result, the corresponding parameter $a_1$ and $a_6$ are required to choose. However, using only these two parameters, perfect damage detection cannot be achieved in the CFRP plate. Parameter $a_2$, $a_3$ and $a_5$ are also required to be included for better damage detection. This may be due to the fact that the signals may have a higher degree of variation in the CFRP plate than the aluminum plate. From Figure \ref{fig:par select composite}(b), it can be observed that after projecting the parameters onto some lower dimensional space, only two parameters are needed for explaining the total variance. However, using only two truncated parameters cannot achieve perfect damage detection in the CFRP plate. As a result, five truncated parameters and the associated covariance matrix were used for the subsequent damage detection using the PCA-based approach. 

\begin{figure}[t!]
    \centering
    \begin{picture}(400,130)
    \put(-40,-50){ \includegraphics[width=0.55\columnwidth]{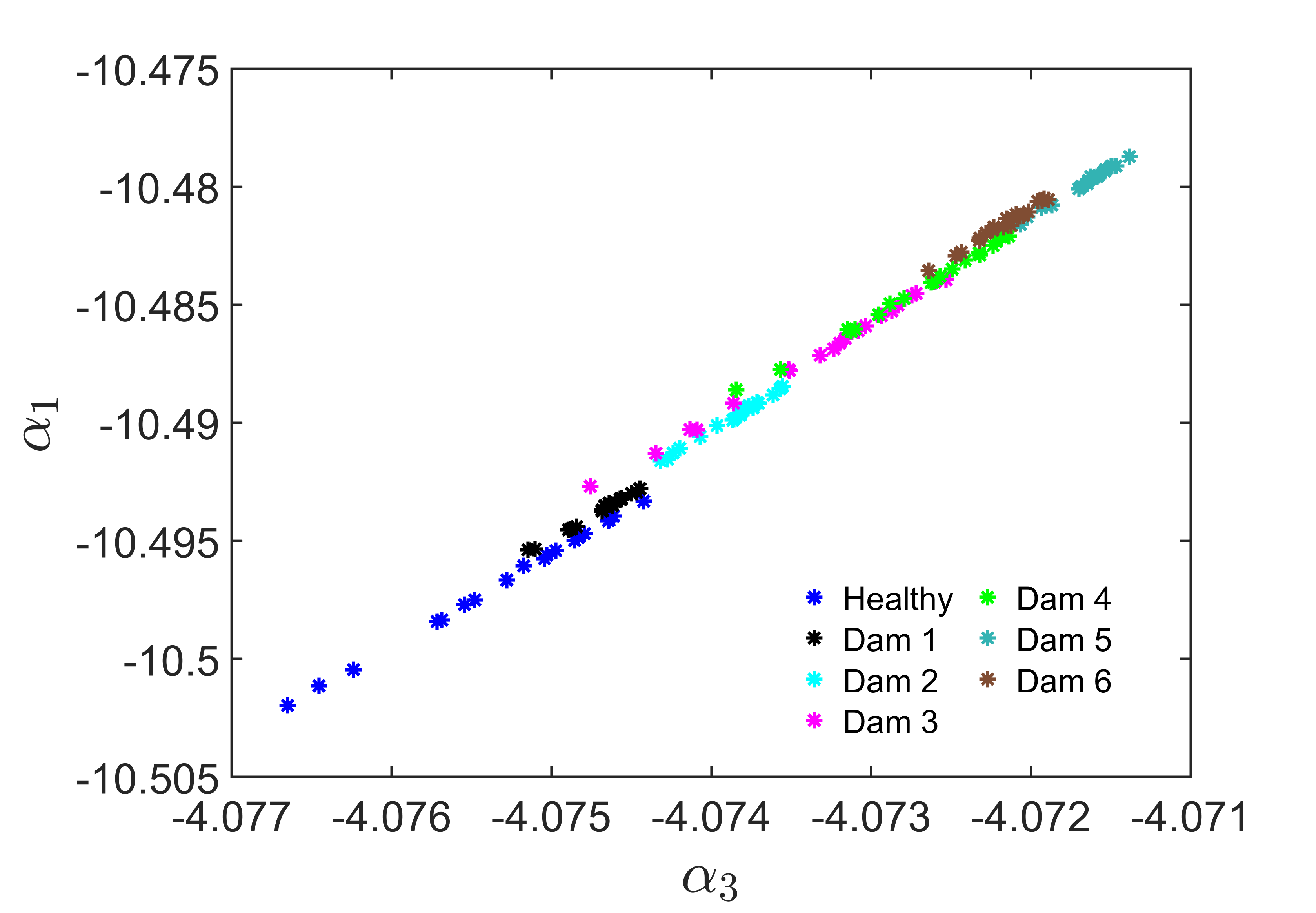}}
    \put(200,-50){\includegraphics[width=0.55\columnwidth]{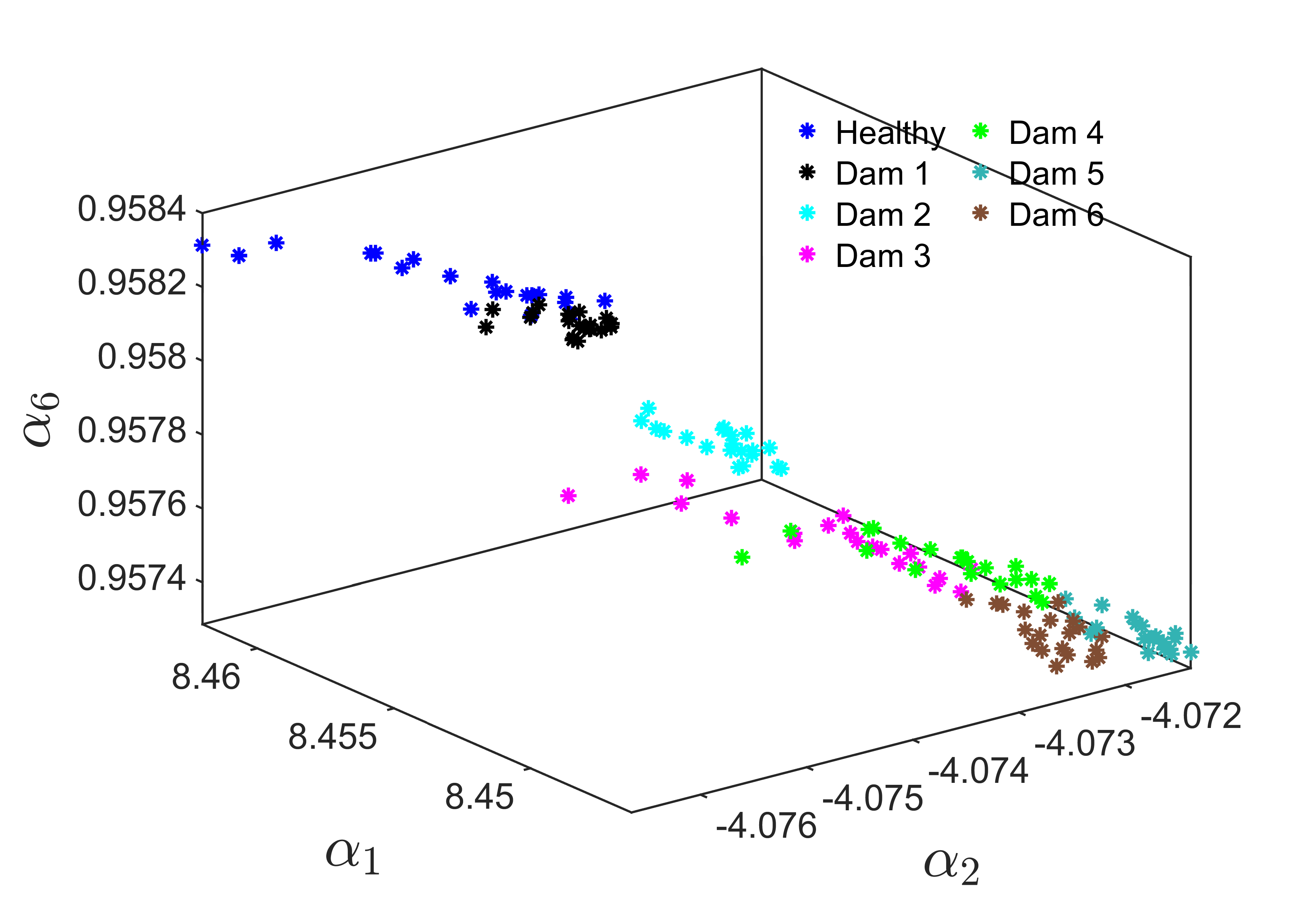}}
     \put(-40,-230){ \includegraphics[width=0.55\columnwidth]{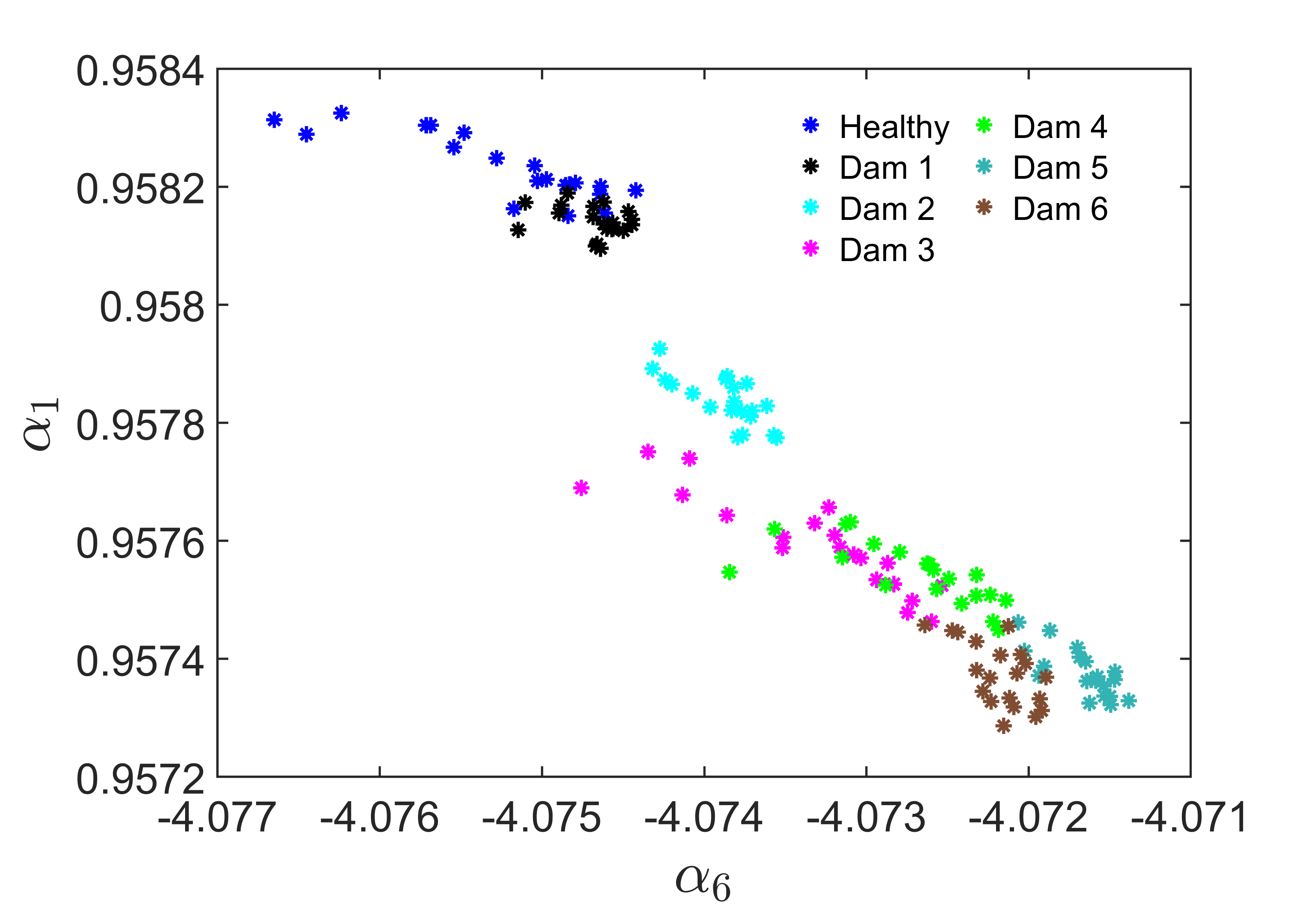}}
    \put(200,-230){\includegraphics[width=0.55\columnwidth]{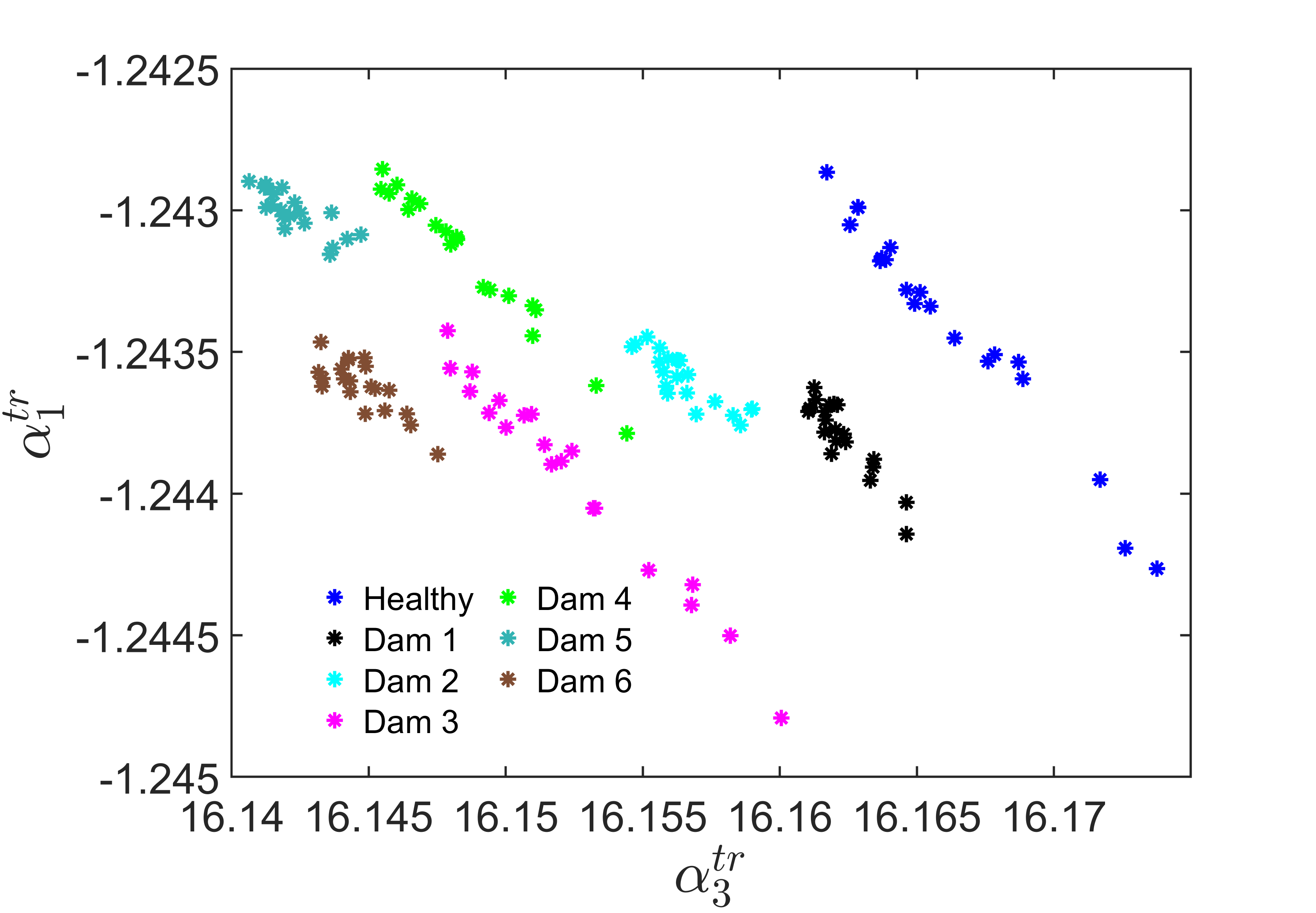}}

    \put(-15,127){\large \textbf{(a)}}
    \put(220,110){ \large \textbf{(b)}}
     \put(-15,-50){\large \textbf{(c)}}
    \put(220,-50){ \large \textbf{(d)}}
    \end{picture}
    \vspace{8cm}
    
    \caption{Indicative AR($6$) model parameters of the CFRP coupon are shown for damage non-intersecting path 1-4: (a) model parameter $\alpha_1$ and $\alpha_2$; (b) model parameter $\alpha_1$, $\alpha_2$ and $\alpha_6$; (c) model parameter to be used in damage detection indicated by SVD-based approach ($\alpha_1$ and $\alpha_6$); (d) truncated model parameters from PCA transformation} 
\label{fig:par par composite} 
\end{figure}
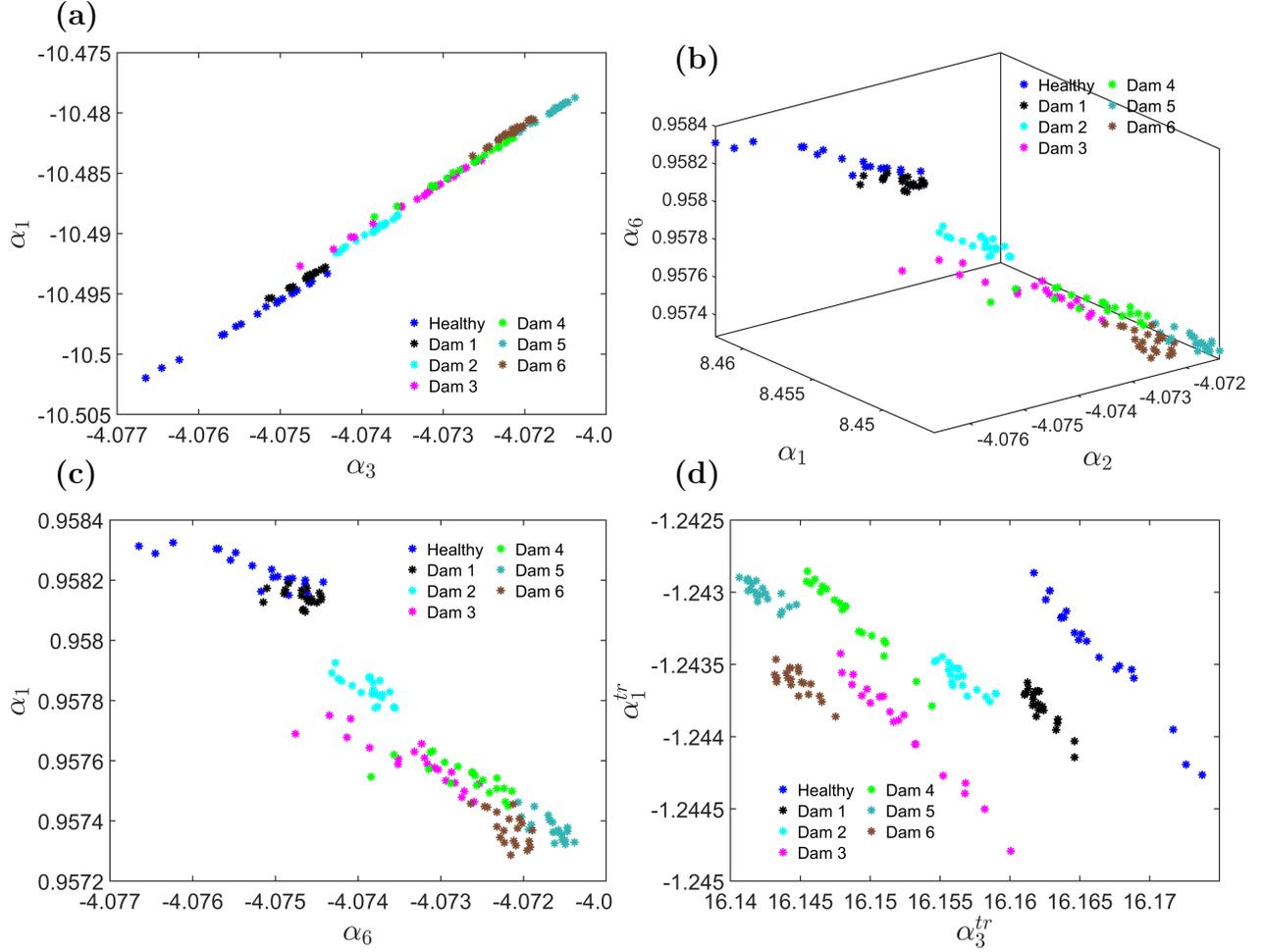

Figure \ref{fig:par par composite}(a) shows the plot of the two parameters $a_1$ and $a_2$ for all structural states for damage non-intersecting path 1-4. Note that the parameters are highly correlated and different structural states overlap on each other. Figure \ref{fig:par par composite}(b) shows a 3-dimensional plot of the three model parameters $\alpha_1$, $\alpha_2$, and $\alpha_6$. Figure \ref{fig:par par composite}(c) shows the plot of the parameter $a_1$ and $a_6$. Note that, similar to the case of the aluminum plate, the parameters are less correlated and different structural states are slightly separated. Figure \ref{fig:par par composite}(d) shows that after performing PCA transformation, the model parameters become uncorrelated and separated for different structural states.

Figure \ref{fig:dam intersect exp composite} shows the damage detection performance of the CFRP plate for damage intersecting path 3-4 using the standard AR, SVD, and PCA-based approach. In this case, the covariance matrix was derived from the 20 experimental healthy signals. It can be observed that for the CFRP plate, the standard AR and the SVD-based approach perform better than the PCA-based approach. Both for SVD and PCA-based approach, 5 parameters were used. Note that for the standard AR and the SVD-based approach, there are 7 instances of missed damage. However, for PCA-based approach, damage level 1 is completely missed. The $\alpha$ level used for the standard AR, SVD, and PCA-based approach was $0.03$,$0.1$, and $0.003$, respectively. Similarly, Figure \ref{fig:dam intersect theory composite} shows the damage detection using the above-mentioned three methods using the covariance matrix derived from the AR($6$) model. The thresholds were manually adjusted for the standard AR and SVD-based approach. The $\alpha$ level used for the PCA-based approach was $0.09$.

\begin{figure}[t!]
    \centering
    \begin{picture}(400,130)
     \put(-46,10){ \includegraphics[width=0.36\columnwidth]{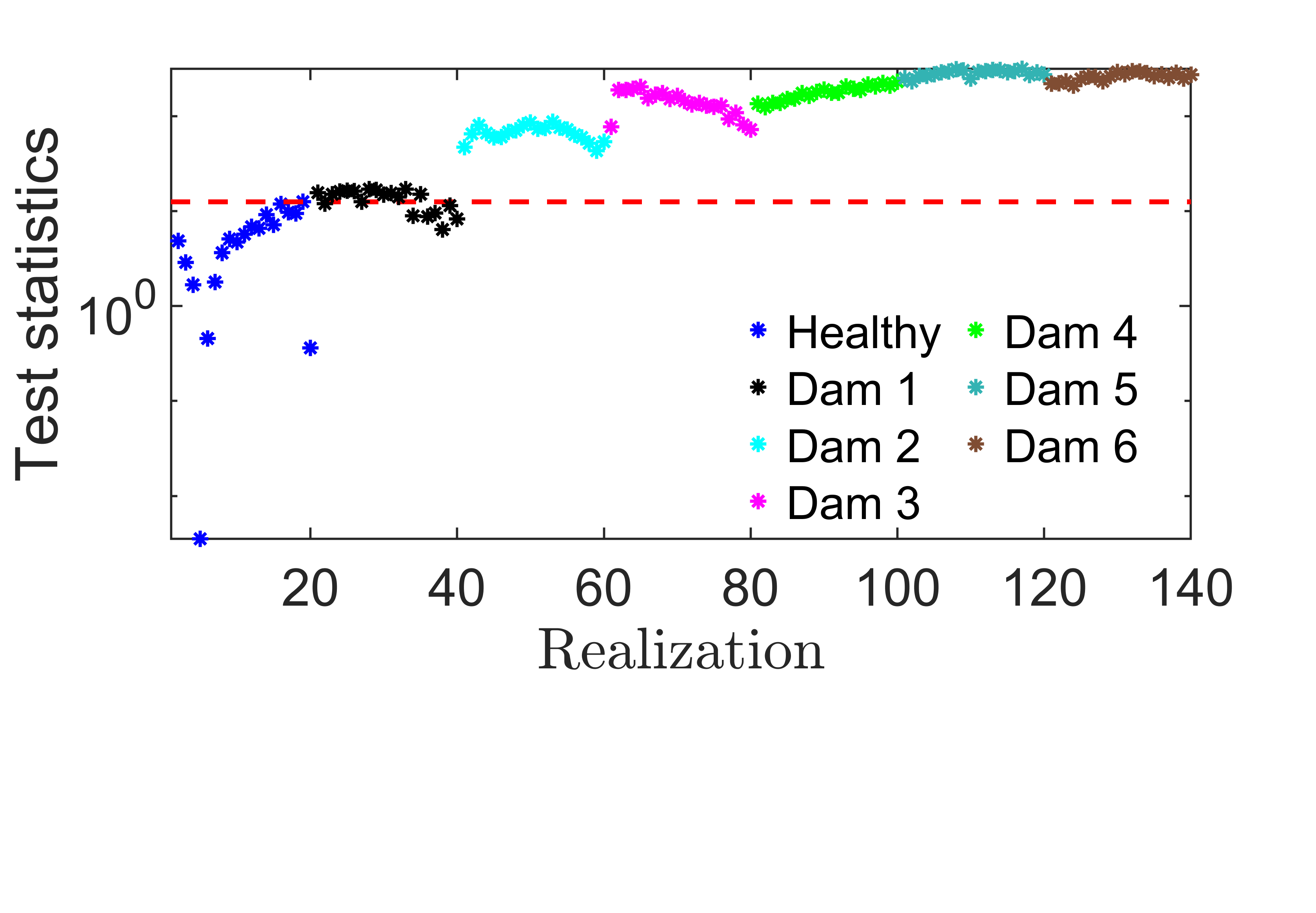}} 
    \put(114,10){ \includegraphics[width=0.36\columnwidth]{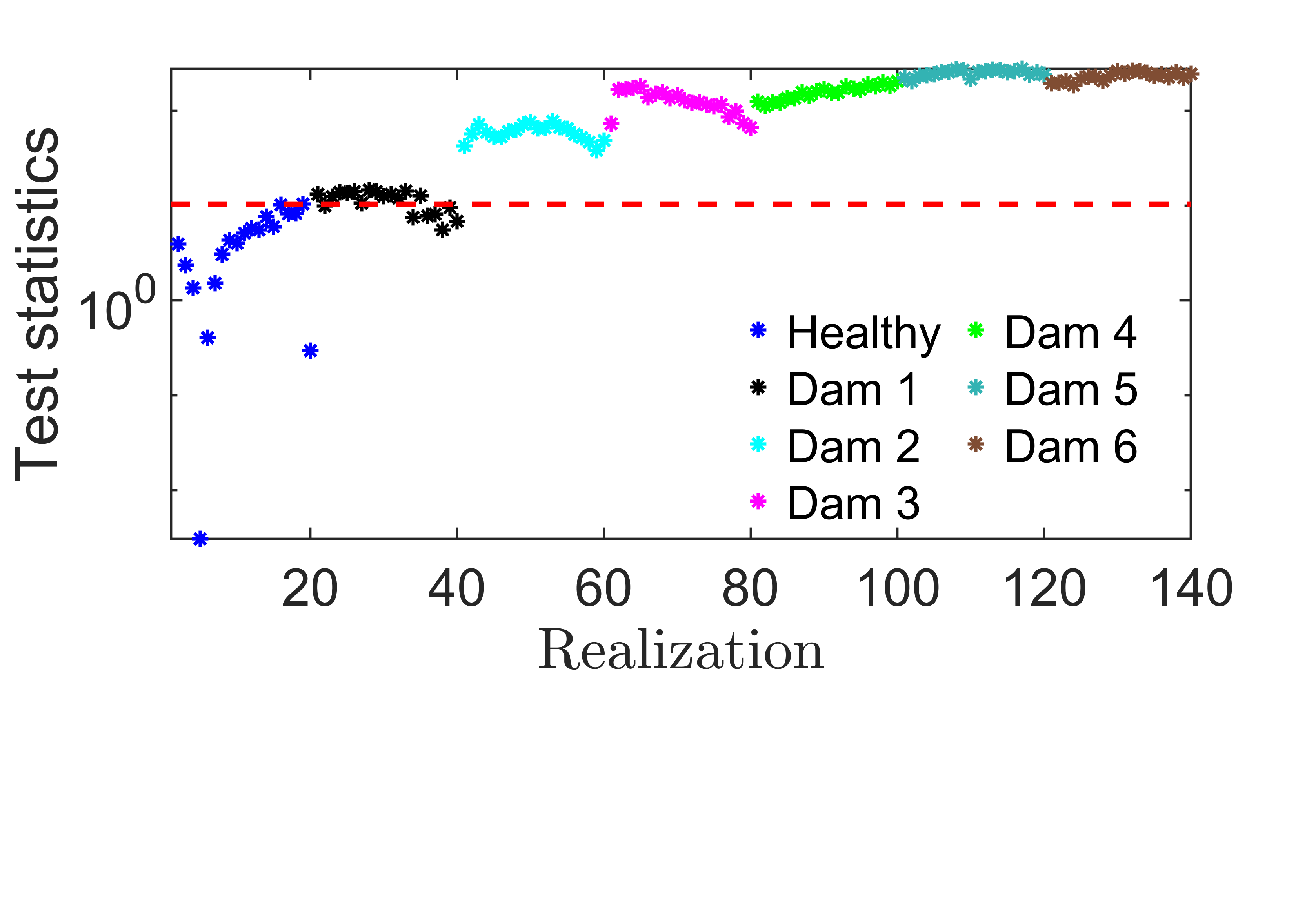}}
    \put(280,10){\includegraphics[width=0.36\columnwidth]{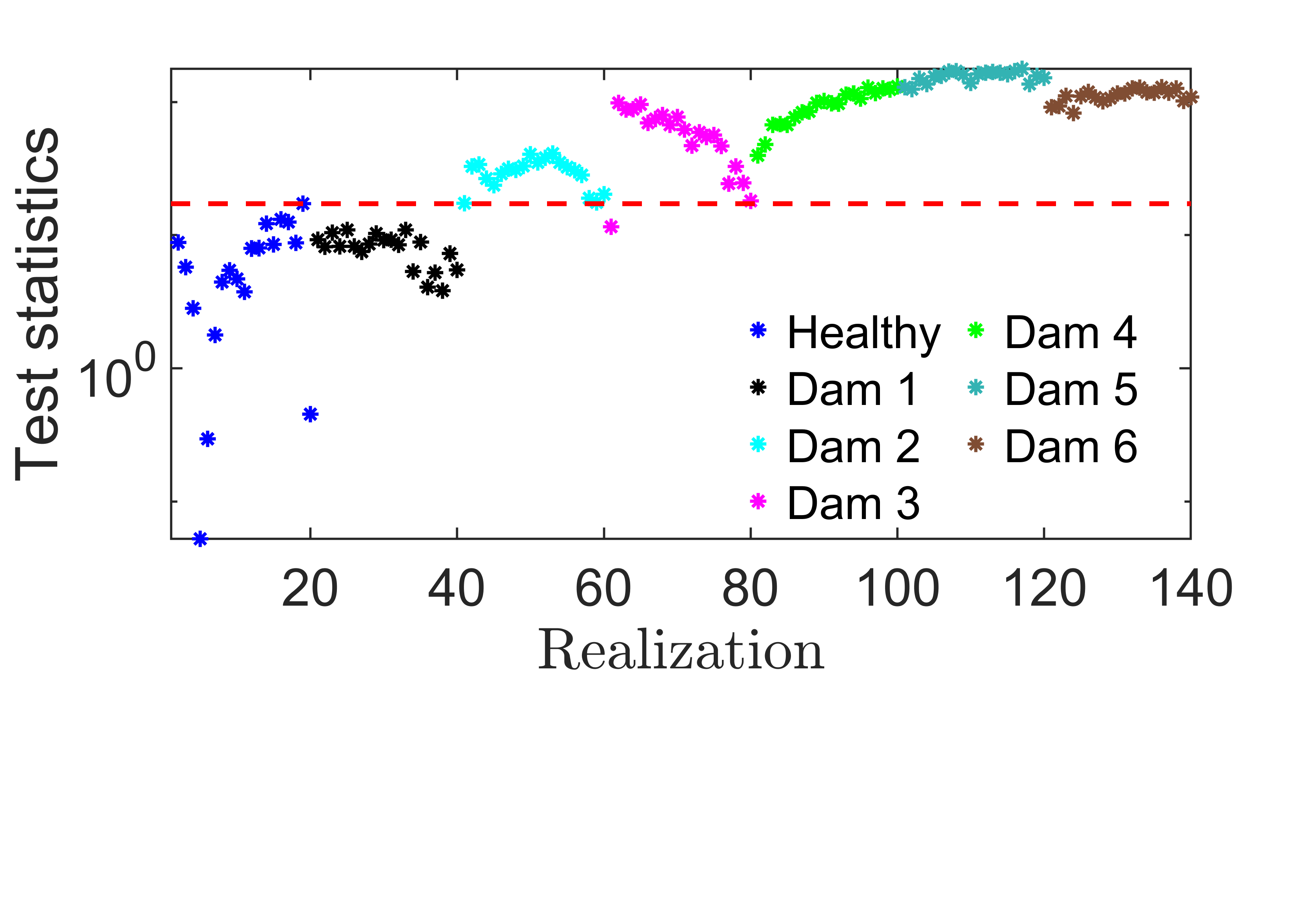}}

    \put(-20,125){\textbf{(a)}}
    \put(140,125){\large \textbf{(b)}}
    \put(290,125){ \large \textbf{(c)}}
    \end{picture}
    \vspace{-1.5cm}
    
    \caption{Damage detection performance comparison for damage intersecting path 3-4 for the CFRP plate using the covariance matrix derived from 20 experimental healthy signals: (a) standard AR approach; (b) SVD-based approach; (c) PCA-based approach. } 
\label{fig:dam intersect exp composite} 
\end{figure}
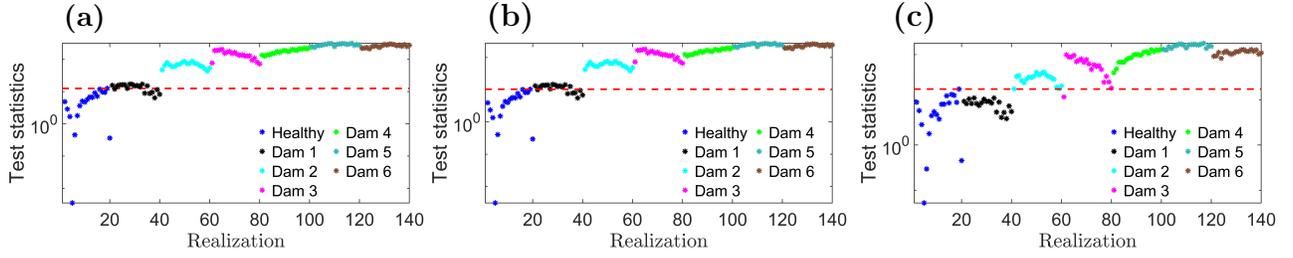 
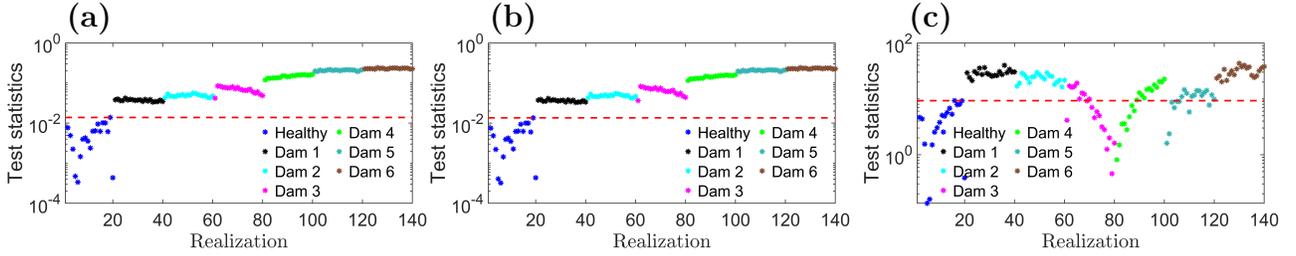
\begin{figure}[t!]
    \centering
    \begin{picture}(400,130)
     \put(-46,10){ \includegraphics[width=0.36\columnwidth]{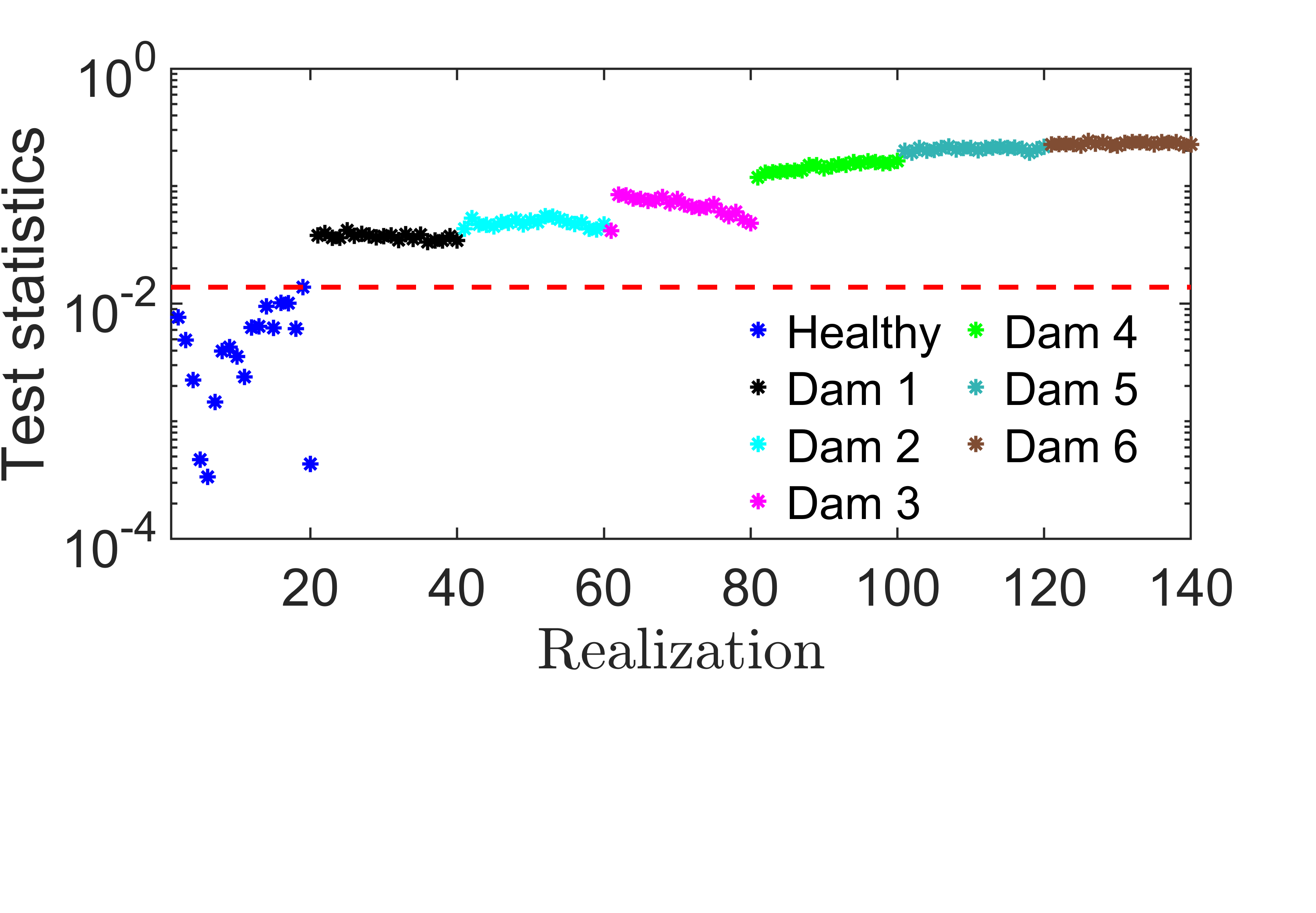}} 
    \put(114,10){ \includegraphics[width=0.36\columnwidth]{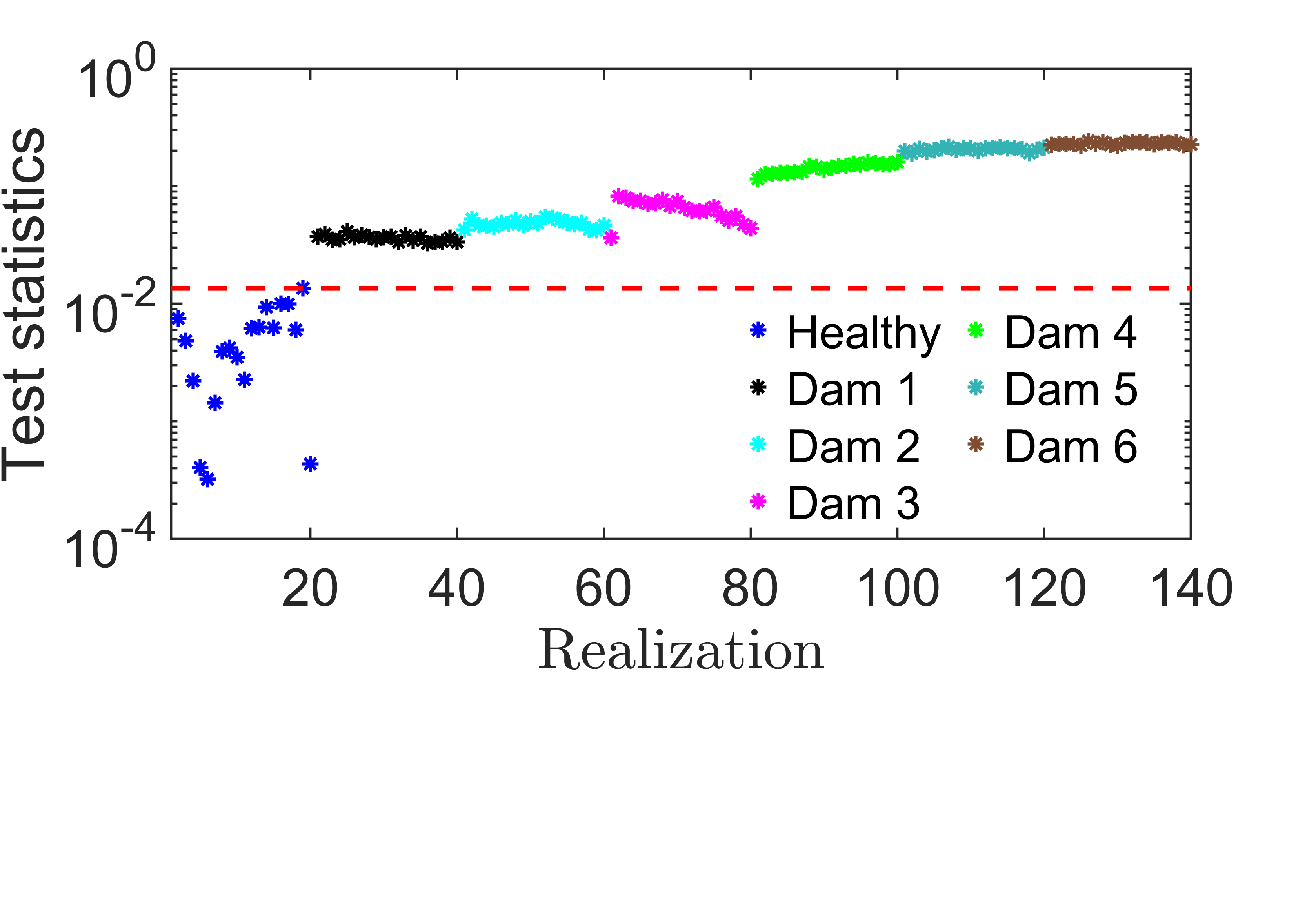}}
    \put(280,10){\includegraphics[width=0.36\columnwidth]{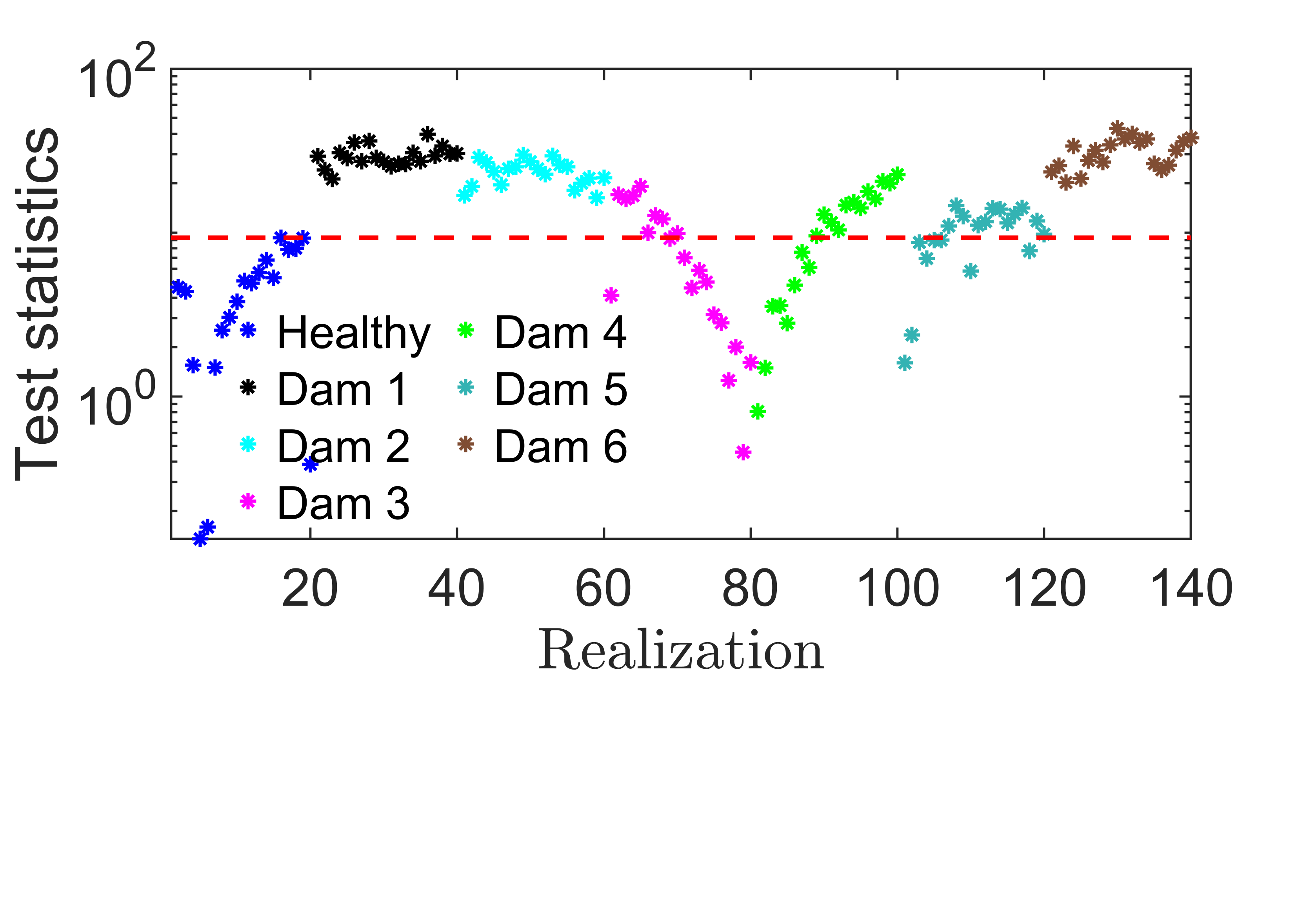}}

     \put(-20,125){\large \textbf{(a)}}
    \put(140,125){\large \textbf{(b)}}
    \put(295,125){ \large \textbf{(c)}}
    \end{picture}
    \vspace{-1.5cm}
    
    \caption{Damage detection performance comparison for damage intersecting path 3-4 for the CFRP plate using the AR($6$)-based covariance matrix: (a) standard AR approach; (b) SVD-based approach; (c) PCA-based approach.} 
\label{fig:dam intersect theory composite} 
\end{figure} 

\begin{figure}[t!]
    \centering
    \begin{picture}(400,130)
     \put(-46,10){ \includegraphics[width=0.36\columnwidth]{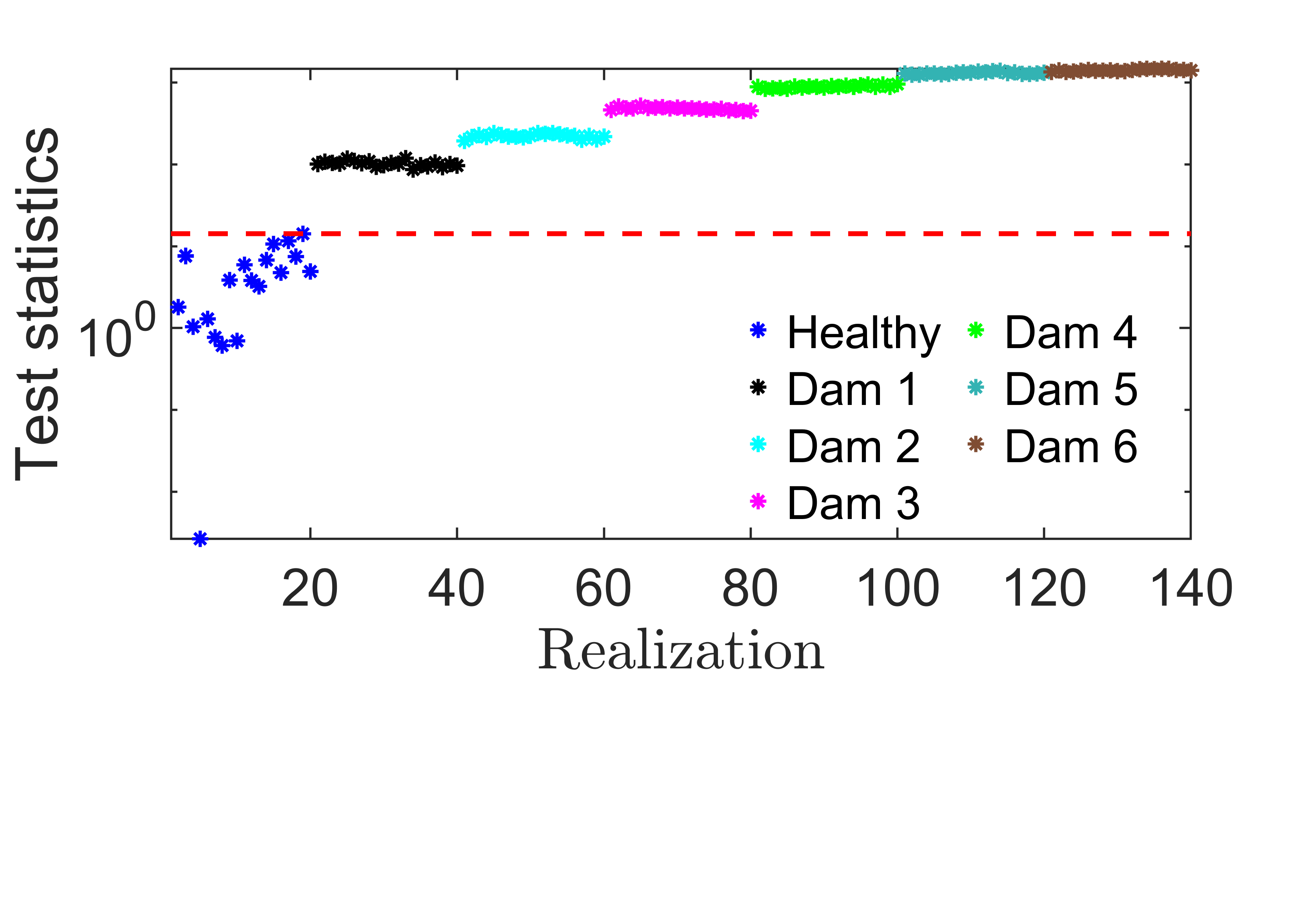}}
    \put(114,10){ \includegraphics[width=0.36\columnwidth]{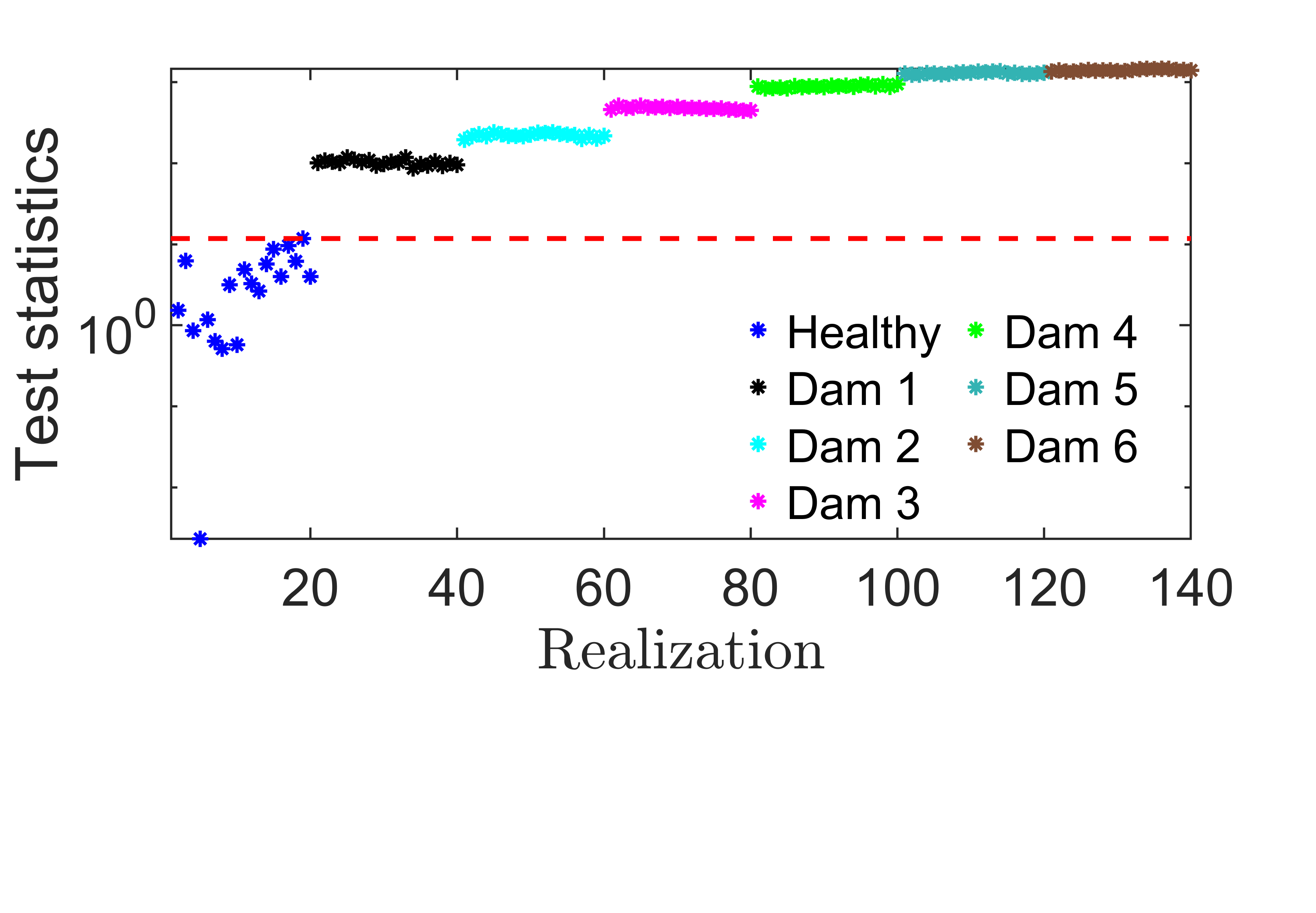}}
    \put(280,10){\includegraphics[width=0.36\columnwidth]{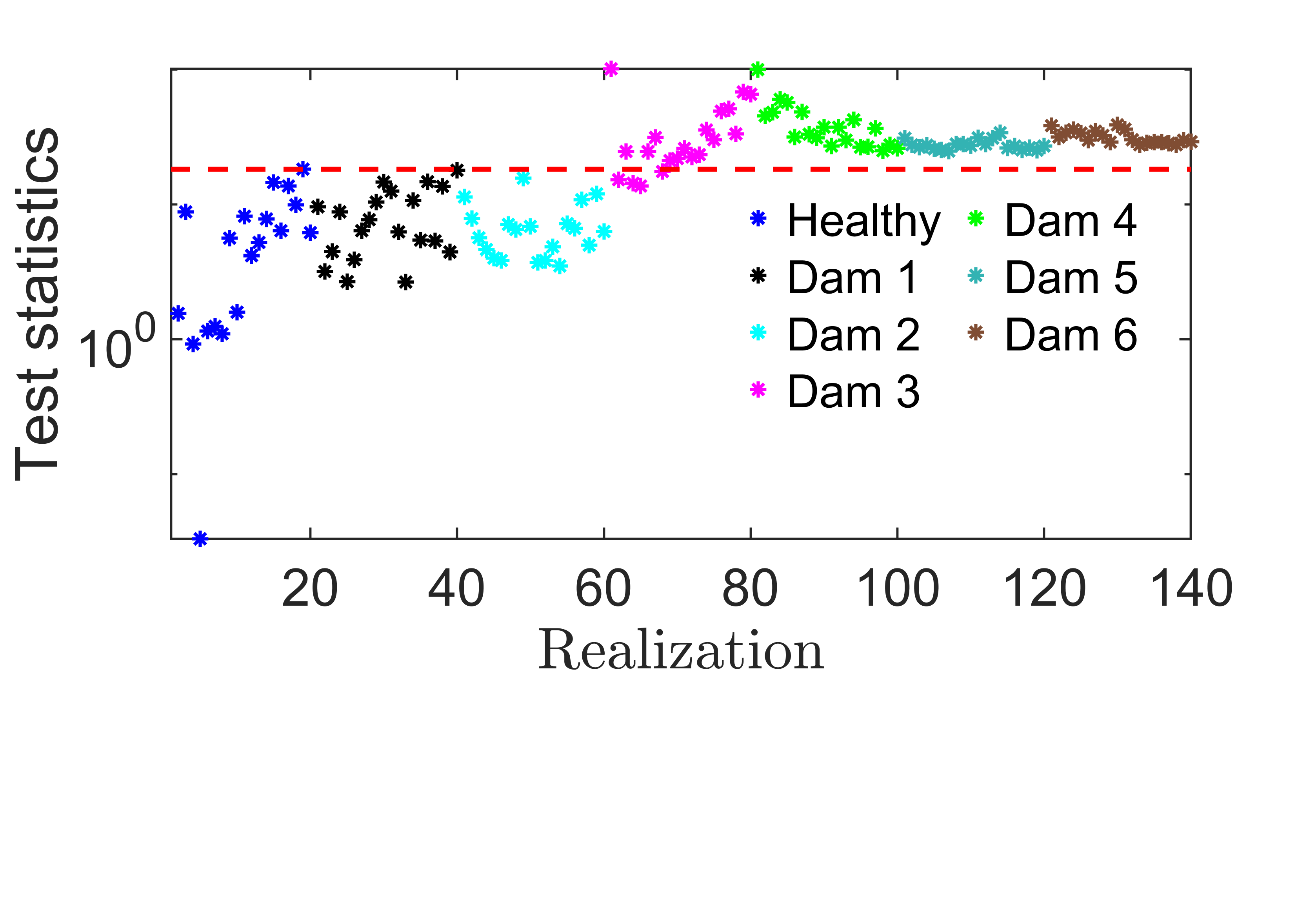}}

     \put(-20,125){\large \textbf{(a)}}
    \put(140,125){\large \textbf{(b)}}
    \put(290,125){ \large \textbf{(c)}}
    \end{picture}
    \vspace{-1.5cm}
    
    \caption{Damage detection performance for damage non-intersecting path 1-4 for the CFRP plate using the covariance matrix derived from 20 experimental healthy signals: (a) standard AR approach; (b) SVD-based approach; (c) PCA-based approach.} 
\label{fig:dam nonintersect exp composite} 
\end{figure}

\begin{figure}[t!]
    \centering
    \begin{picture}(400,130)
     \put(-46,10){ \includegraphics[width=0.36\columnwidth]{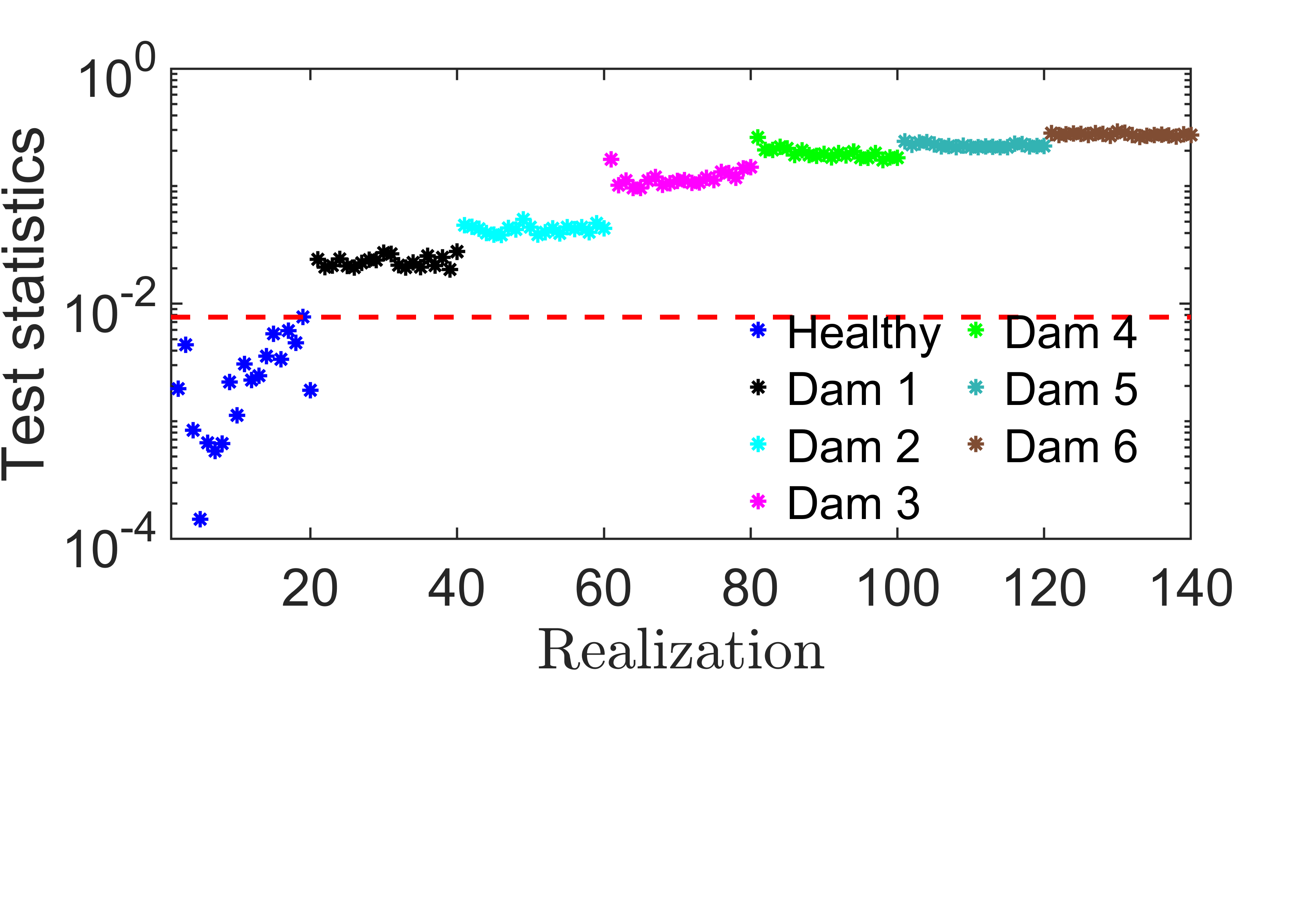}} 
    \put(114,10){ \includegraphics[width=0.36\columnwidth]{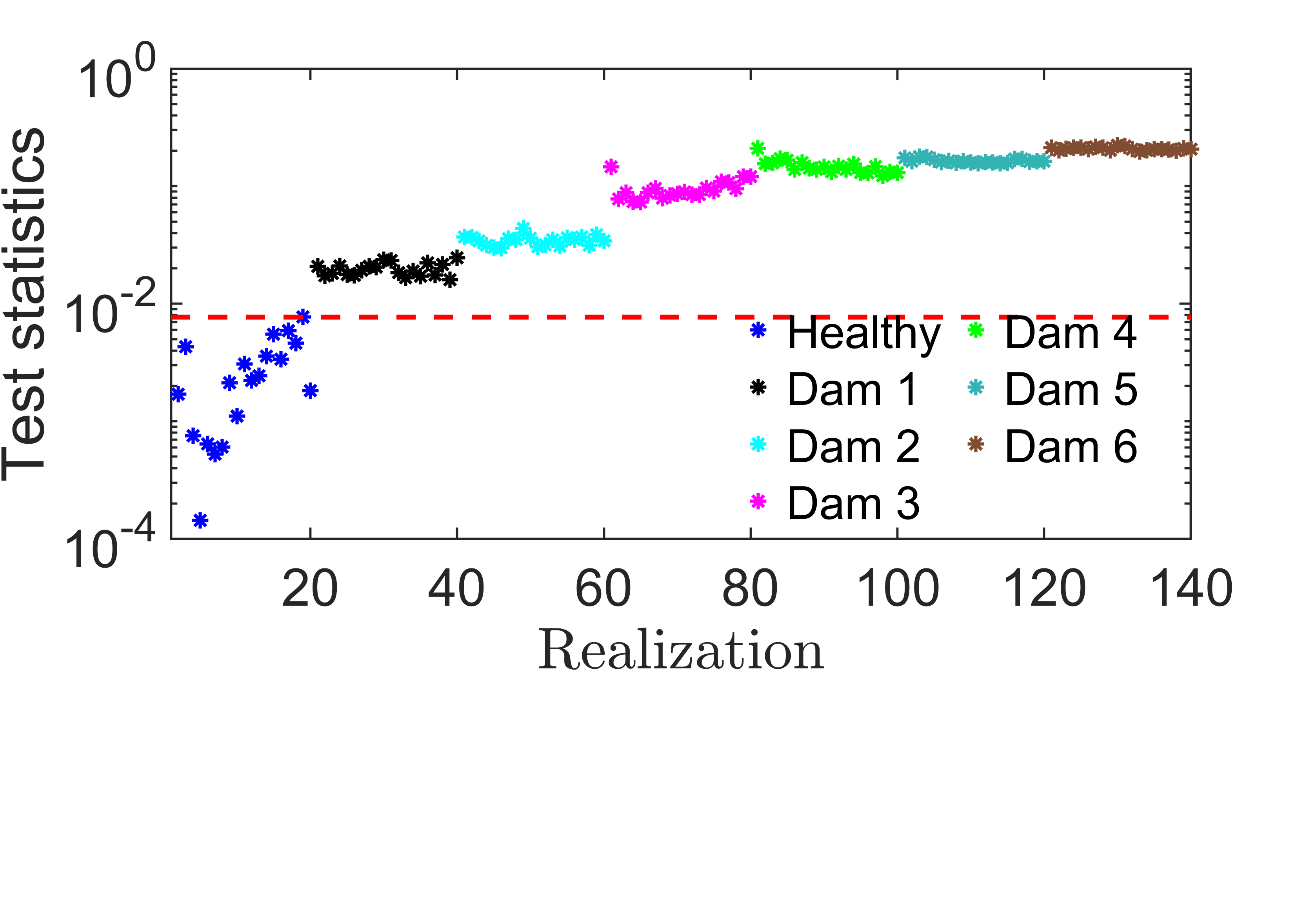}}
    \put(280,10){\includegraphics[width=0.36\columnwidth]{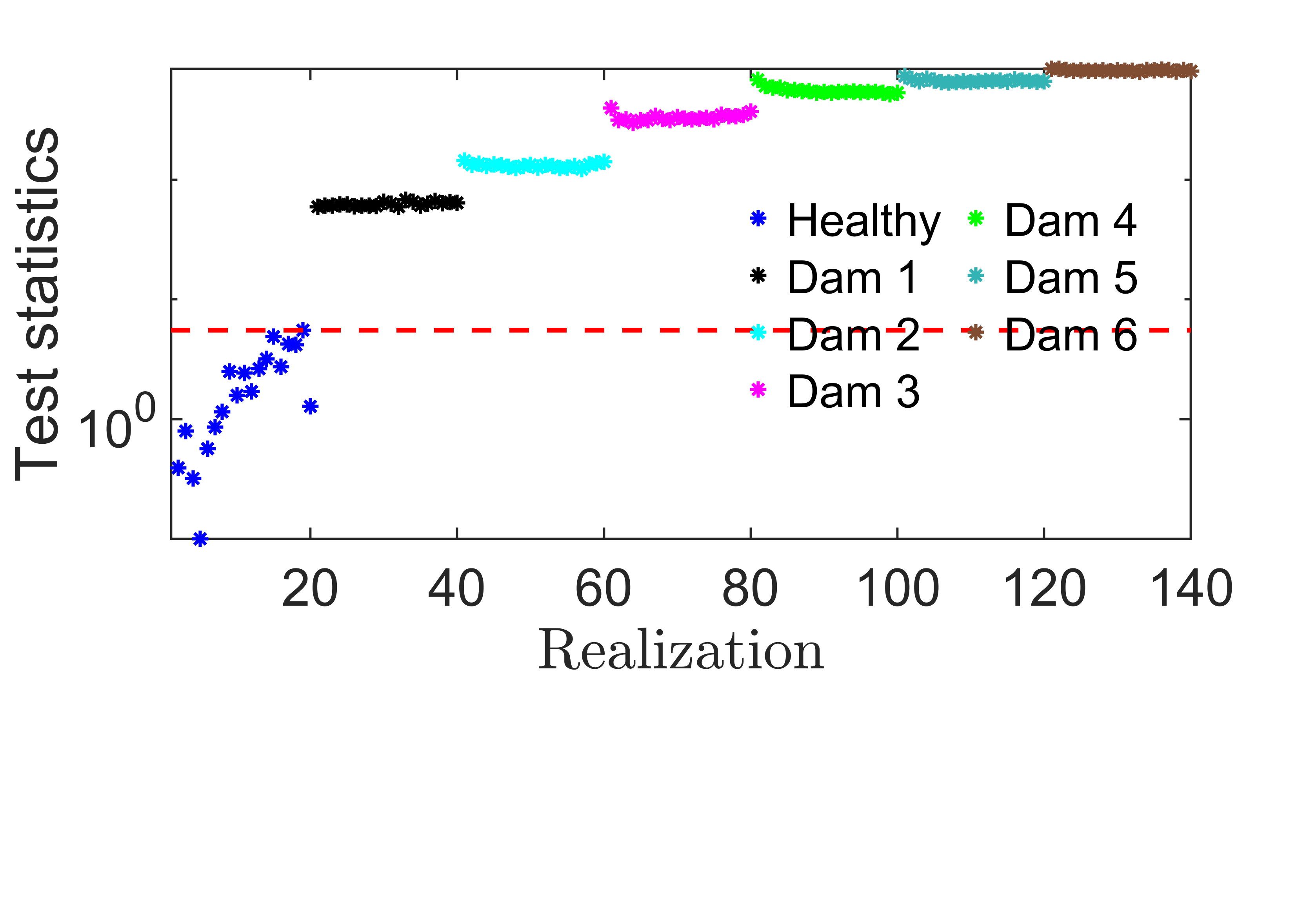}}

     \put(-20,125){\large \textbf{(a)}}
    \put(140,125){\large \textbf{(b)}}
    \put(290,125){ \large \textbf{(c)}}
    \end{picture}
   \vspace{-1.5cm}
    
    \caption{Damage detection performance comparison for damage non-intersecting path 1-4 for the CFRP plate using the AR($6$)-based covariance matrix: (a) standard AR approach; (b) SVD-based approach; (c) PCA-based approach.} 
\label{fig:dam nonintersect theory composite} 
\end{figure}
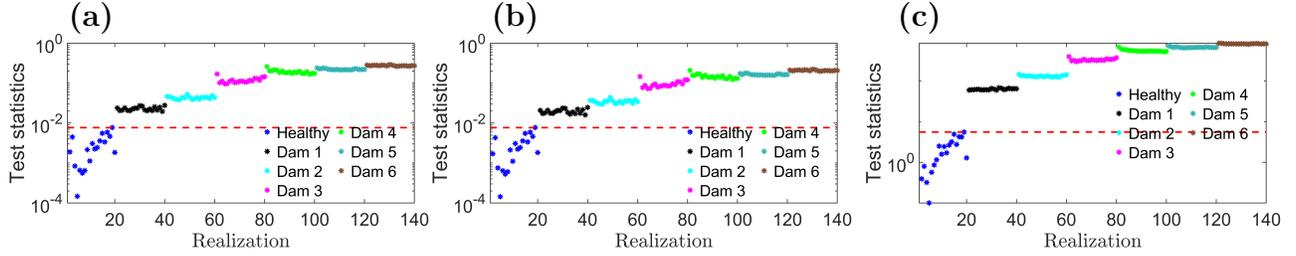

Figure \ref{fig:dam nonintersect exp composite} shows the damage detection performance of the damage non-intersecting path 1-4 using the standard AR, SVD, and PCA-based approach. In this case, the covariance matrix was derived from the 20 experimental healthy signals. It can be observed that, for the case of the CFRP plate, perfect damage detection (all damage states) was achieved for the damage non-intersecting path 1-4 for the standard AR and the SVD-based approach. The PCA-based approach shows poor performance in this case. The $\alpha$ level used for the standard AR, SVD, and the PCA-based case was $0.01$, $0.01$, and $1 \times 10^{-4}$, respectively. Figure \ref{fig:dam nonintersect theory composite}(a),(b) and (c) show the damage detection performance of the damage non-intersecting path 1-4 using the standard AR, SVD, and PCA-based approach using the AR($6$)-based covariance. It can be observed that perfect detection was achieved for all three cases. The thresholds were manually adjusted when using AR($6$)-based covariance. Table \ref{tab: alpha compo} compactly shows the different $\alpha$-level used in the composite plate study.

\begin{table}[t!]
\caption{$\alpha$-level chart for composite plate}\label{tab: alpha compo}
\centering
{\footnotesize
\centering
\begin{tabular}{lllc}
\hline
  
Method & Path & Covariance & $\alpha$-level  \\

 \hline
 Standard AR & 1-4 & Experiment &  $0.01$ \\
 & 1-4 & Theory & manual \\
  & 2-6 & Experiment & $0.03$  \\
  & 2-6 & Theory & manual \\
\hline
SVD-based & 1-4 & Experiment & $0.01$ \\
 & 1-4 & Theory & manual \\
  & 2-6 & Experiment & $0.1$\\
  & 2-6 & Theory & manual \\
\hline
PCA-based & 1-4 & Experiment & $1 \times 10^{-4}$ \\

 & 1-4 & Theory & manual\\
 
 & 2-6 & Experiment & $0.003$  \\
 & 2-6 & Theory & $0.09$ \\

\hline
\end{tabular} }
\end{table}

\begin{figure}[t!]
    \centering
    \begin{picture}(400,130)
    \put(-40,-50){ \includegraphics[width=0.55\columnwidth]{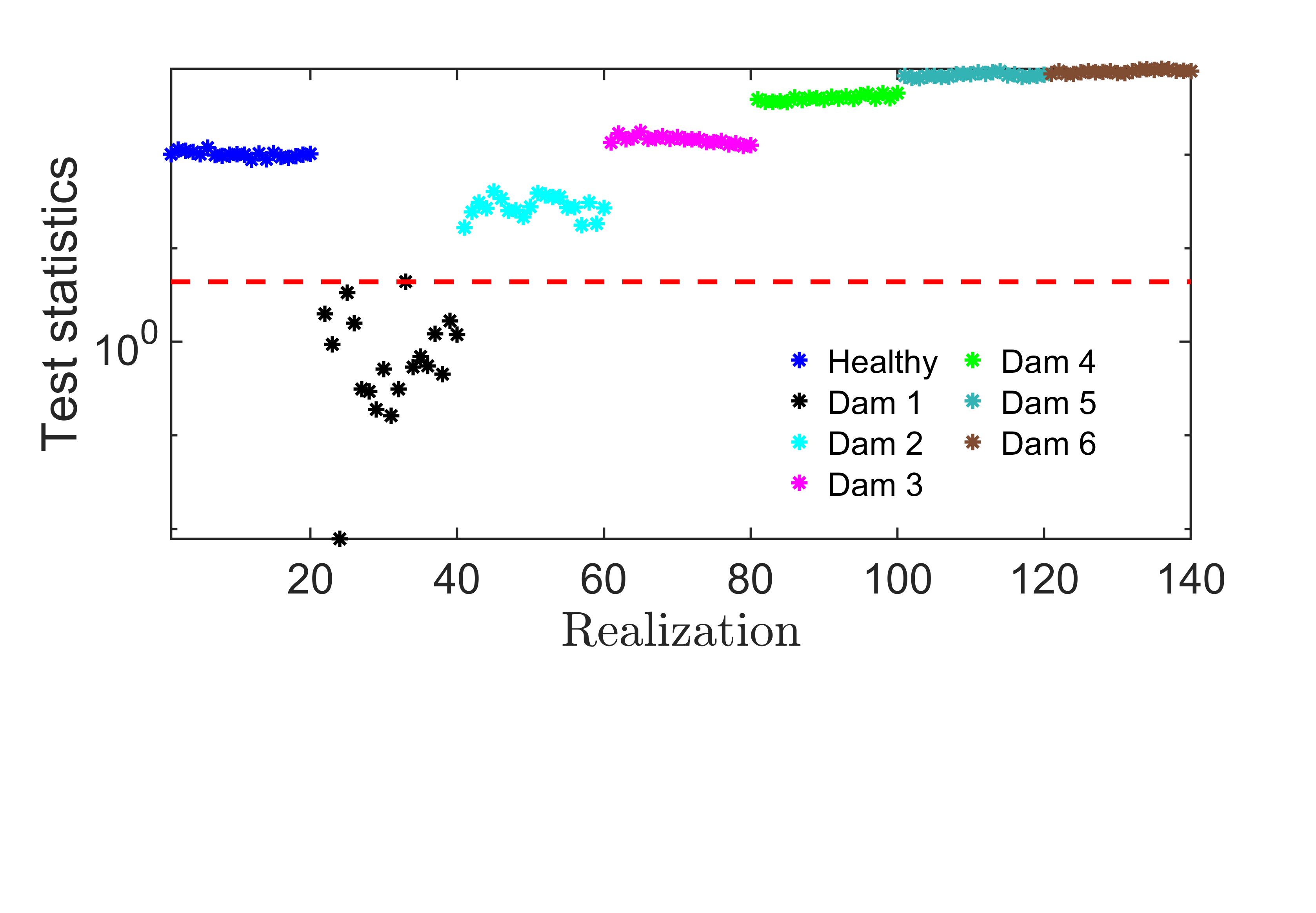}}
    \put(200,-50){\includegraphics[width=0.55\columnwidth]{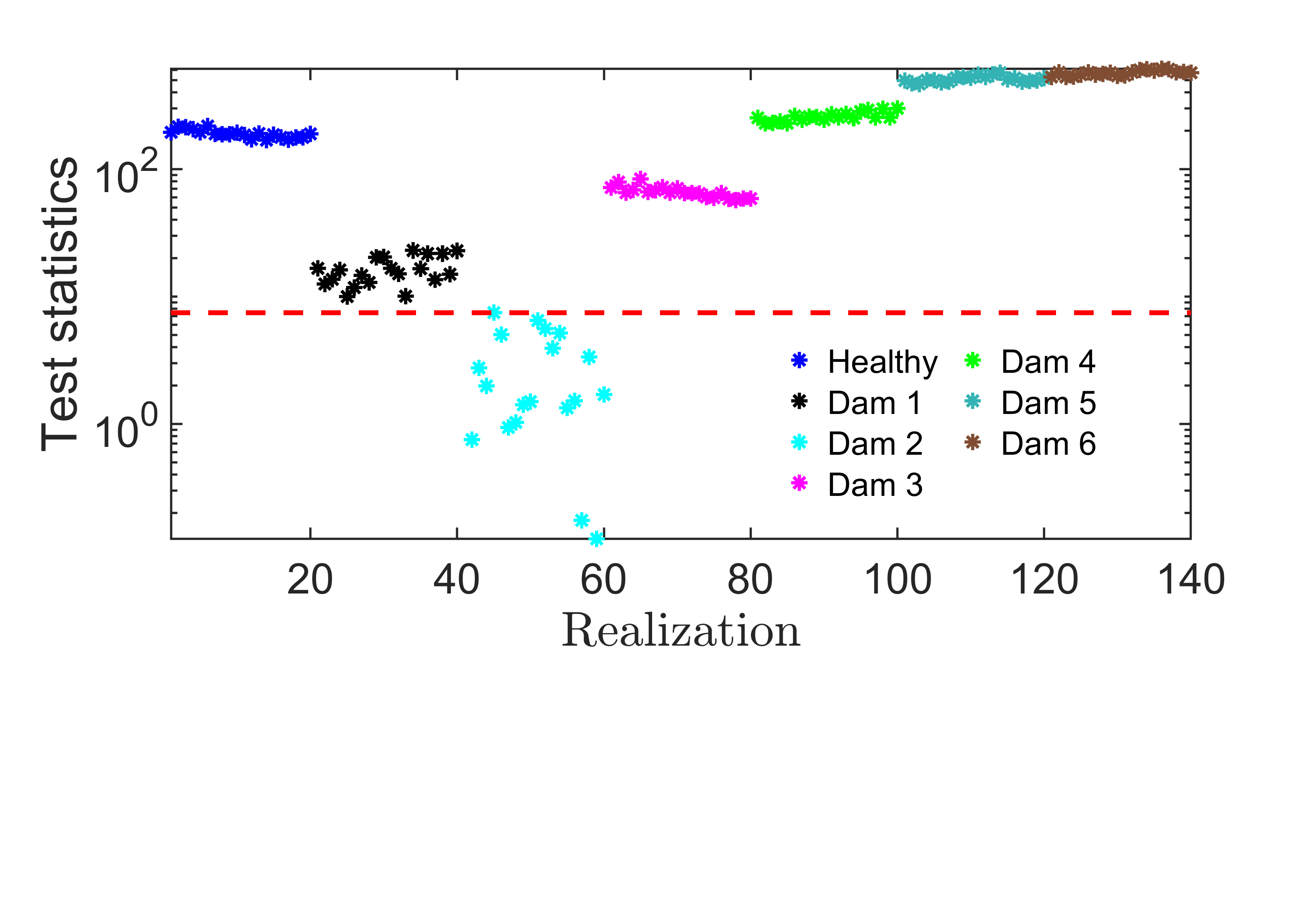}}
    
    \put(-40,-185){ \includegraphics[width=0.55\columnwidth]{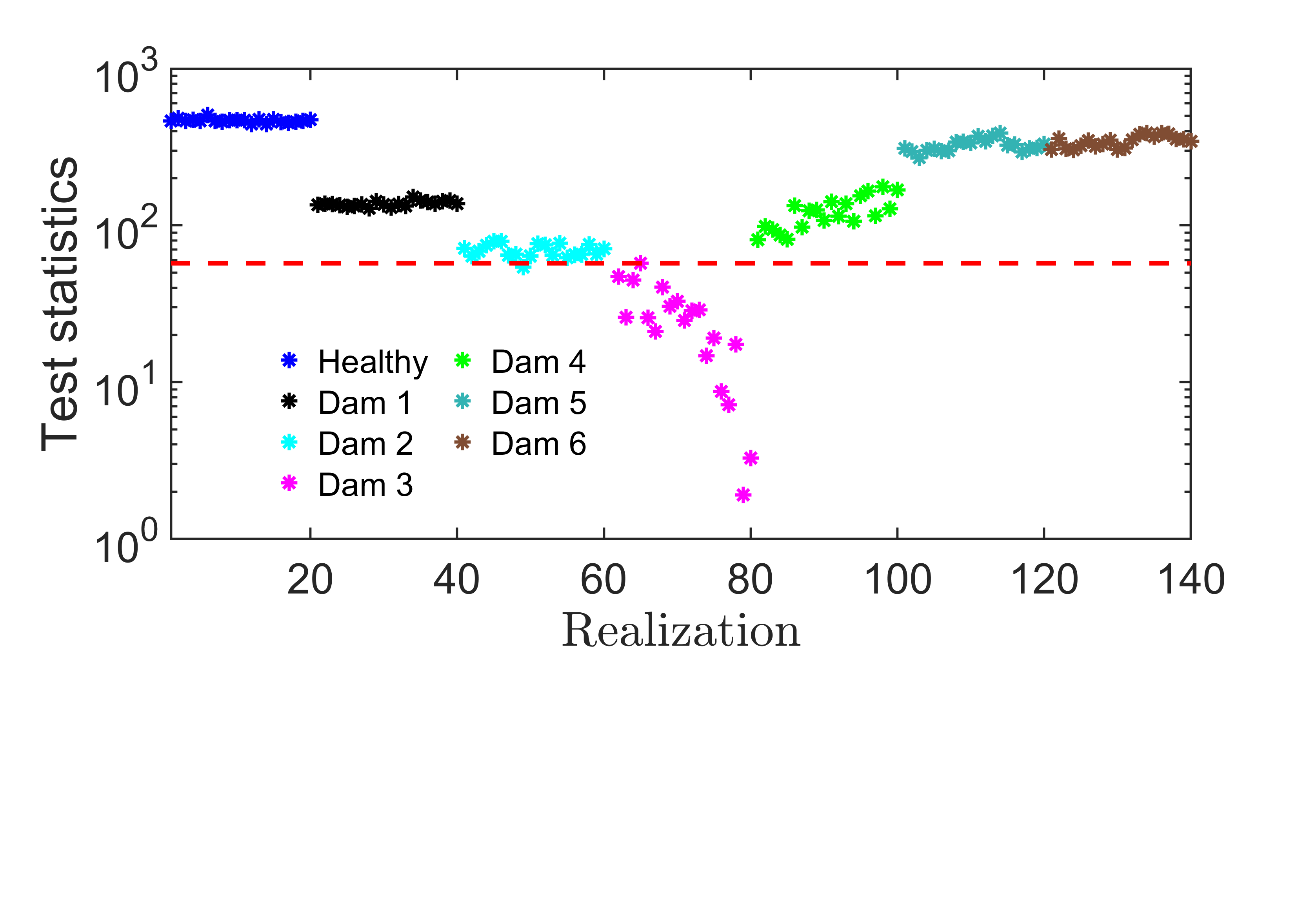}}
    \put(200,-185){\includegraphics[width=0.55\columnwidth]{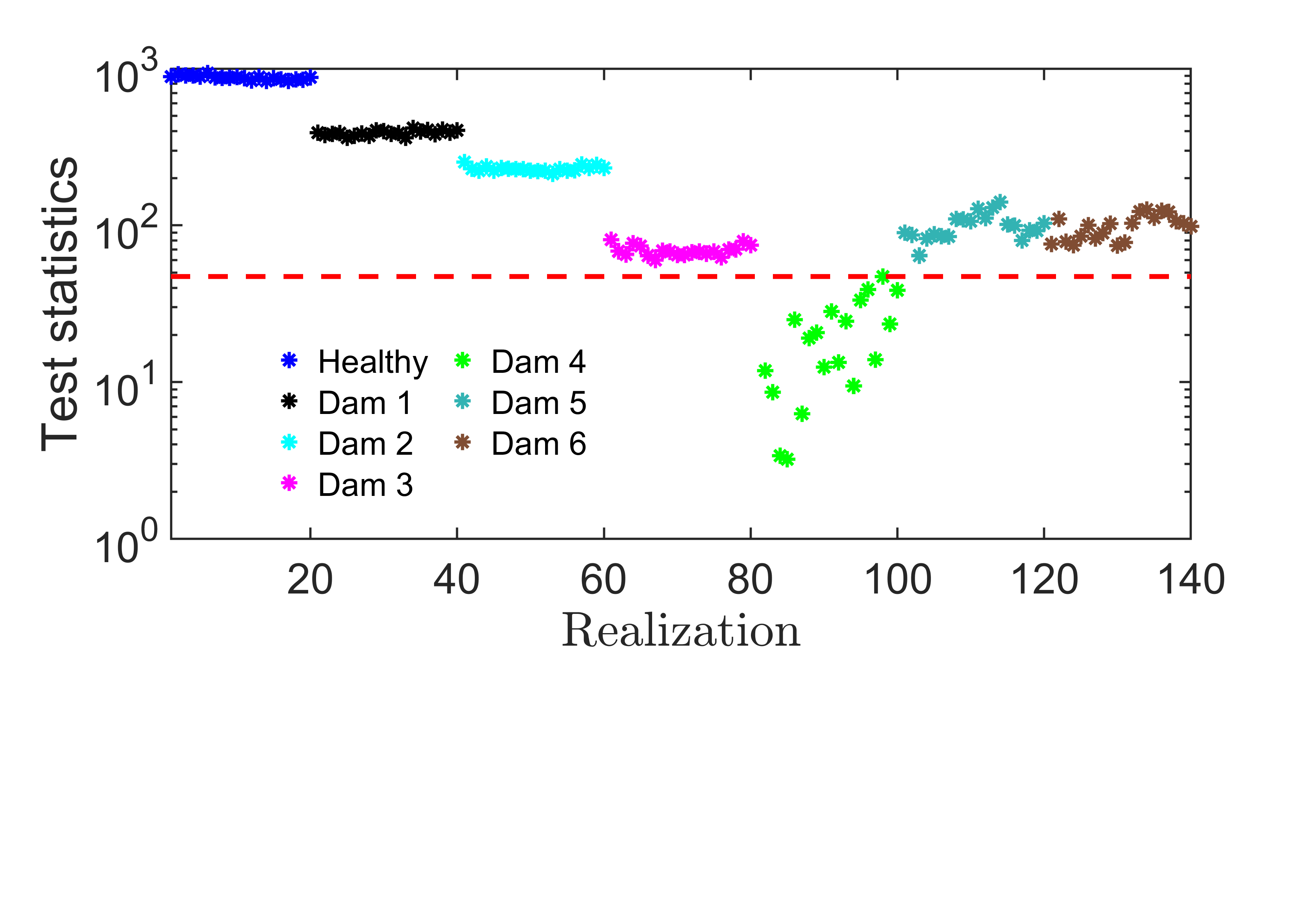}}
    
    \put(-40,-320){ \includegraphics[width=0.55\columnwidth]{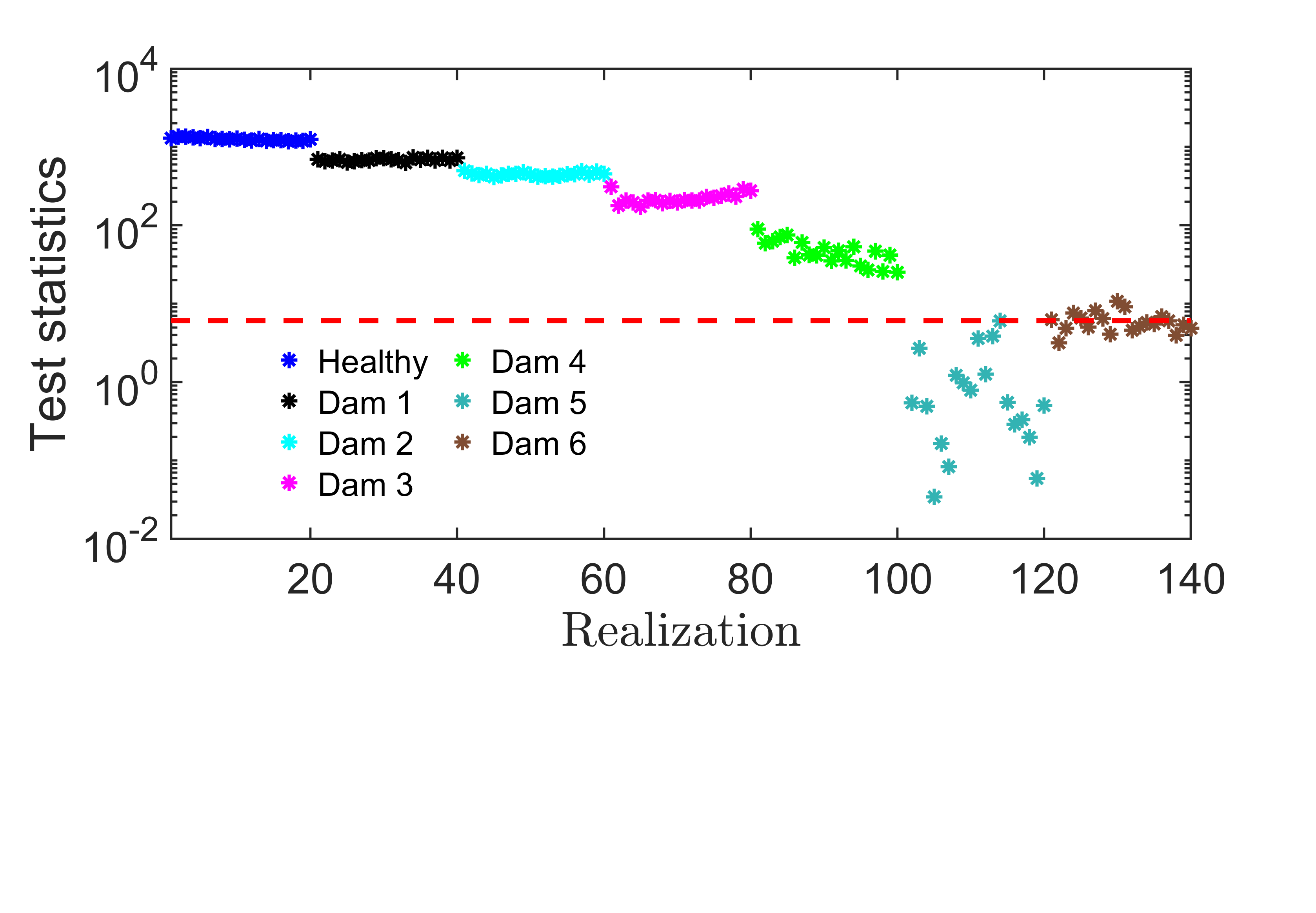}}
    \put(200,-320){\includegraphics[width=0.55\columnwidth]{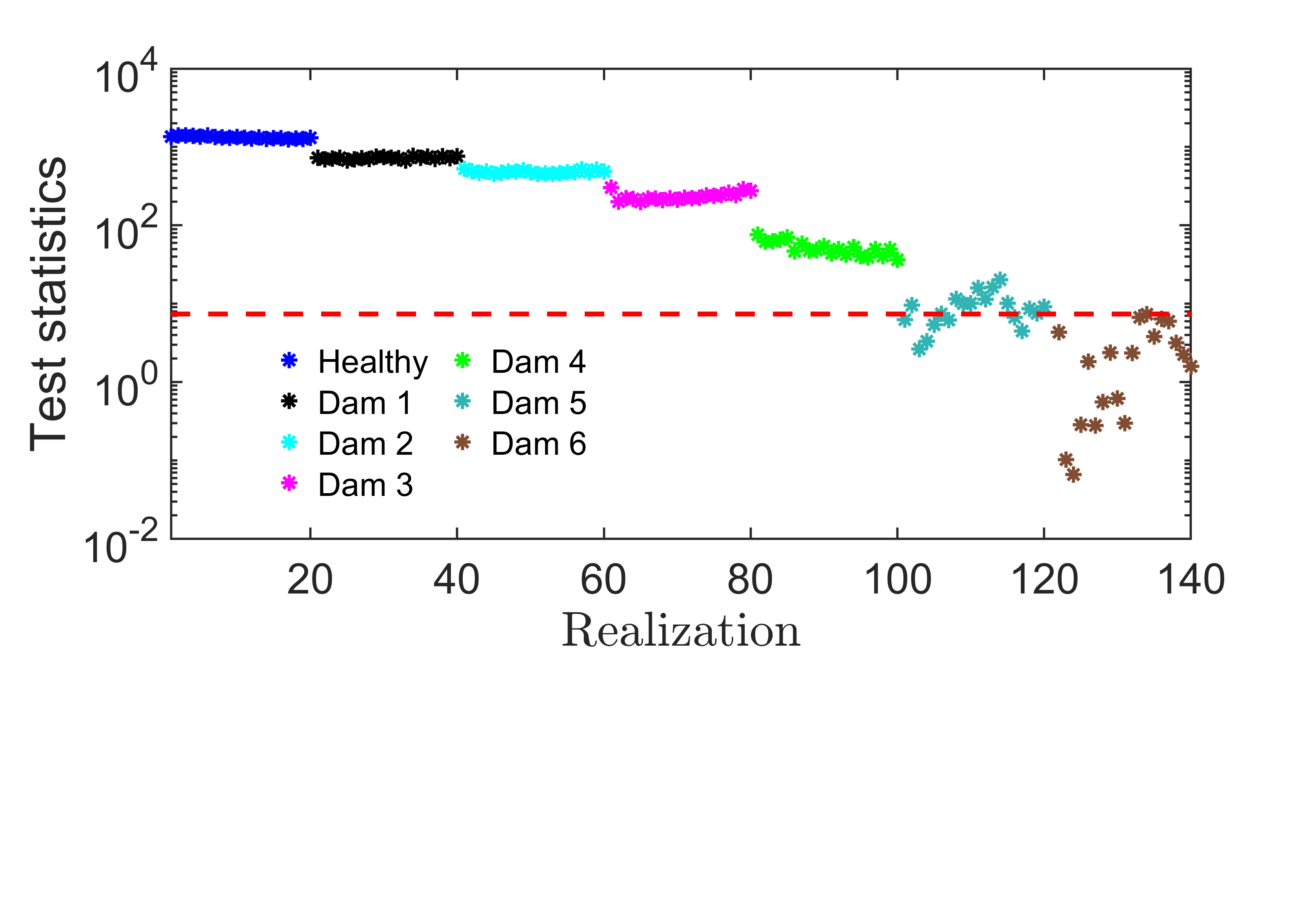}}

    \put(-15,127){\large \textbf{(a)}}
    \put(220,127){ \large \textbf{(b)}}
    \put(-15,-9){\large \textbf{(c)}}
    \put(220,-9){ \large \textbf{(d)}}
    \put(-15,-142){\large \textbf{(e)}}
    \put(220,-142){ \large \textbf{(f)}}
    \end{picture}
    \vspace{9cm}
    \caption{Damage identification results for the CFRP plate using damage non-intersecting path 1-4 for the standard AR-based approach and using the covariance matrix derived from 20 healthy signals: (a) damage level 1; (b) damage level 2; (c) damage level 3; (d) damage level 4; (e) damage level 5; (f) damage level 6.} 
\label{fig:dam iden theory composites} 
\end{figure}
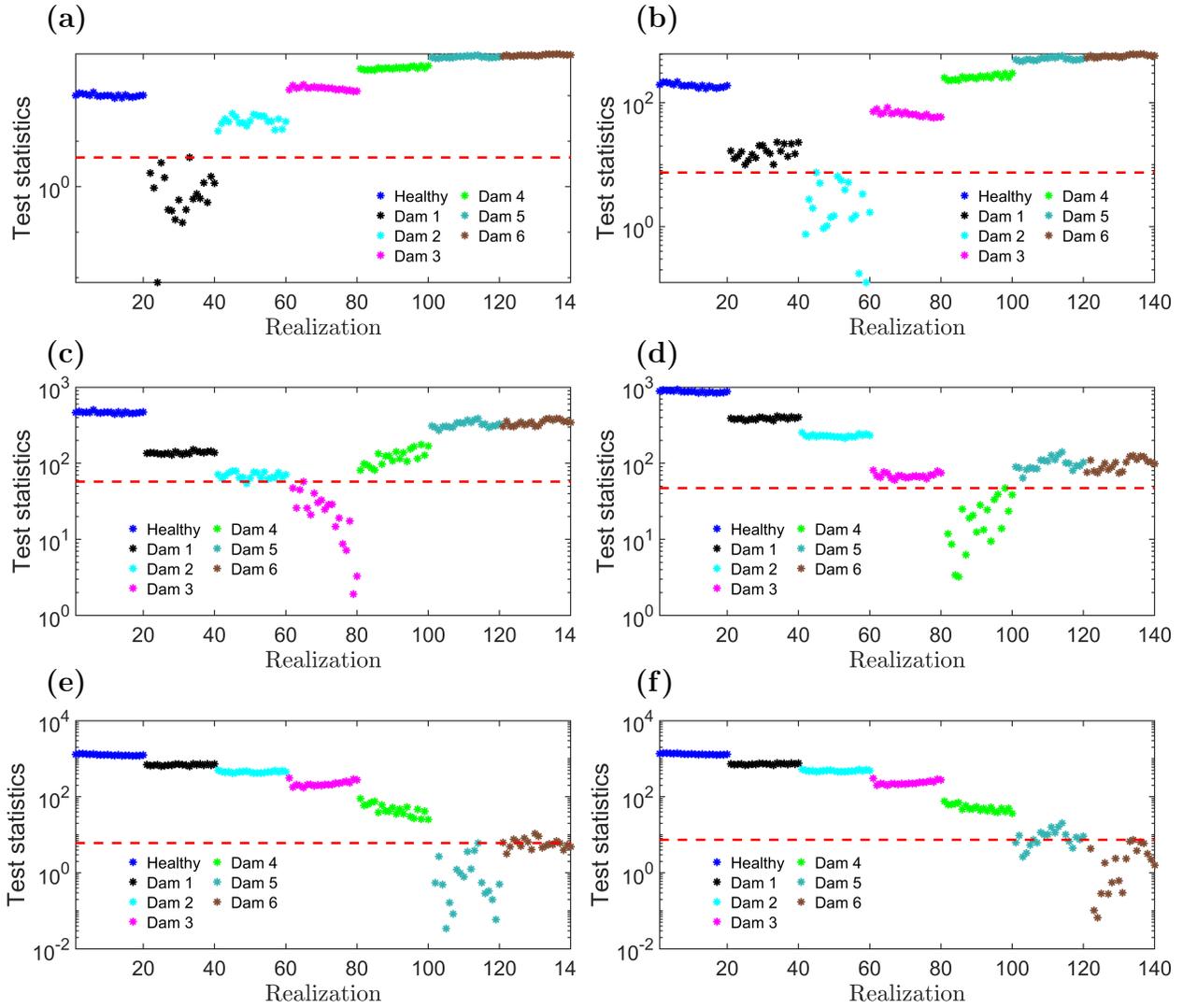

Figure \ref{fig:dam iden theory composites} shows the damage identification results for the damage non-intersecting path 1-4 for the CFRP plate using the standard AR-based approach and using the covariance matrix derived from 20 experimental healthy signals. It can be observed that damage level 1,2,3, and 4 were identified perfectly. On the other hand, damage level 5 and 6 were identified with missed classification.

\begin{table}[t!]
\caption{Damage detection and identification summary results for the CFRP plate.}\label{tab:sum_res composite}
\centering
\renewcommand{\arraystretch}{1.2}
{\footnotesize
\centering
\hspace*{-1cm}
\begin{tabular}{lllccccccc}
\hline
   &  &  &  \multicolumn{5}{|c}{\bf Damage Detection} \\  
 \cline{4-10}
Method & Path & Covariance& \multicolumn{1}{|c}{False alarms} & \multicolumn{6}{|c}{Missed damage} \\
\cline{4-10}
 \multicolumn{3}{c|}{}  &Healthy & Damage 1 & Damage 2 & Damage 3 & Damage 4 & Damage 5& Damage 6 \\
 \hline
 Standard AR & 1-4 & Experiment &0/20 & 0/20& 0/20&0/20 &0/20 &0/20 &0/20 \\
 & 1-4 & Theory &0/20 &0/20 &0/20 & 0/20& 0/20 &0/20 &0/20\\
  & 3-4 & Experiment &0/20 &7/20 &0/20 &0/20 &0/20 &0/20 &0/20 \\
  & 3-4 & Theory & 0/20&0/20 &0/20 &0/20 &0/20 & 0/20&0/20 \\
\hline
SVD-based & 1-4 & Experiment &0/20 & 0/20& 0/20&0/20 &0/20 &0/20 &0/20 \\
 & 1-4 & Theory &0/20 &0/20 &0/20 & 0/20& 0/20 &0/20 &0/20\\
  & 3-4 & Experiment &0/20 &7/20 &0/20 &0/20 &0/20 &0/20 &0/20 \\
  & 3-4 & Theory & 0/20&0/20 &0/20 &0/20 &0/20 & 0/20&0/20 \\
\hline
PCA-based & 1-4 & Experiment &0/20 &20/20 &20/20 &4/20 &0/20 &0/20 &0/20 \\

 & 1-4 & Theory &0/20 &0/20 &0/20 &0/20 &0/20 &0/20 &0/20 \\
 
 & 3-4 & Experiment &0/20 &0/20 &1/20 &0/20 &0/20 &0/20 &0/20 \\
 & 3-4 & Theory & 0/20&0/20 &0/20 &12/20 &7/20 &8/20 &0/20 \\

\hline
\end{tabular} 
\hspace*{-1.7cm}
\begin{tabular}{lllcccccc}

   &  &  &  \multicolumn{4}{|c}{\bf Damage Identification} \\  
 \cline{4-9}
Method & Path & Covariance& \multicolumn{5}{|c}{Missed damage} \\
\cline{4-9}
 \multicolumn{3}{c|}{}  & Damage 1 & Damage 2 & Damage 3 & Damage 4 & Damage 5 & Damage 6 \\
 \hline
 Standard AR & 1-4 & Experiment & (--,0,0,0,0,0)&(0,--,0,0,0,0) & (0,0,--,0,0,0)&(0,0,0,--,0,0) &(0,0,0,0,--,12) & (0,0,0,0,8,--) \\
 & 1-4 & Theory &(--,0,0,0,0,0) &(0,--,0,0,0,0) &(0,0,--,20,0,0) &(0,0,4,--,0,0) &(0,0,0,0,--,0) &(0,0,0,0,2,--) \\
  & 3-4 & Experiment  &(--,0,0,0,0,0) &(0,--,2,0,0,0) &(16,20,--,13,0,0) &(0,14,20,--,0,4) &(0,0,0,6,--,20) & (0,0,7,14,11,--)\\
  & 3-4 & Theory & (--,12,0,0,0,0)& (20,--,0,0,0,0)& (20,20,--,2,0,0)&(0,0,5,--,0,0) &(0,0,0,0,--,5) &(0,0,0,0,0,--) \\
\hline
SVD-based & 1-4 & Experiment & (--,0,0,0,0,0)&(0,--,0,0,0,0) & (0,0,--,0,0,0)&(0,0,0,--,0,0) &(0,0,0,0,--,7) & (0,0,0,0,7,--) \\
 & 1-4 & Theory &(--,0,0,0,0,0) &(0,--,0,0,0,0) &(0,0,--,20,0,0) &(0,0,8,--,0,0) &(0,0,0,0,--,0) &(0,0,0,0,5,--) \\
  & 3-4 & Experiment  &(--,0,0,0,0,0) &(0,--,2,0,0,0) &(7,20,--,13,0,0) &(0,11,20,--,0,4) &(0,0,1,6,--,20) & (0,0,8,14,12,--)\\
  & 3-4 & Theory & (--,12,0,0,0,0)& (20,--,0,0,0,0)& (20,20,--,2,0,0)&(0,1,6,--,0,0) &(0,0,0,0,--,9) &(0,0,0,0,0,--) \\
\hline
PCA-based & 1-4 & Experiment & (--,14,0,0,0,0) &(20,--,14,0,0,0) & (18,13,--,17,1,0)&(17,14,20,--,1,0) &(0,0,3,9,--,17) &(0,0,13,17,4,--) \\

 & 1-4 & Theory &(--,0,0,0,0,0) & (0,--,0,0,0,0) &(0,0,--,0,0,0) &(0,0,0,--,1,0) &(0,0,0,0,--,16) &(0,0,0,0,0,--) \\
 
 & 3-4 & Experiment & (--,4,1,0,0,0)&(18,--,5,1,0,0) &(20,20,--,8,0,5) &(20,20,20,--,2,20) &(0,0,3,12,--,17) &(0,0,2,2,0,--) \\
 & 3-4 & Theory &(--,6,0,0,0,13) &(20,--,1,0,0,19) &(20,20,--,16,20,19) &(1,8,20,--,20,3) &(0,0,15,12,--,0) &(20,18,0,0,0,--) \\

\hline
\end{tabular} }
\end{table}

\begin{figure}[t!]
    \centering
    \begin{picture}(400,130)
    \put(-40,-50){ \includegraphics[width=0.5\columnwidth]{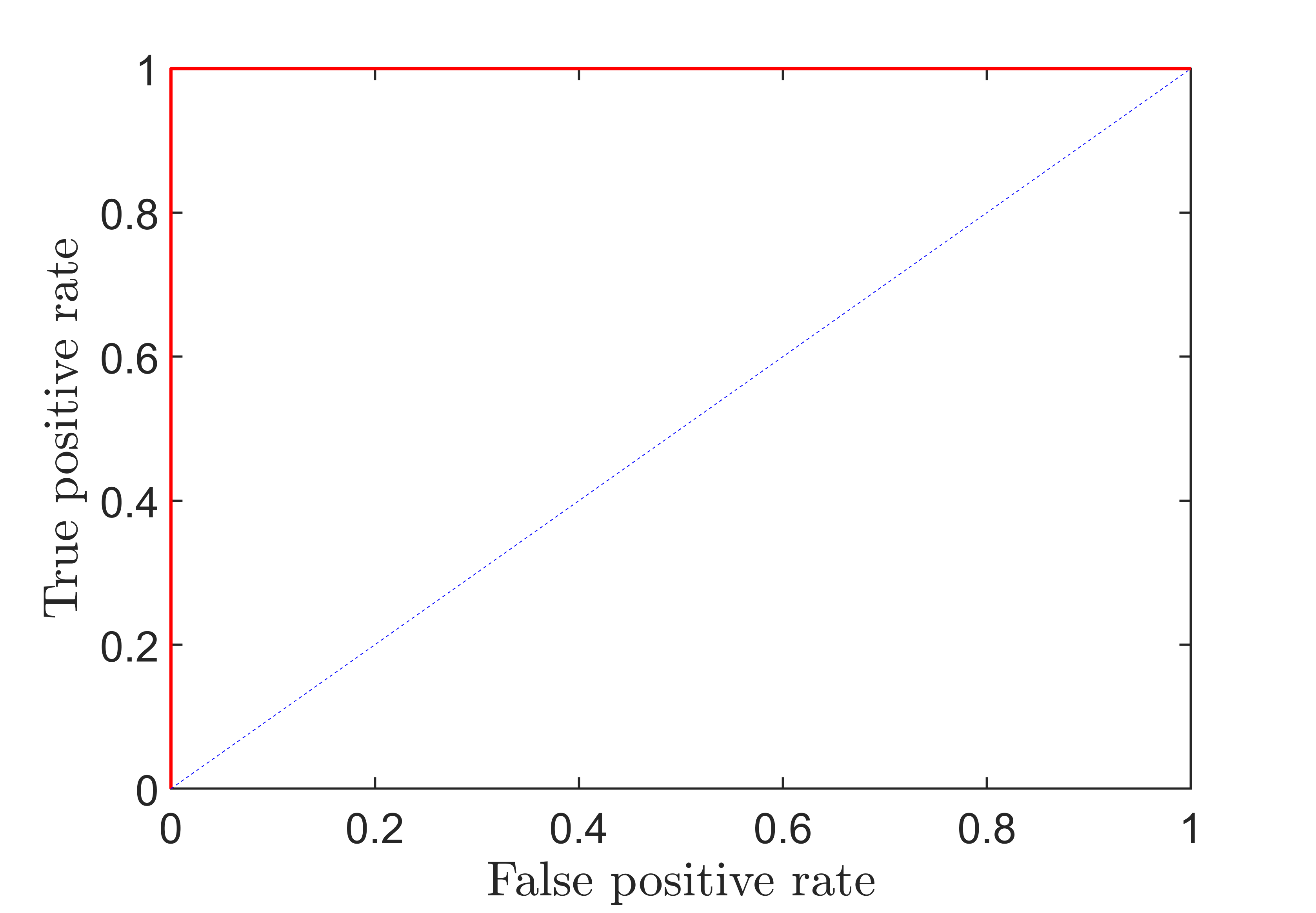}}
    \put(200,-50){\includegraphics[width=0.5\columnwidth]{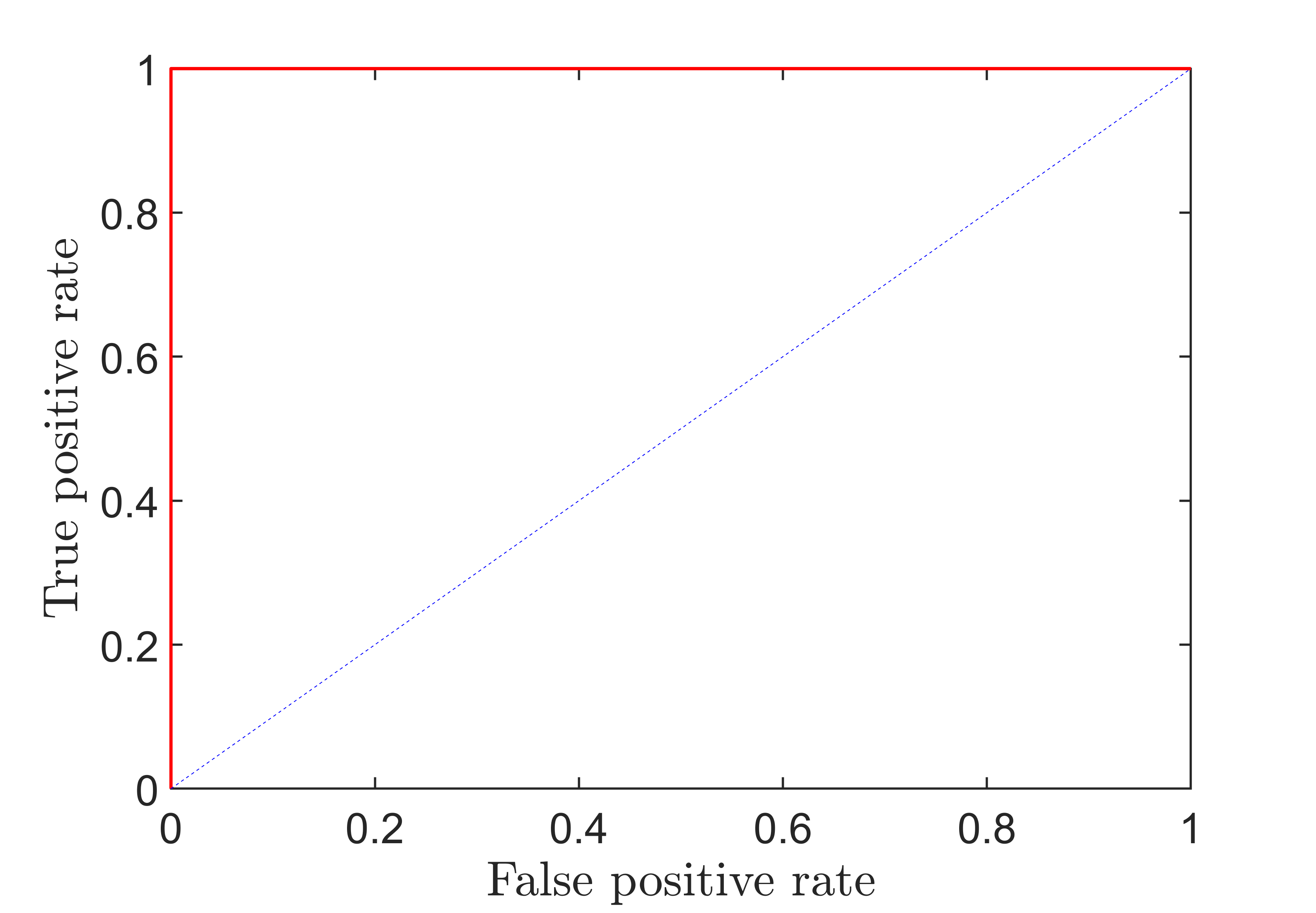}}
    
    \put(-40,-230){ \includegraphics[width=0.5\columnwidth]{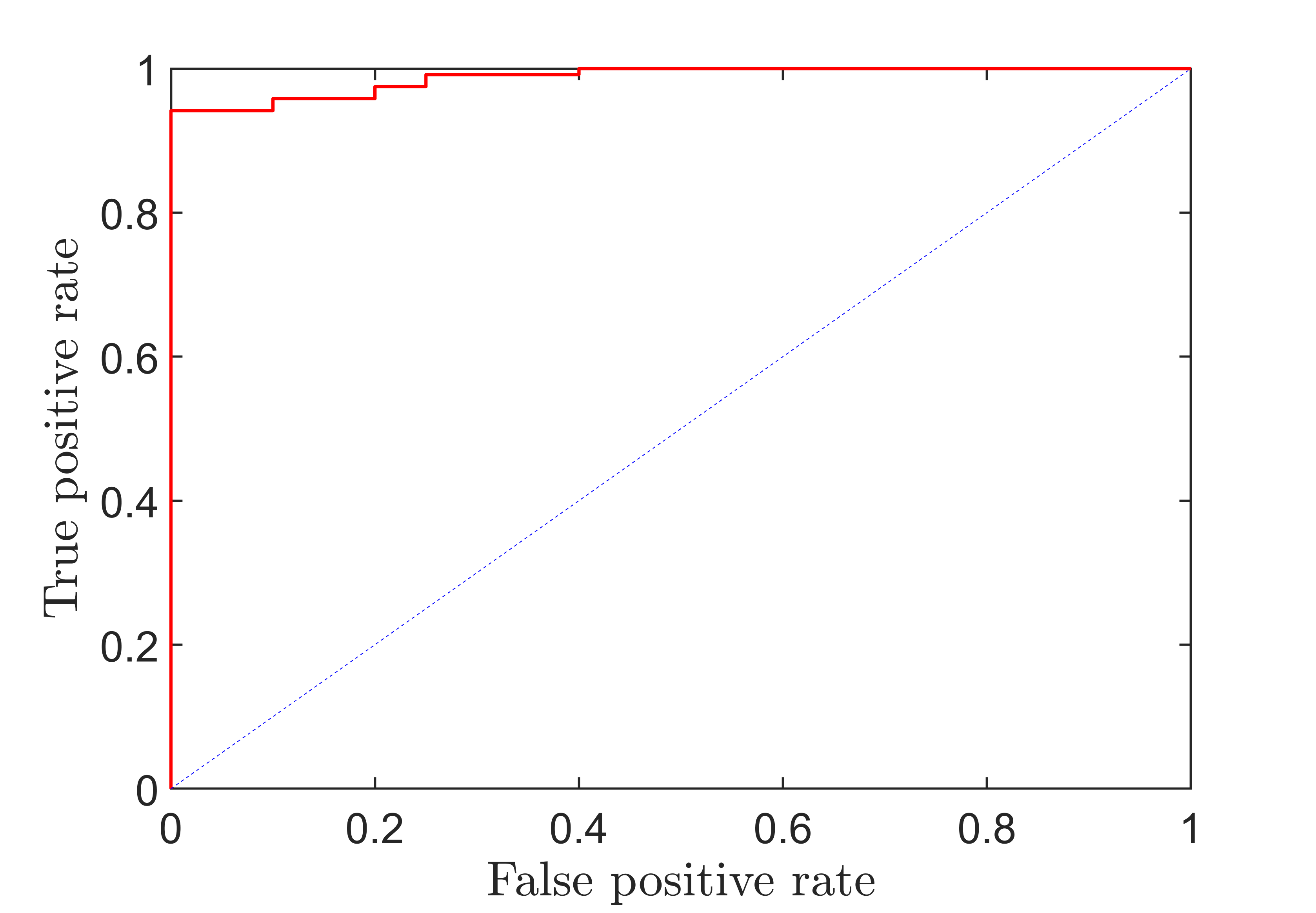}}
    \put(200,-230){\includegraphics[width=0.5\columnwidth]{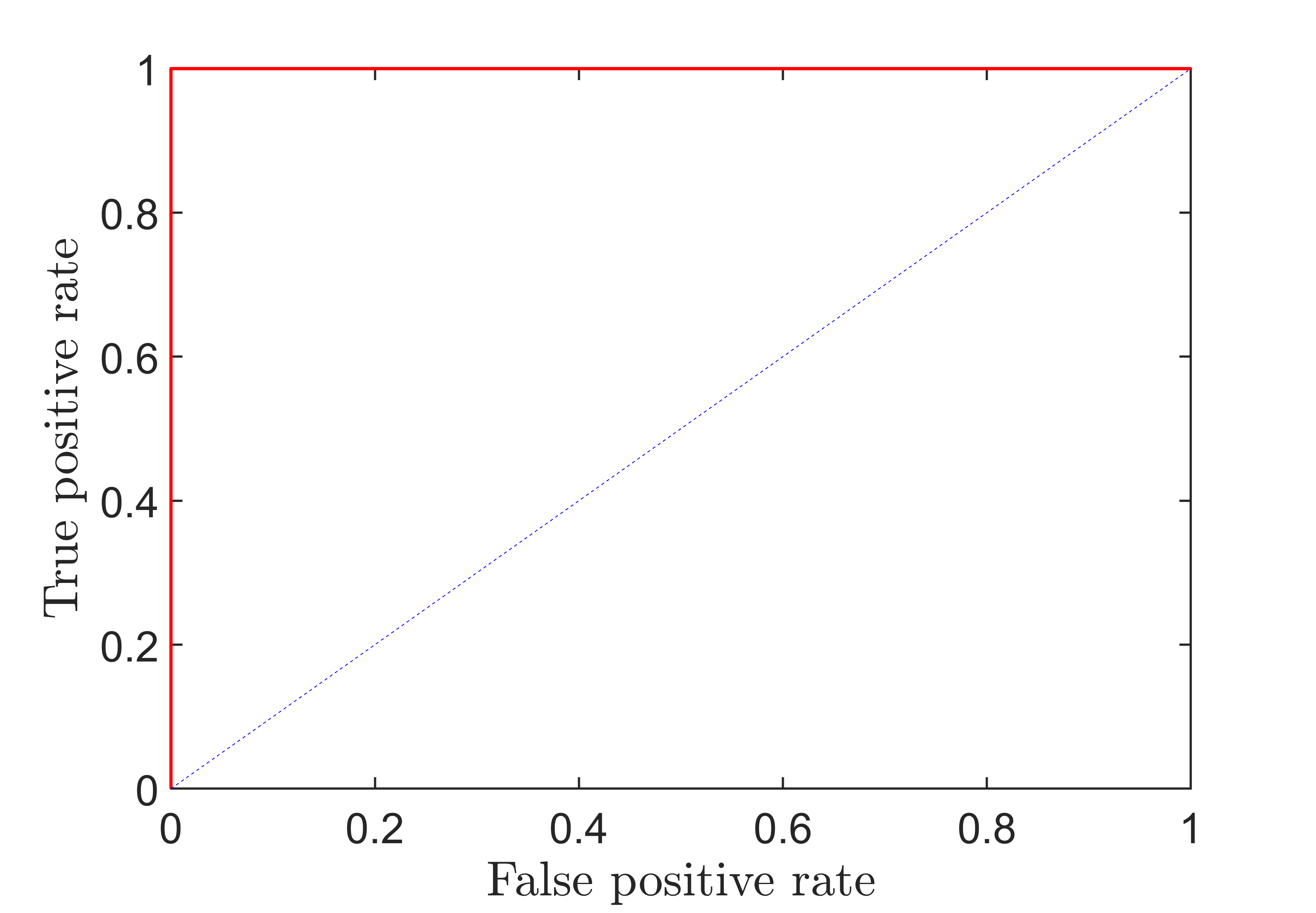}}

    \put(-15,110){\large \textbf{(a)}}
    \put(220,110){ \large \textbf{(b)}}
     \put(-15,-70){\large \textbf{(c)}}
    \put(220,-70){ \large \textbf{(d)}}
    \end{picture}
    \vspace{8cm}
    \caption{Receiver operating characteristic (ROC) plots comparing the SVD-based damage detection methods for the CFRP plate for different paths and covariance used: (a) path 1–4 with the experimental covariance; (b) path 1–4 with the AR($6$)-based covariance; (c) path 3–4 with the experimental covariance; and (d) path 3–4 with the AR($6$)-based covariance.} 
\label{fig:ROC composite} 
\end{figure} 

Figures \ref{fig:ROC composite} show the ROC plots of the SVD-based method for the damage non-intersecting and the damage intersecting case using the experimental as well as the theoretical covariance matrix. In constructing each plot, the threshold of the Q-statistics was varied from -100 to $10^5$ with an increment of 1, that is covering all possible values. It can be observed that for the CFRP plate, in terms of the damage detection, the damage non-intersecting path 1-4 performs better than the damage intersecting path 3-4.
\section{Conclusions}

The objective of this work was the formulation and numerical assessment of a probabilistic damage diagnostic scheme in the context of ultrasonic guided wave-based damage diagnosis using stationary AR models. These are output-only stochastic models and automatically account for uncertainties. In addition to using standard AR-based damage detection and identification where all the parameters were used without any modifications, two additional methods were also investigated where the model parameters were modified to simplify the damage detection algorithm. One approach was an SVD-based approach where model parameters were sorted according to the magnitude of the eigenvalues obtained from the parameter matrix. Another approach was a PCA-based truncation approach where model parameters were projected onto a lower dimensional subspace. These methods were tested on an aluminum plate as well as on a composite coupon for different damage scenarios and different paths. It was found that for the case of the aluminum plate, all three methods work for damage detection and identification. On the other hand, for the CFRP plate, the methods presented in the paper partially work. Perfect damage identification remains challenging with the current framework for composites. This issue will be resolved in the future paper by the authors using time-varying time series models.

The methods presented in the current paper for damage detection and identification for guided wave-based SHM are mathematically simple, computationally inexpensive, probabilistic in nature, and easy to use. In addition, these methods can potentially be automated with an aim of developing smart and intelligent structural systems.
\section*{Acknowledgment}

This work is supported by the U.S. Air Force Office of Scientific Research (AFOSR) grant ``Formal Verification of Stochastic State Awareness for Dynamic Data-Driven Intelligent Aerospace Systems'' (FA9550-19-1-0054) with Program Officer Dr. Erik Blasch.


\bibliographystyle{ieeetr}

\bibliography{references}

\appendix

\section{Material Properties and Equations}

\begin{table}[t] \begin{minipage}{\columnwidth} 
\centering
\caption{Nominal material property values at $25 \ ^0$C}
\label{tbl:matprop}
\begin{tabular}{lll} \hline \hline
Materials & Property name & Values \\ \hline
 
Piezo-electric: PZT-5A & Density ($\rho$) & $7750$ $kg/m^3$ \\
 & Young's modulus (GPa) &$E_{11} =E_{22} = 60.97$\\ $E_{33}=53.19$  \\
 & Poisson ratio & $\nu_{13} = \nu_{23} = 0.4402, \nu_{12}=0.35$ \\
 & \makecell{Piezo-elctric charge constant (m/V)} & \makecell{$d_{31}= d_{32}= 171e-12, d_{33}=374e-12,$\\ $d_{15}=d_{24}=558e-12$} \\
 & Dielectric constant &$\epsilon_{11}=\epsilon_{22}=15.32e-9,\epsilon_{33}=15e-9$ \\
Aluminum & Density ($\rho$) ($Kg/m^3$)  & $2700$ \\
& Young's modulus ($E$ GPa)& $68.9$ \\
& Poisson ratio & $0.33$ \\
Adhesive & Density ($\rho$) ($Kg/m^3$)  & $1100$ \\
& Young's modulus ($E$ GPa) & $2.19$ \\
& Poisson ratio & $0.30$ \\ \hline \hline
\end{tabular} \end{minipage}
\end{table}

The established functional relationships are outlined in the following. 

Properties of piezoelectric materials:
$$\frac{\partial \rho} {\partial T} = 7751.80-7.26e-02T$$
$$E_{11}=E_{22} = 60.45+2.09e-02T$$
$$E_{33}=52.95+9.8e-03T$$
$$\nu_{13}=\nu_{23}=0.43+3e-04T-3e-06T^2-1e-09T^3$$
$$\nu_{12}=0.35+2e-04T-8e-07T^2+2e-09T^3$$
$$d_{31}=d_{32}=170.78-7.1e-03T+6e-04T^2+2e-16T^3$$
$$d_{33}=369.12+1.49e-01T+1.9e-03T^2-4e-09T^3$$
$$d_{15}=d_{24}=556+4.9e-02T+2e-06T^2-2e-09T^3$$
$$ \varepsilon_{11}=\varepsilon_{22}\\=14.9e-09+1.42e-11T+9.74e-14T^2+4.43e-17T^3$$
$$ \varepsilon_{33}\\=14.60e-09+1.47e-11T+1.18e-13T^2-5.31e-18T^3 $$

Properties of aluminum:
$$E_{Al} = 69.62-2.63e-02T$$ 
$$ \frac{\partial \rho} {\partial x} = 2794.60 - 1.84e-01T$$
$$ \nu_{Al} = 0.32+3e-04T$$
Properties of the adhesive:
$$E_{adh}=3.2-0.065T+1.18e-03T^2-7.72e-06T^3$$
$$G_{adh}=1+0.001T-4e-05T^2$$
 \section{Additional Figures}
 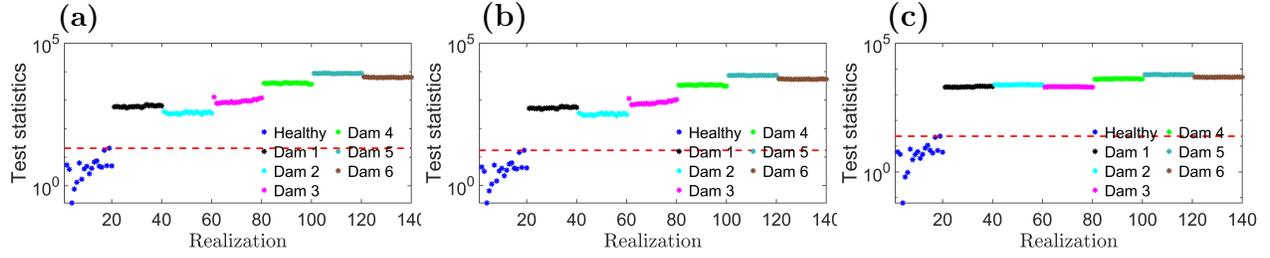
\begin{figure}[t!]
    \centering
    \begin{picture}(400,130)
     \put(-43,10){ \includegraphics[width=0.36\columnwidth]{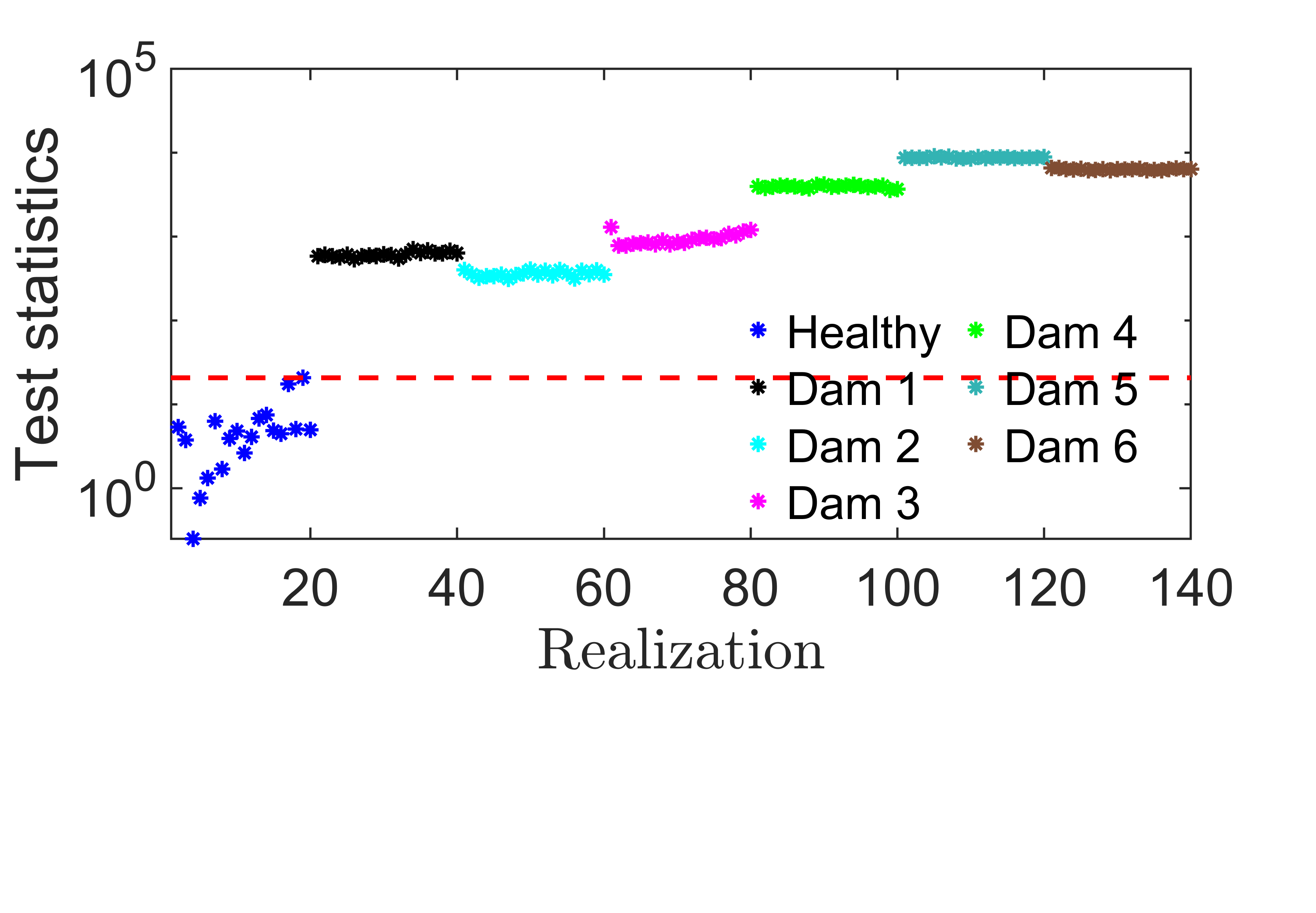}} 
    \put(114,10){ \includegraphics[width=0.36\columnwidth]{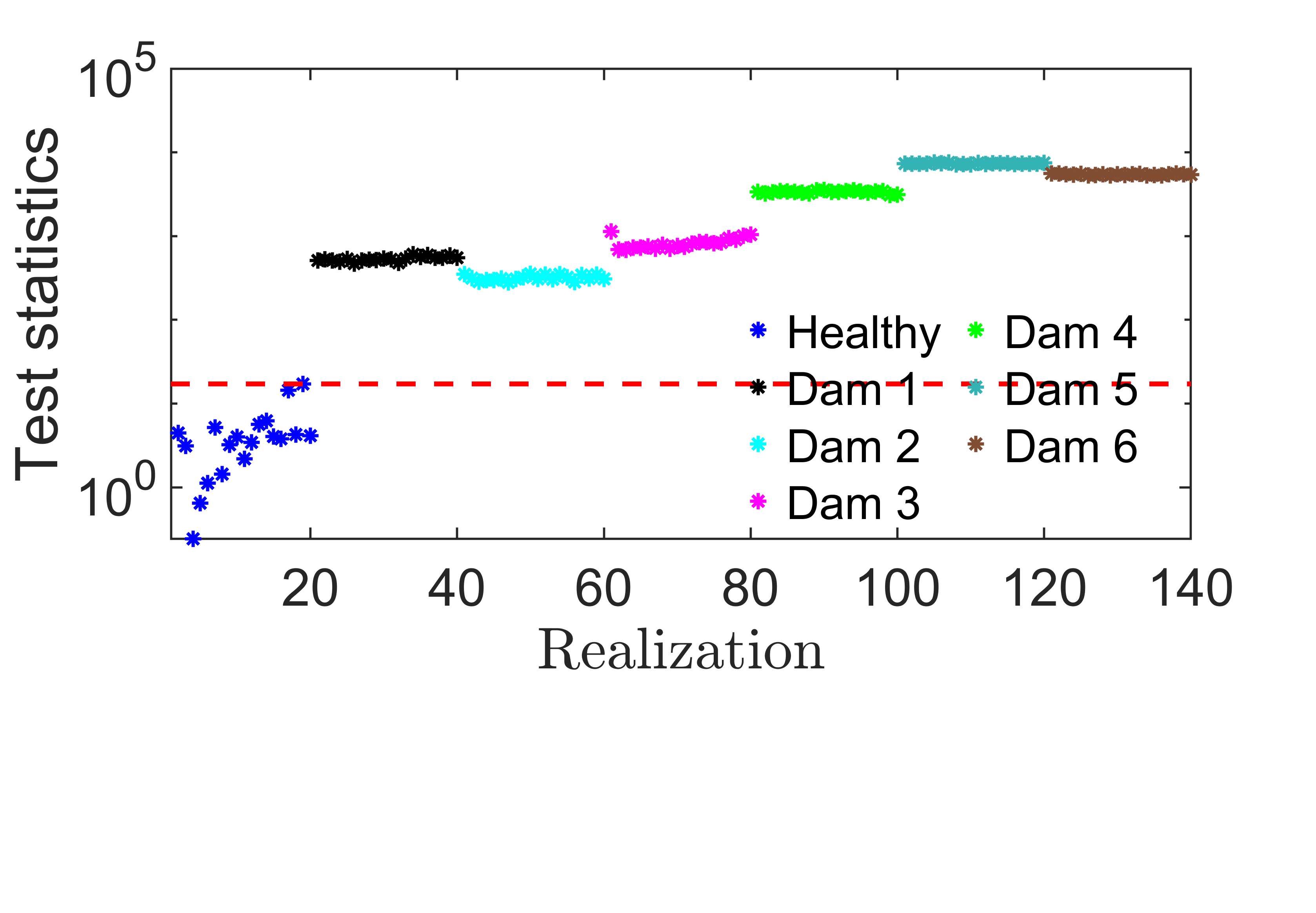}}
    \put(275,10){\includegraphics[width=0.36\columnwidth]{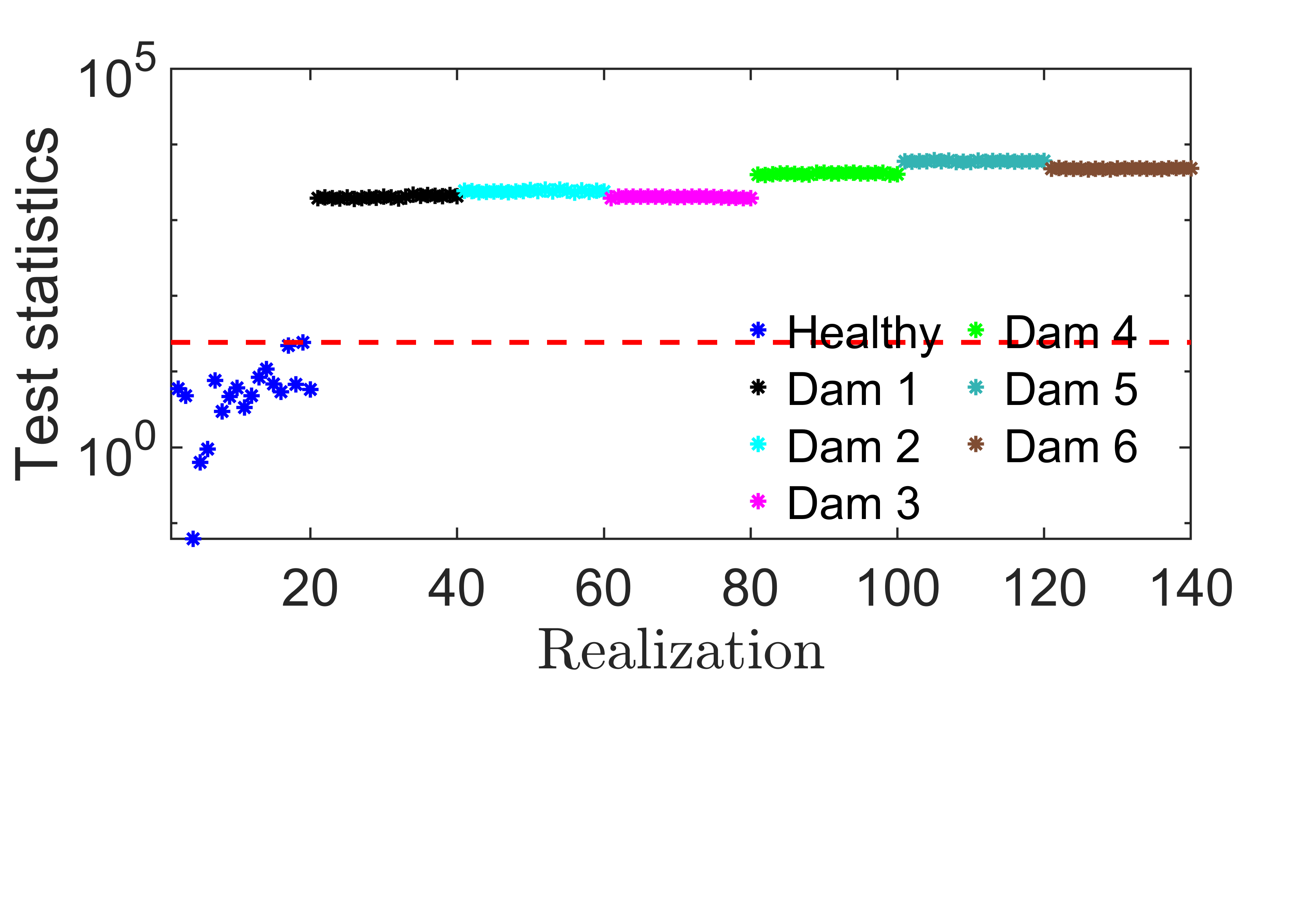}}

    \put(-20,125){\textbf{(a)}}
    \put(140,125){\large \textbf{(b)}}
    \put(290,125){ \large \textbf{(c)}}
    \end{picture}
    \vspace{-1.5cm}
    
    \caption{Damage detection performance comparison for damage intersecting path 5-2 for the CFRP plate using the covariance matrix derived from 20 experimental healthy signals: (a) standard AR approach; (b) SVD-based approach; (c) PCA-based approach. } 
\label{fig:dam intersect exp composite P23} 
\end{figure} 

\begin{figure}[t!]
    \centering
    \begin{picture}(400,130)
     \put(-43,10){ \includegraphics[width=0.36\columnwidth]{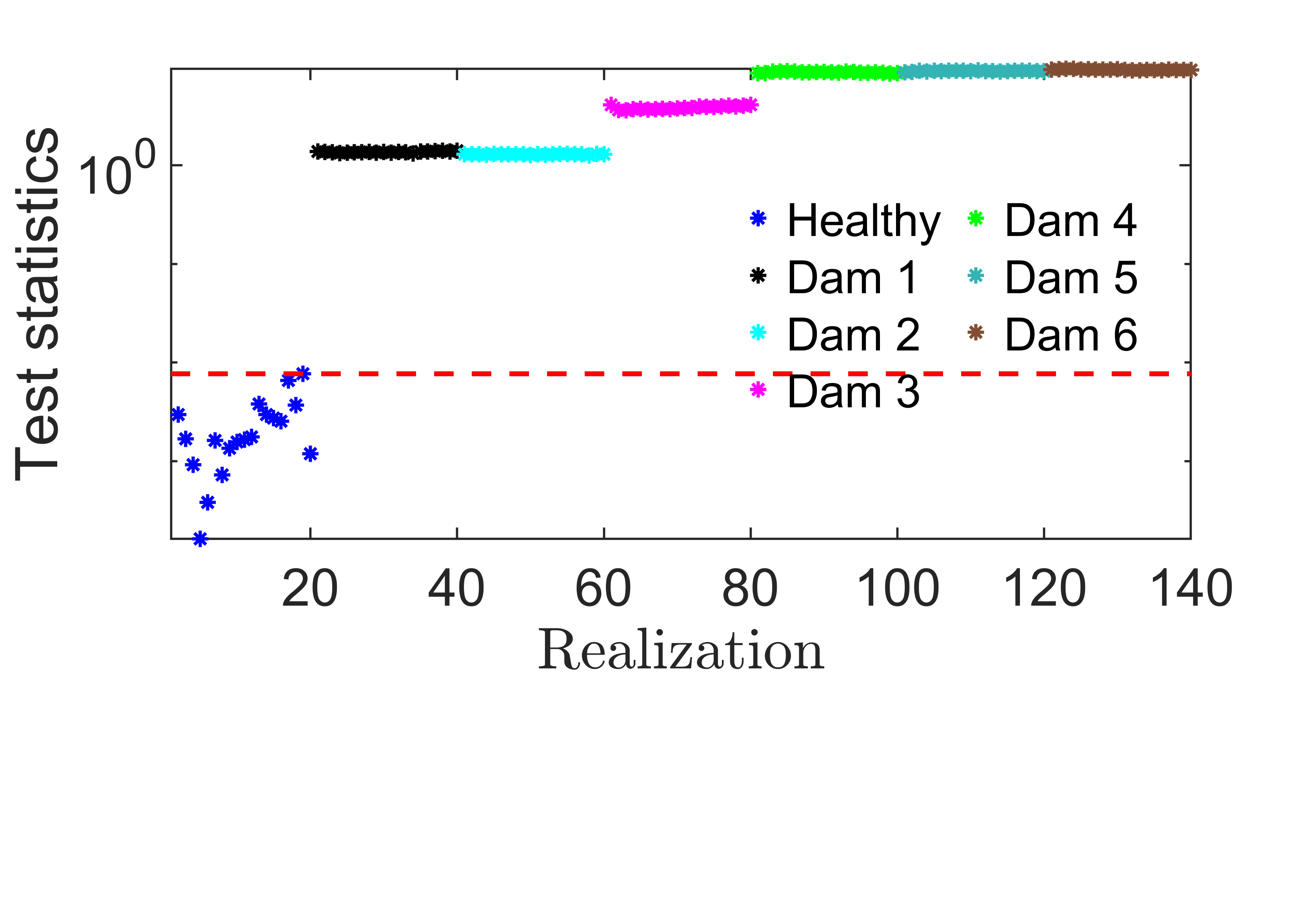}} 
    \put(114,10){ \includegraphics[width=0.36\columnwidth]{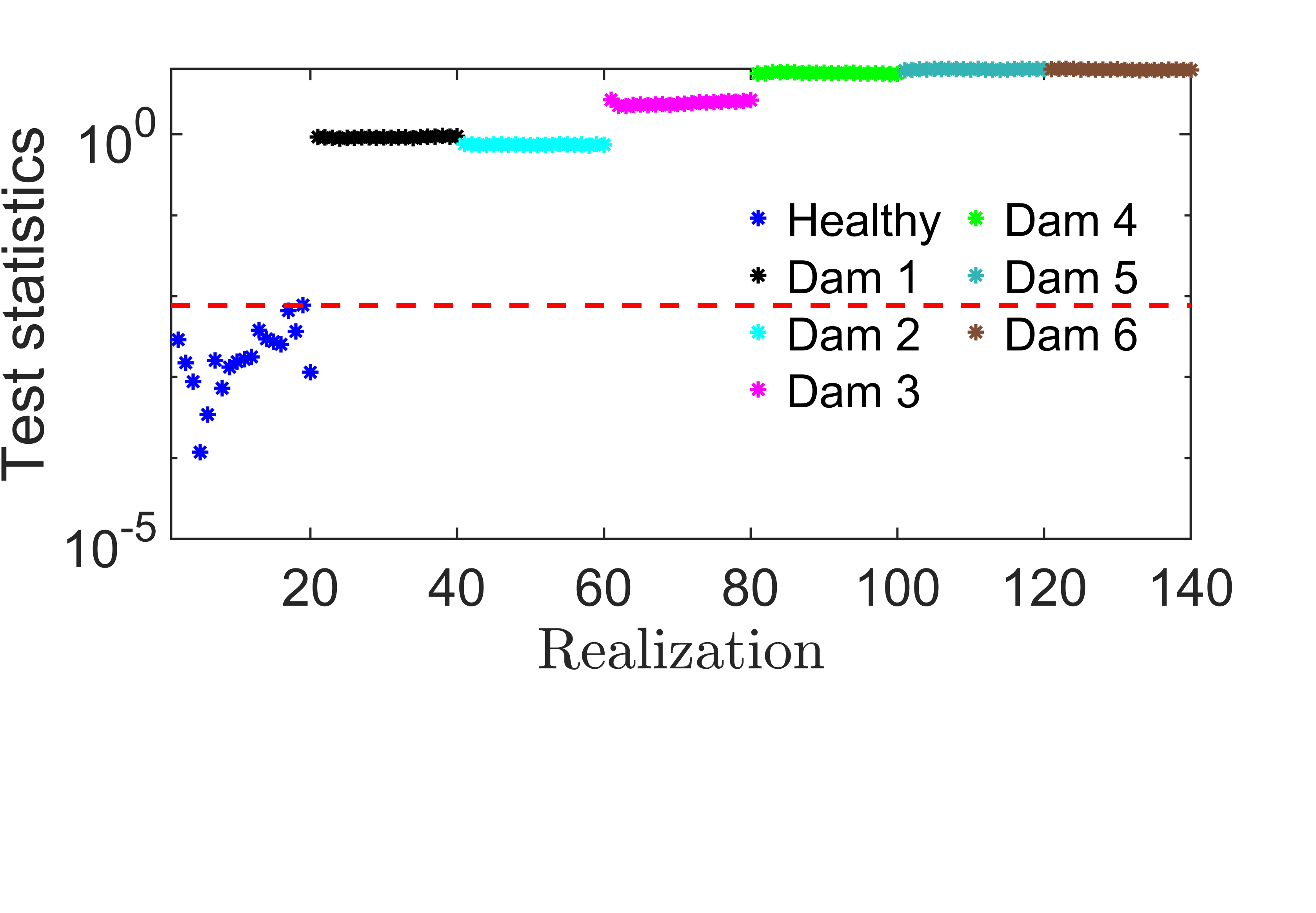}}
    \put(275,10){\includegraphics[width=0.36\columnwidth]{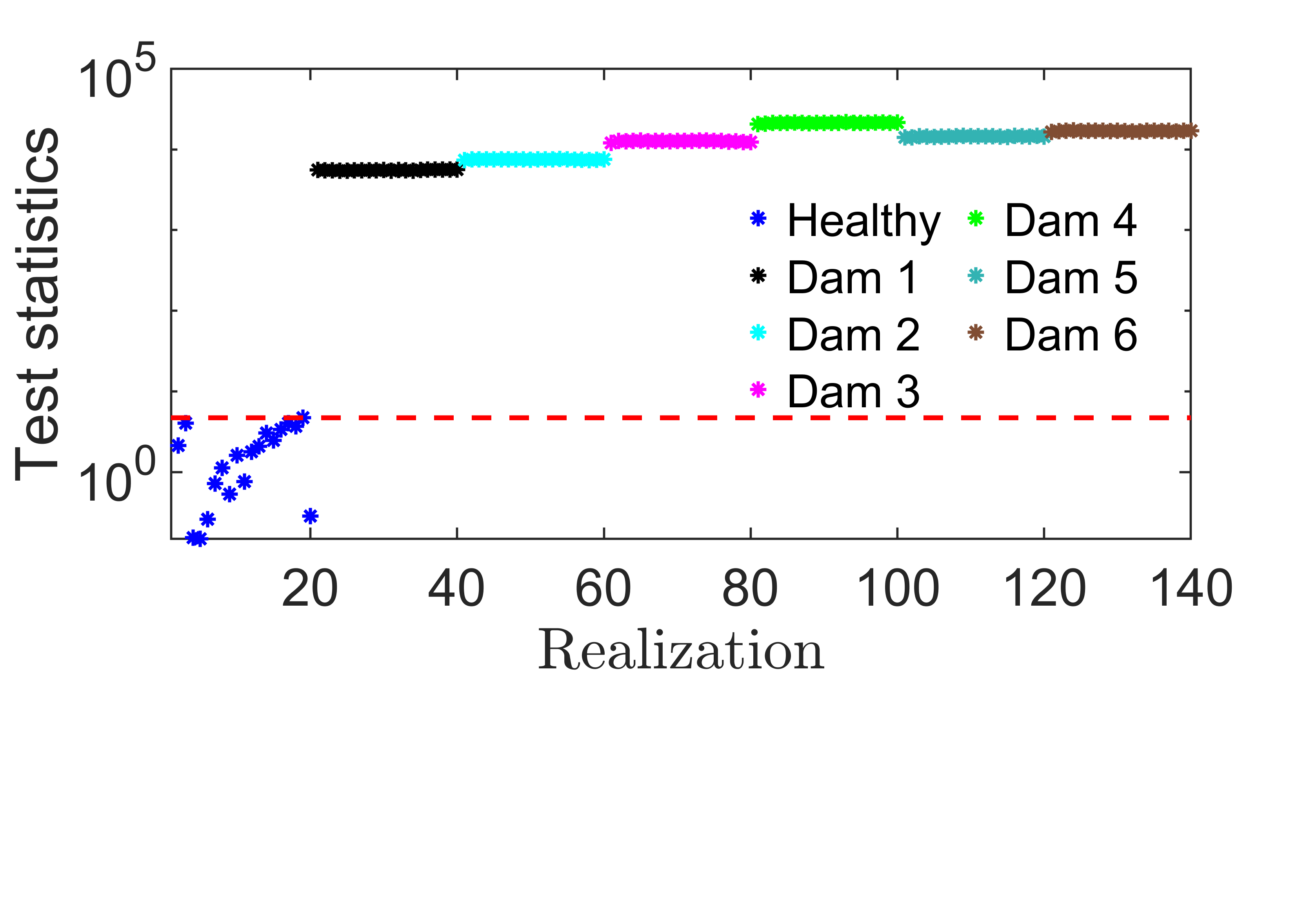}}

    \put(-20,125){\textbf{(a)}}
    \put(140,125){\large \textbf{(b)}}
    \put(290,125){ \large \textbf{(c)}}
    \end{picture}
    \vspace{-1.5cm}
    
    \caption{Damage detection performance comparison for damage intersecting path 5-2 for the CFRP plate using the AR($6$)-based covariance matrix: (a) standard AR approach; (b) SVD-based approach; (c) PCA-based approach. } 
\label{fig:dam intersect theory composite P23} 
\end{figure} 

\begin{figure}[t!]
    \centering
    \begin{picture}(400,130)
    \put(-40,-50){ \includegraphics[width=0.55\columnwidth]{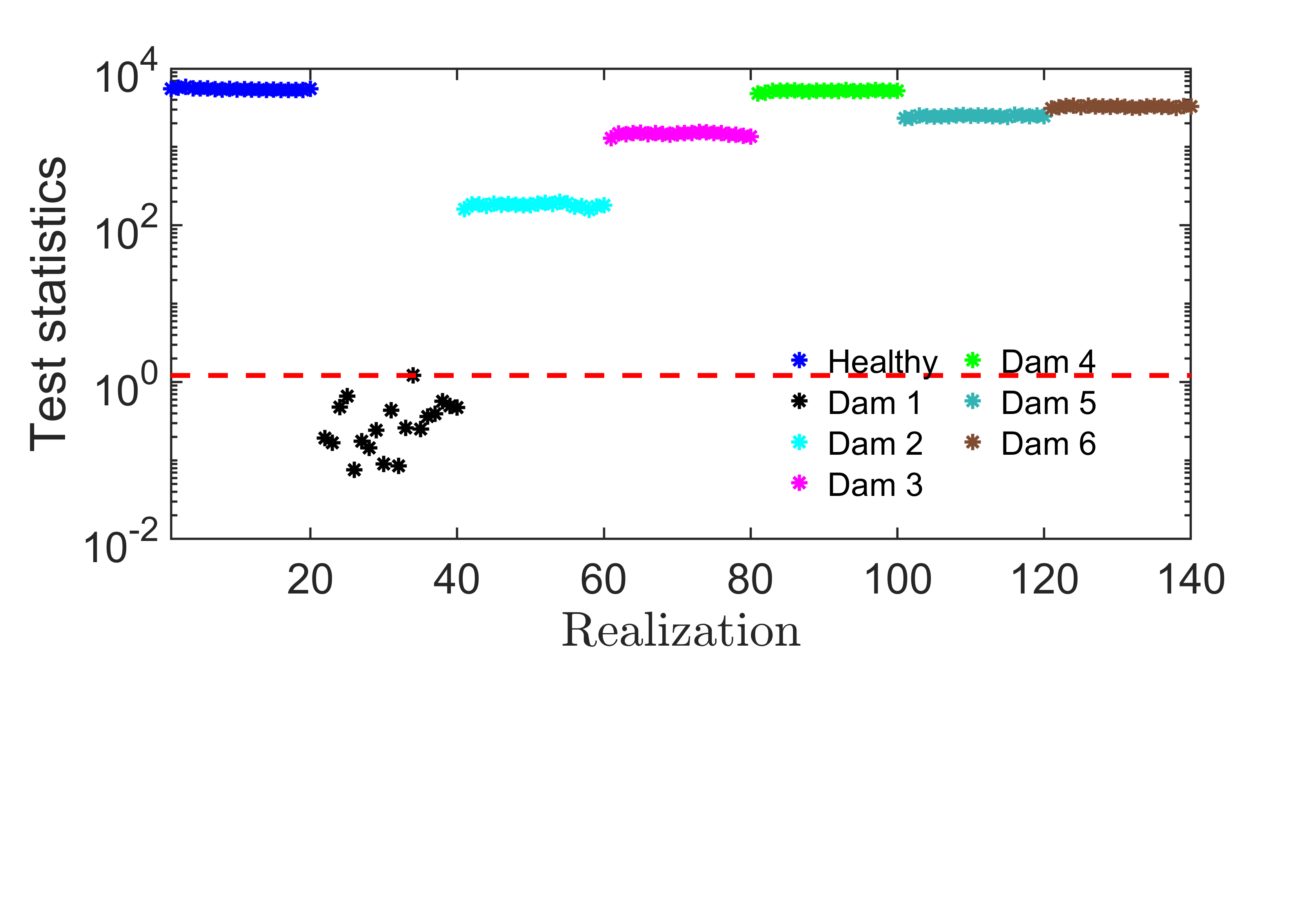}}
    \put(200,-50){\includegraphics[width=0.55\columnwidth]{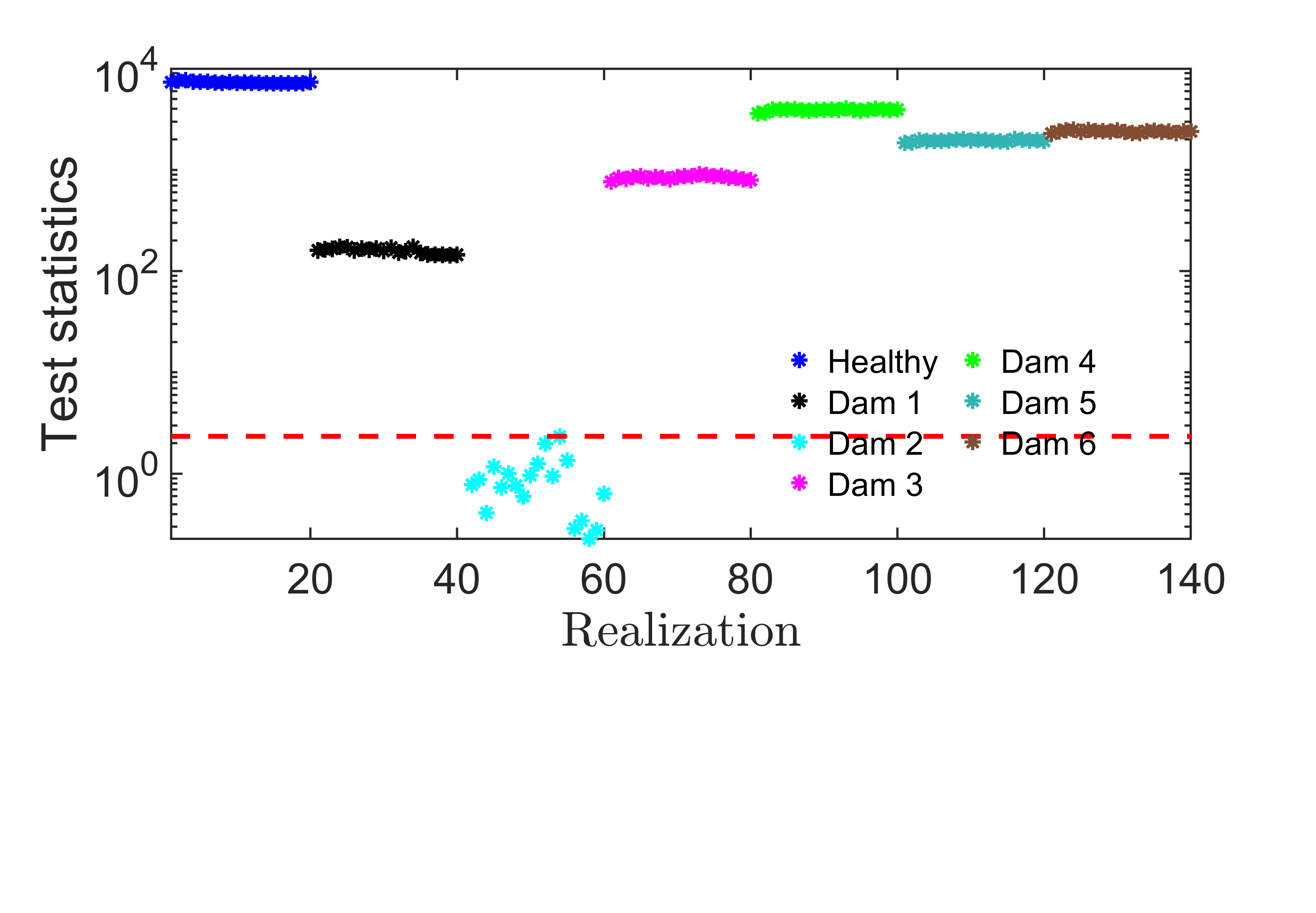}}
    
    \put(-40,-185){ \includegraphics[width=0.55\columnwidth]{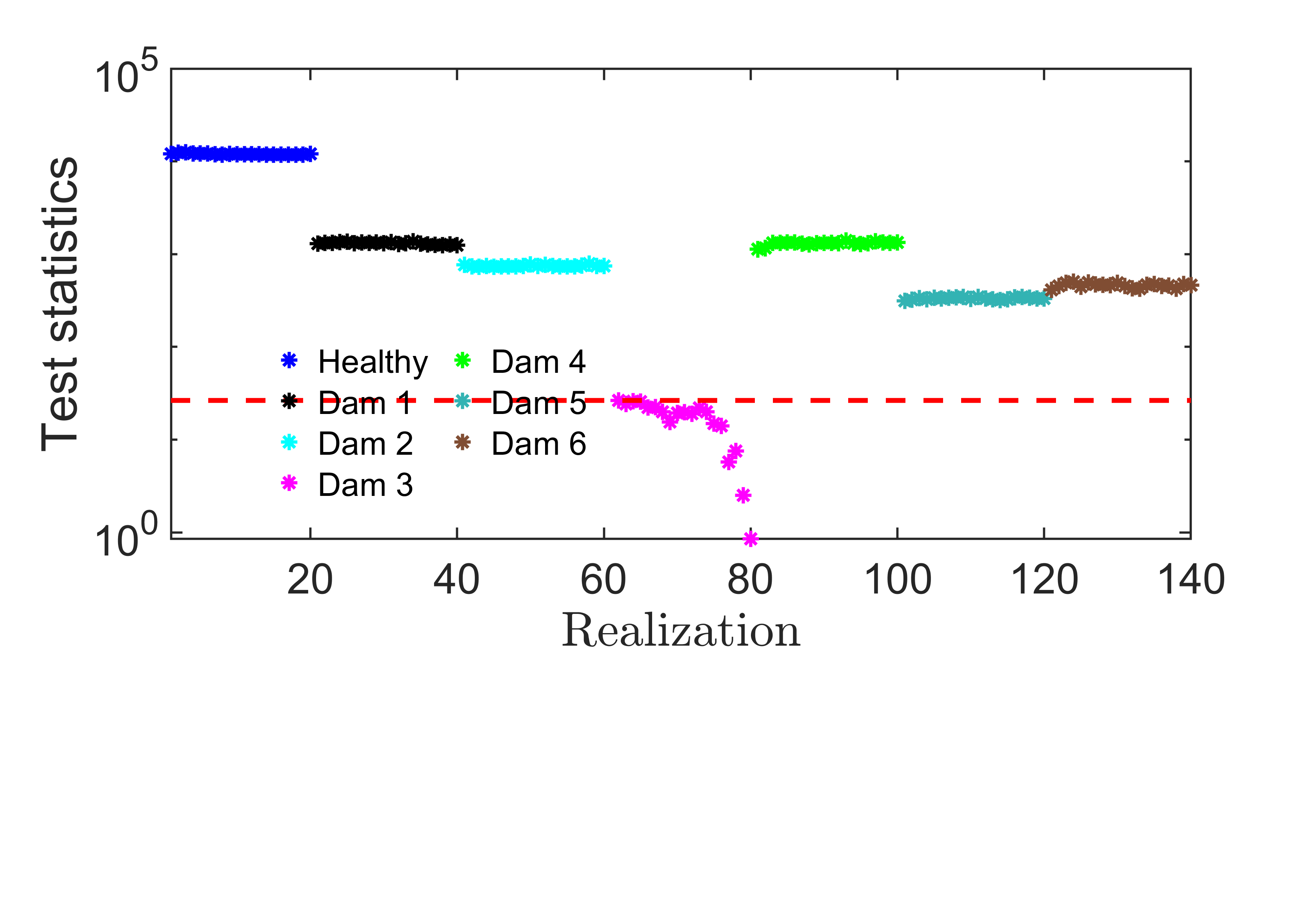}}
    \put(200,-185){\includegraphics[width=0.55\columnwidth]{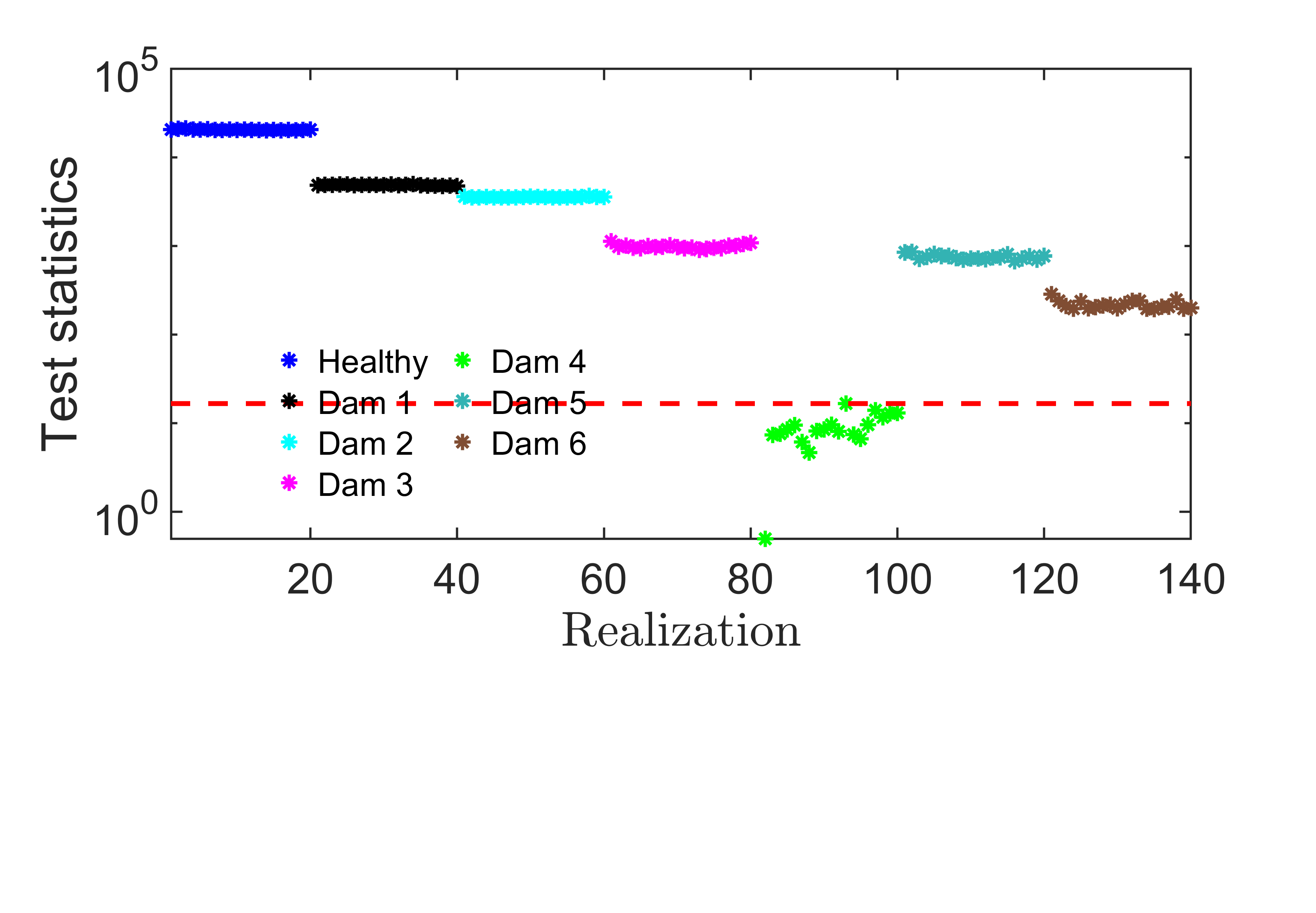}}
    
    \put(-40,-320){ \includegraphics[width=0.55\columnwidth]{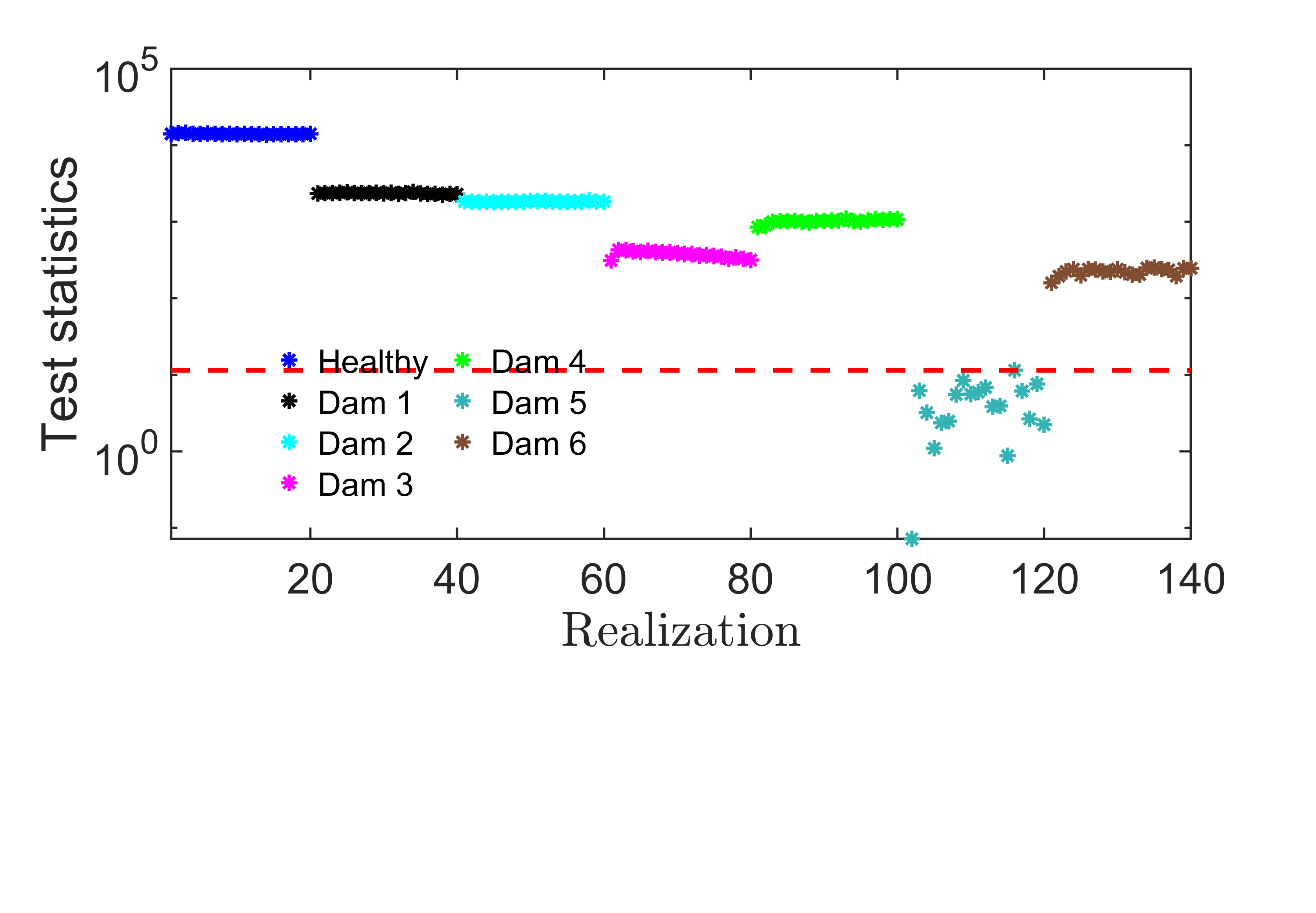}}
    \put(200,-320){\includegraphics[width=0.55\columnwidth]{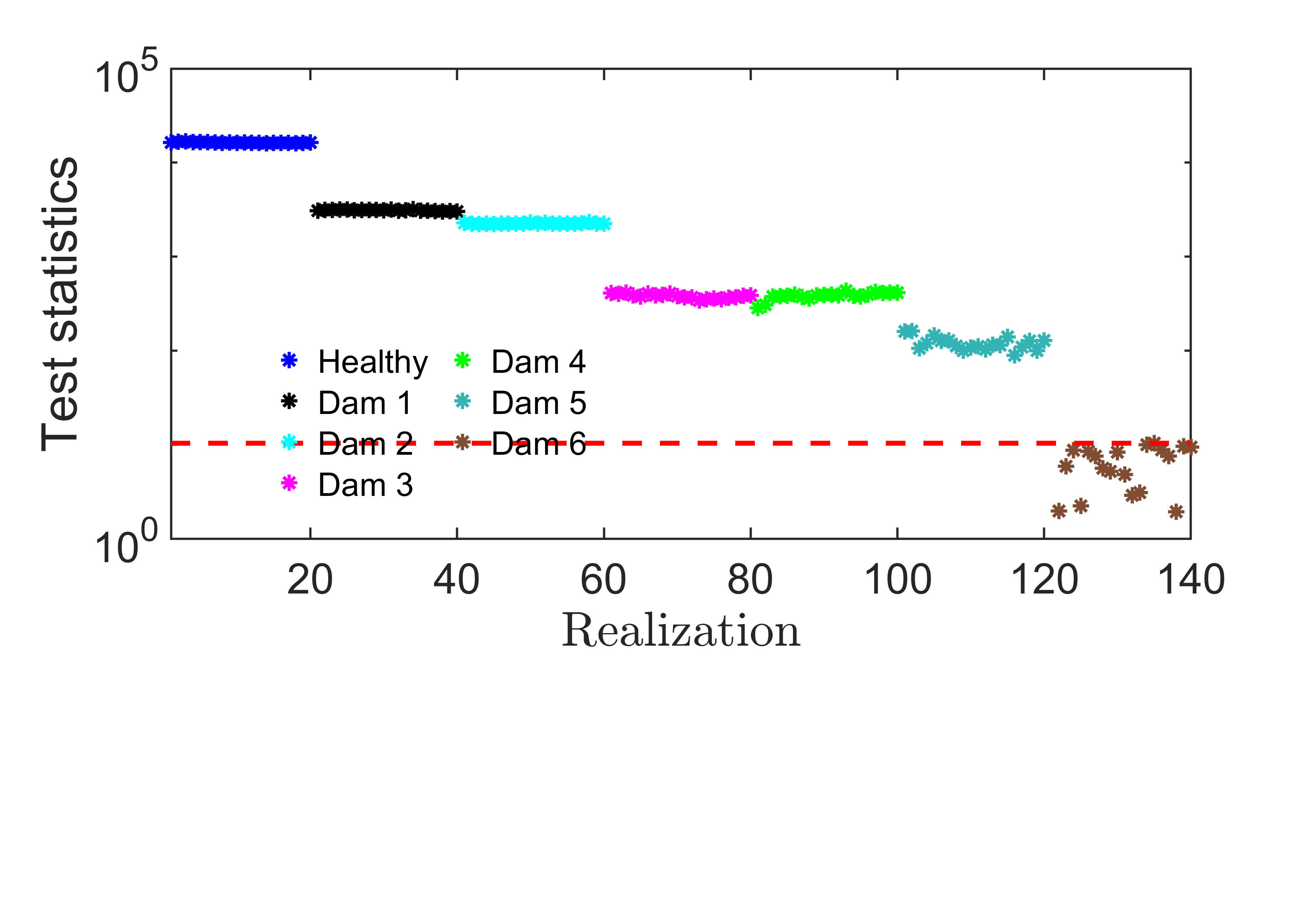}}

    \put(-15,127){\large \textbf{(a)}}
    \put(220,127){ \large \textbf{(b)}}
    \put(-15,-9){\large \textbf{(c)}}
    \put(220,-9){ \large \textbf{(d)}}
    \put(-15,-142){\large \textbf{(e)}}
    \put(220,-142){ \large \textbf{(f)}}
    \end{picture}
    \vspace{9cm}
    \caption{Damage identification results for the CFRP plate using damage intersecting path 5-2 for the PCA-based approach and using AR($6$)-based covariance: (a) damage level 1; (b) damage level 2; (c) damage level 3; (d) damage level 4; (e) damage level 5; (f) damage level 6.} 
\label{fig:dam iden theory composites P23} 
\end{figure}
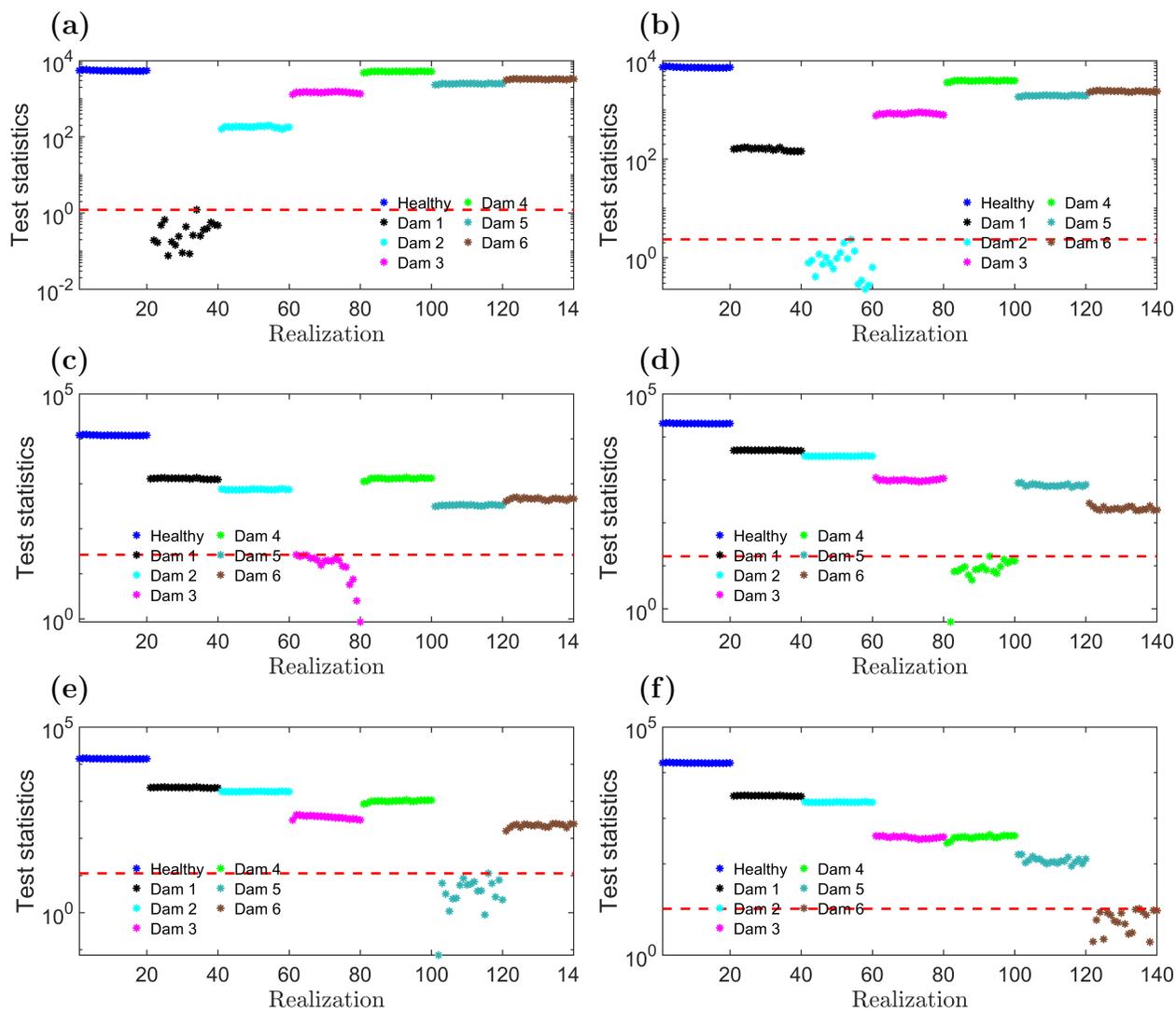

 \end{document}